\documentclass[11pt]{article}
\usepackage{jheppub}
\usepackage{graphicx}
\usepackage[space]{grffile}
\usepackage{latexsym}
\usepackage{amsfonts,amsmath,amssymb}
\usepackage{url}
\usepackage{soul}
\usepackage[utf8]{inputenc}
\usepackage{hyperref}
\hypersetup{colorlinks=true,pdfborder={0 0 0}}
\usepackage{textcomp}
\usepackage{longtable}
\usepackage{multirow,booktabs}
\usepackage{amsfonts}
\usepackage{datetime}
\usepackage{caption}
\usepackage{subcaption}

\usepackage{tikz}
\usetikzlibrary{%
    decorations.pathreplacing,%
    decorations.pathmorphing%
}
\usepackage{mathtools}
\usepackage{enumitem}
\usetikzlibrary{calc}

\newcommand{\lsim}{\mathrel{\hbox{\rlap{\lower.75ex \hbox{$\sim$}} \kern-.3em \raise.4ex \hbox{$<$}}}}
\newcommand{\gsim}{\mathrel{\hbox{\rlap{\lower.75ex \hbox{$\sim$}} \kern-.3em \raise.4ex \hbox{$>$}}}}
\begin{document}

\hfill{\tt BNL-223115-2022-JAAM}

\title{Parton distribution function uncertainties in theoretical predictions for far-forward tau neutrinos  at the Large Hadron Collider}

\author[a]{Weidong Bai,}
\author[b]{Milind Diwan,}
\author[c]{Maria Vittoria Garzelli,}
\author[d]{Yu Seon Jeong,}
\author[b,e]{Karan Kumar,}
\author[f]{Mary Hall Reno}

\affiliation[a]{School of Physics, Sun Yat-sen University, 
Guangzhou, Guangdong 510275, P. R. China}
\affiliation[b]{Brookhaven National Laboratory, Upton, New York, USA}
\affiliation[c]{Universit\"at Hamburg, II Institut f\"ur Theoretische Physik, \\
Luruper Chaussee 149, D-22761, Hamburg, Germany}
\affiliation[d]{High Energy Physics Center, Chung-Ang University,
Dongjak-gu, Seoul 06974, Republic of Korea}
\affiliation[e]{Department of Physics and Astronomy, Stony Brook University, Stony Brook, NY 11794, USA}
\affiliation[f]{Department of Physics and Astronomy, University of Iowa, Iowa City, IA 52242, USA}

\emailAdd{baiwd3@mail.sysu.edu.cn}
\emailAdd{diwan@bnl.gov}
\emailAdd{maria.vittoria.garzelli@desy.de}
\emailAdd{yusjeong@cau.ac.kr}
\emailAdd{fnu.karankumar@stonybrook.edu}
\emailAdd{mary-hall-reno@uiowa.edu}

\date{\today}
\abstract{
New experiments dealing with neutrinos in the far-forward region at the Large Hadron Collider (LHC) are under design or already 
in preparation. Two of them, FASER$\nu$ and SND@LHC, 
are expected to be active during Run 3 and have the potential to detect the interactions of $\nu$ and $\bar{\nu}$ that come from high-energy collisions in one of the LHC interaction points, extracted along the direction tangent to the beam line.
Tau neutrinos and antineutrinos come pre\-do\-mi\-nantly from $D_s^\pm$ production in $pp$ collisions, followed by the leptonic decay of these mesons.
Neutrino pseudorapidities in the range of $\eta>6.9$ and $\eta>8.9$ are relevant to these future experiments. At such pseudorapidities at high energies, QCD theoretical predictions for the flux of $\nu_\tau$ plus $\bar{\nu}_\tau$ rely on parton distribution functions (PDFs) in a combination of very small and large parton$-x$ values. 
We evaluate PDF uncertainties 
affecting
the flux of $\nu_\tau$ + $\bar{\nu}_\tau$ produced by $D_s^\pm$ decay in the far forward region at the LHC. Next-to-leading order (NLO) QCD uncertainties are included in the calculation of $D_s^\pm$ production and NLO PDF sets are used for consistency.
The theoretical uncertainty associated with the 40 PDF sets of the PROSA19 group amounts to $\pm (20-30)$\% for the ($\nu_\tau$ + $\bar{\nu}_\tau$) number of charged-current (CC) events. Scale uncertainties are much larger, resulting in a range of CC event predictions from $\sim 70\%$  lower to $\sim 90\%$ higher than the central prediction. A comparison of the predictions with those obtained using as input the central PDFs from the 3-flavour NLO PDF sets of the CT14, ABMP16 and NNPDF3.1 collaborations show that far-forward neutrino energy distributions vary by as much as a factor of $\sim 2-4$ relative to the PROSA19 predictions at TeV neutrino energies. The Forward Physics Facility in the high luminosity LHC era will  provide data capable of constraining NLO QCD evaluations with these PDF sets. 
}
\dedicated{\today}
\maketitle

\flushbottom
\section{Introduction}

Forward production of neutrinos at high-energy hadron colliders has been under discussion for several decades~\cite{DeRujula:1984pg,Winter:1990ry,DeRujula:1992sn,Vannucci:1993ud}. In particular, $pp$ collisions at the Large Hadron Collider (LHC) interaction points can be regarded as sources of beams of neutrinos of different flavours, with an energy spectrum peaked in the hundreds of GeV - TeV energy range, the highest energies ever reached for a neutrino beam in a human-made accelerator. Measurements of the interactions of these neutrinos in detectors placed 100's of meters from the LHC interaction points are on their way to become a reality with recent proposals for LHC experiments, for example, by the FASER$\nu$~\cite{Abreu:2019yak,Abreu:2020ddv}, XSEN~\cite{Buontempo:2018gta,Beni:2019pyp} and SND@LHC~\cite{Ahdida:2020evc} collaborations. The FASER$\nu$ and SND@LHC experiments were both approved and are expected to take data during Run~3, in the two service tunnels TI12 and TI18 
located at a distance of $\sim$~480~m from the ATLAS interaction point (IP), on opposite sides.
Already a 29 kg pilot detector was placed in TI18 for four weeks in 2018. With 12.2 fb$^{-1}$ of data from $pp$ collisions at $\sqrt{s}=13$ TeV, neutrino interaction candidates were claimed to be observed \cite{FASER:2021mtu}, with muon neutrino interactions expected to dominate inside the sample.
Beyond the pilot and the first phase of 
experiments, the idea of a forthcoming Forward Physics Facility (FPF), capable of hosting a suite of far-forward experiments active during the HL-LHC phase, has been recently raised, gaining increasing attention \cite{Anchordoqui:2021ghd}. First studies in this direction are ongoing, considering the options of either enhancing the dimensions of one of the two aforementioned tunnels by enlarging it with alcoves, or building a new bigger cavern and access tunnel at approximately $\sim$ 500-600 m 
from the ATLAS IP \cite{Feng:2022inv}. In that cavern there would be enough space to host various experiments using different technologies at different distances from the beam-axis, significantly larger detectors, and with the completion of the HL-LHC phase, 20 times the integrated luminosity expected during Run 3. All these experiments will be sensitive to neutrinos and antineutrinos arising from the decay of particles and hadrons produced in the ATLAS IP, and propagating along the tangent to the accelerator arc. 
In this paper, we focus on the tau neutrino plus antineutrino flavour case. Differently from other neutrino flavours, tau neutrinos and antineutrinos are not, or are only very rarely, produced in pion and kaon decays because $\tau^-\bar{\nu}_\tau$ final states 
are kinematically forbidden;  $K\to \pi\nu_\tau\bar{\nu}_\tau$ decays require  flavour-changing neutral currents, and two body meson decays to $\nu_\tau\bar\nu_\tau$ are suppressed by angular momentum considerations \cite{E949:2005efl,NA62:2021zjw}. They are instead produced predominantly via $D_s^+$ (and charge conjugate)  production and its leptonic decay $D_s^+\to \tau^+\nu_\tau$. The tau neutrino plus antineutrino flux along the beam line will provide the opportunity to make direct tests of lepton flavour universality and to constrain new physics signals in tau neutrino oscillations over the considered baselines of several hundred meters \cite{Bai:2020ukz}. In what follows, we refer to both neutrinos and antineutrinos generically as ``neutrinos."

Making a robust prediction of charm hadron production and the corresponding tau neutrino fluxes at very large rapidities can be regarded as a theoretical challenge
\cite{Park:2011gh,Bai:2020ukz}. The initial FASER$\nu$ experiment is planned for $\eta_\nu>8.9$ \cite{Abreu:2020ddv}. The XSEN proposal covers the neutrino rapidity ranges $7.4-8.1$ and $8.0-8.6$ \cite{Beni:2020yfy}, whereas the SND@LHC experiment explores the rapidity range $7.2-8.6$ \cite{Ahdida:2020evc}. These rapidity values follow from space considerations and the limited dimensions of the already available caverns. The FPF could extend the range towards lower rapidities, to an extent that will depend on its dimensions, on the dimension of the experiments included in it and on the distance from the IP. In the aforementioned hypothesis of a new purpose-built cavern, it is foreseen that neutrino rapidities down to at least 6.5 can be covered with sufficient statistics. Lower rapidities down to $\sim$ 5.6 could be covered as well by some of the experiments, however at the price of  reduced statistics, considering an  area of $\sim$ 1 m$^2$ foreseen for the detectors that will be put off-axis. 

On the other hand, charm hadron production has been measured by LHCb in rapidity intervals in the range $2.0 < y < 4.5$ for different center-of-mass energies ($\sqrt{s}$ = 5, 7 and 13 TeV) \cite{Aaij:2016jht,LHCb:2013xam,Aaij:2015bpa}. As we will see in the following, this ensures some overlap with the rapidity range seen by the far-forward experiments, considering that neutrinos at high-rapidity come from the decay of $D$-mesons from a range of rapidities, including lower rapidities with respect to the neutrino rapidity. Using next-to-leading order (NLO) QCD \cite{Nason:1987xz,Nason:1989zy,Mangano:1991jk} and a phenomenologically motivated transverse momentum smearing function, theoretical predictions for open heavy-flavour hadroproduction have been compared with the LHCb data \cite{Aaij:2015bpa}, to extract the optimal values of some of the input parameters to be used in the theoretical calculations. Theoretical calculations with input parameters constrained by the LHCb results, have then been used to make predictions in the whole rapidity range explored by far-forward experiments, including higher rapidities \cite{Bai:2020ukz}. This can be regarded as an approximate extrapolation procedure.

Very forward heavy-flavour production evaluated in the QCD-improved parton model probes both small- and large-$x$ parton distribution functions (PDFs), where $x$ is the fraction of longitudinal momentum of the proton that goes to the interacting parton. Theoretical uncertainties in the PDFs are particularly large in both these $x$ regimes, considering that the bulk of HERA $ep$ DIS data that form the backbone of PDF fits are limited to the $x$ range $10^{-4} \lesssim x \lesssim 10^{-1}$ \cite{H1:2015ubc,H1:2018flt}. Some additional data concerning the $F_L$ structure functions have also been obtained at HERA. In principle, this would allow to extend the $x$-coverage
down to $x\sim 10^{-5}$. However, due to the limited statistics of these data, they are either not used in the PDF fits or have a very small constraining power. 

In this paper, we evaluate PDF uncertainties in the NLO QCD calculation of $pp\to D_s^\pm X$ with decays to $\nu_\tau$ and $\bar{\nu}_\tau$, using as a basis the 40 PDF sets of the most recent PDF fit delivered by the PROSA collaboration~\cite{Zenaiev:2019ktw}, an update including more sets of experimental data and updated theory input with respect to the first PROSA PDF fit~\cite{PROSA:2015yid}.
In particular, this fit, besides HERA data, includes LHCb data on open-heavy flavour hadroproduction in $pp$ collisions at center-of-mass energies up to $\sqrt{s} = 13$ TeV, which has allowed to extend the $x$ range down to $x \sim 10^{-6}$. These data have also an impact on the large $x$ range (i.e., $10^{-1} < x < 1$). Additionally, it includes ALICE $D$-meson production data, with impact on more central $x$ values, very useful to cross-check the constraints of DIS data + sum rules, playing a role in the same region. The PROSA fit has been first performed in the decoupling factorization and renormalization scheme, considering three active flavours, which is appropriate in the kinematical regime covered by the bulk of the LHCb and HERA charm production cross-sections.
Besides the fit in the decoupling scheme, a PROSA Variable Flavour Number Scheme PDF set has also been developed and released in Ref.~\cite{Zenaiev:2019ktw}. For the sake of completeness, we observe that a large fraction of the LHCb $D$-meson production data included in the PROSA PDF fit have also been accounted for in independent PDF sets~\cite{Gauld:2015kvh, Gauld:2016kpd, Bertone:2018dse}, built by applying a reweighting procedure on top of the NNPDF3.0~\cite{NNPDF:2014otw} and NNPDF3.1~\cite{Ball:2017nwa} NLO PDF fits. The PROSA collaboration has explicitly checked the consistency, within uncertainties, of the gluon PDFs in the last version of their fit with the gluon PDFs in the most updated version of this independent PDF set, for factorization scales of the order of $Q^2 \sim  10$~GeV$^2$ and low $x$ values, as relevant for LHCb charm production. The outcome of these checks supports the robustness of these works and of the underlying heterogeneous methodologies.

In the following, for the process we are interested in, namely forward $\nu_\tau$ and $\bar{\nu}_\tau$ production through $D_s$ 
production and decay in $pp$ collisions at the LHC, the PDF uncertainties are compared with the uncertainties associated with renormalization and factorization scale variation. The energy distributions of $\nu_\tau+\bar{\nu}_\tau$ in different rapidity ranges are compared with those obtained using central sets of other NLO PDF fits as well \cite{Dulat:2015mca,Alekhin:2018pai,Ball:2017nwa,Harland-Lang:2015qea}, which do not include any constraint from charm hadroproduction data. We also evaluate the associated uncertainties affecting the $\nu_\tau N$ and $\bar{\nu}_\tau N$ charged current DIS cross sections, considering that in the foreseen far-forward experiments $\nu_\tau$ and $\bar{\nu}_\tau$ will be detected thanks to their DIS interactions with appropriate targets. 
This work aims at the twofold purpose of a) updating the predictions for the number of events in far-forward experiments presented in our previous work~\cite{Bai:2020ukz}, by considering more reliable experimental setups, better motivated inputs for the theory calculations, and an improved description of fragmentation; b) providing for the first time an evaluation of PDF uncertainties affecting these predictions, comparing their size to those of other important QCD uncertainties. 

The outline of this paper is as follows: in Section~\ref{sec:input} the ingredients of the QCD theoretical calculation of charm hadron production are presented, with predictions including NLO radiative corrections compared to the available LHCb data at $\sqrt{s}=13$ TeV. We also make quantitative comparisons of charm production in different rapidity ranges, corresponding to different parton-$x$ ranges. The tau neutrino plus antineutrino rapidity and energy distributions are presented in Section \ref{sec:neutrinos}, with a focus on PDF uncertainties and the role of scale choice and transverse momentum smearing.    Section~\ref{sec:events} shows the number of events as a function of energy and in total for different rapidity ranges and detector masses, with an assessment of PDF uncertainties associated with neutrino interactions assuming a tungsten target. 
Finally, in Section~\ref{sec:conclu} we draw our conclusions and discuss the outlook for future developments. Appendix 
\ref{sec:errorbands} describes the procedure for combining PDF uncertainties from various sources. Appendix \ref{sec:tables} reports numerical tables for the tau neutrino plus antineutrino energy distributions for several neutrino rapidity ranges, including QCD scale and PDF uncertainties, and tau neutrino and antineutrino charged-current cross sections per nucleon for a tungsten target. These tables are available in ascii format in the ancillary files of the preprint in the arXiv web archive\footnote{\href{https://arxiv.org/src/2112.11605v1/anc}{https://arxiv.org/src/2112.11605v1/anc}}.

\section{Charm hadron production at the LHC}
\label{sec:input}

\subsection{Input factors}

A detailed description of the inputs to the NLO QCD evaluation of charm hadron production and decay at the LHC is provided in Ref. \cite{Bai:2020ukz}. For completeness, we summarize the main features here. 
Our QCD evaluation of charm production is performed at NLO using the one-particle inclusive results and formulas for the parton-level hard-scattering cross sections 
first published by Nason, Dawson and Ellis in Ref. \cite{Nason:1989zy}. 

Transverse momentum smearing is implemented phenomenologically by a Gaussian smearing function according to 
\begin{equation}
\frac{d^2\sigma({\rm NLO})}{dp_T\, dy } = \int d^2 \vec{k}_T\, \frac{1}{\pi\langle k_T^2\rangle}\exp[{-k_T^2/\langle k_T^2\rangle}] \frac{d^2\sigma({\rm NLO})}{dq_T\, dy}\Biggl|_{q_T=|\vec{p}_T-\vec{k}_T|} \, ,
\end{equation}
where $\langle k_T^2\rangle= 4\langle k_T\rangle^2/\pi$ is the effective transverse momentum squared that accounts both for intrinsic transverse momentum and as a phenomenological stand-in for higher order QCD effects.
Our default is $\langle k_T\rangle=0.7$ GeV \cite{Bai:2020ukz}, 
a value that makes our predictions for $D$-meson energy spectra very similar to those of the POWHEG \cite{Frixione:2007vw}+PYTHIA \cite{Sjostrand:2019zhc} implementation used in Ref.~\cite{Bai:2020ukz}.

\begin{figure}
\begin{centering}
\includegraphics[width=0.49\textwidth]{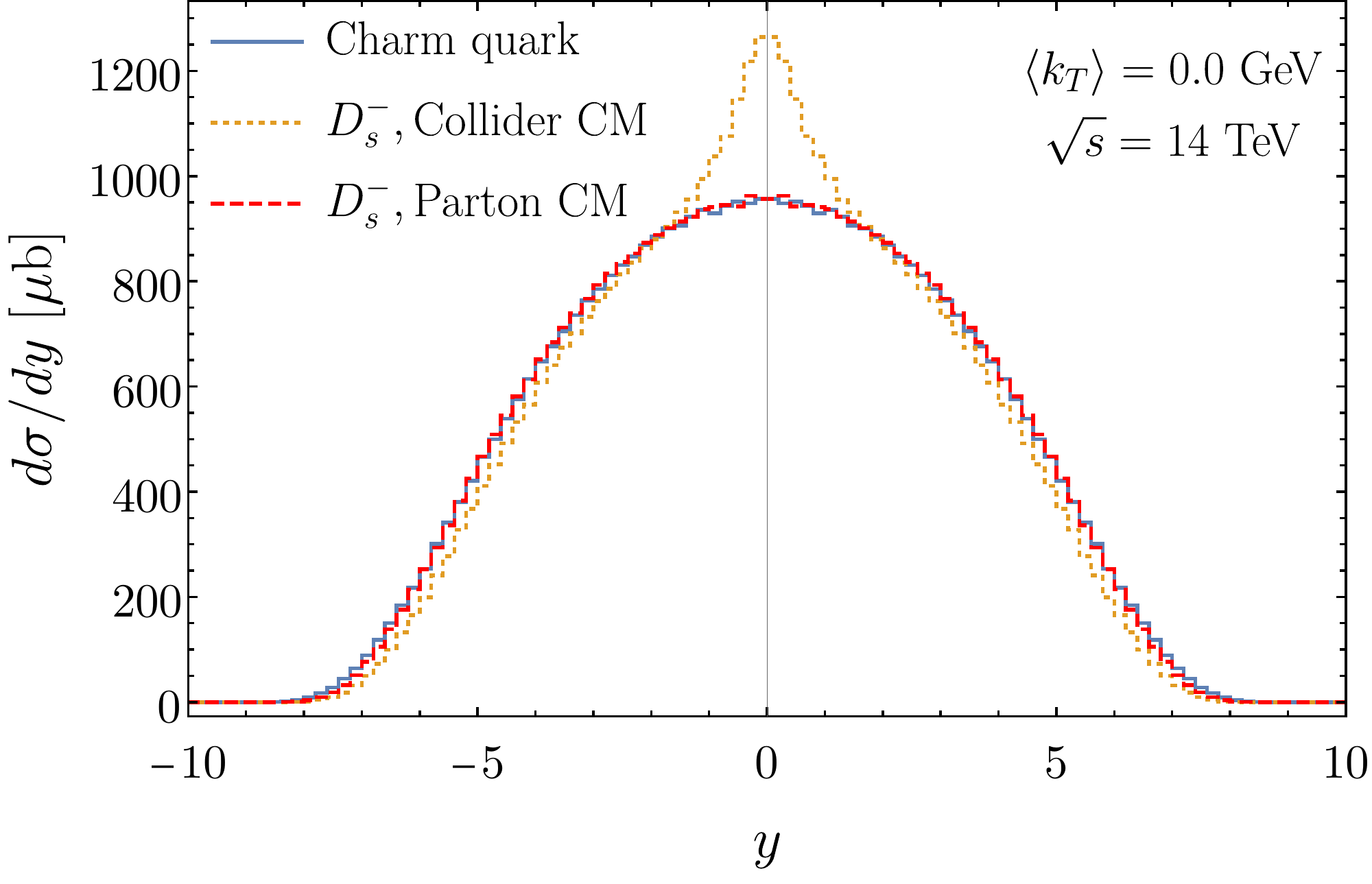}
\includegraphics[width=0.49\textwidth]{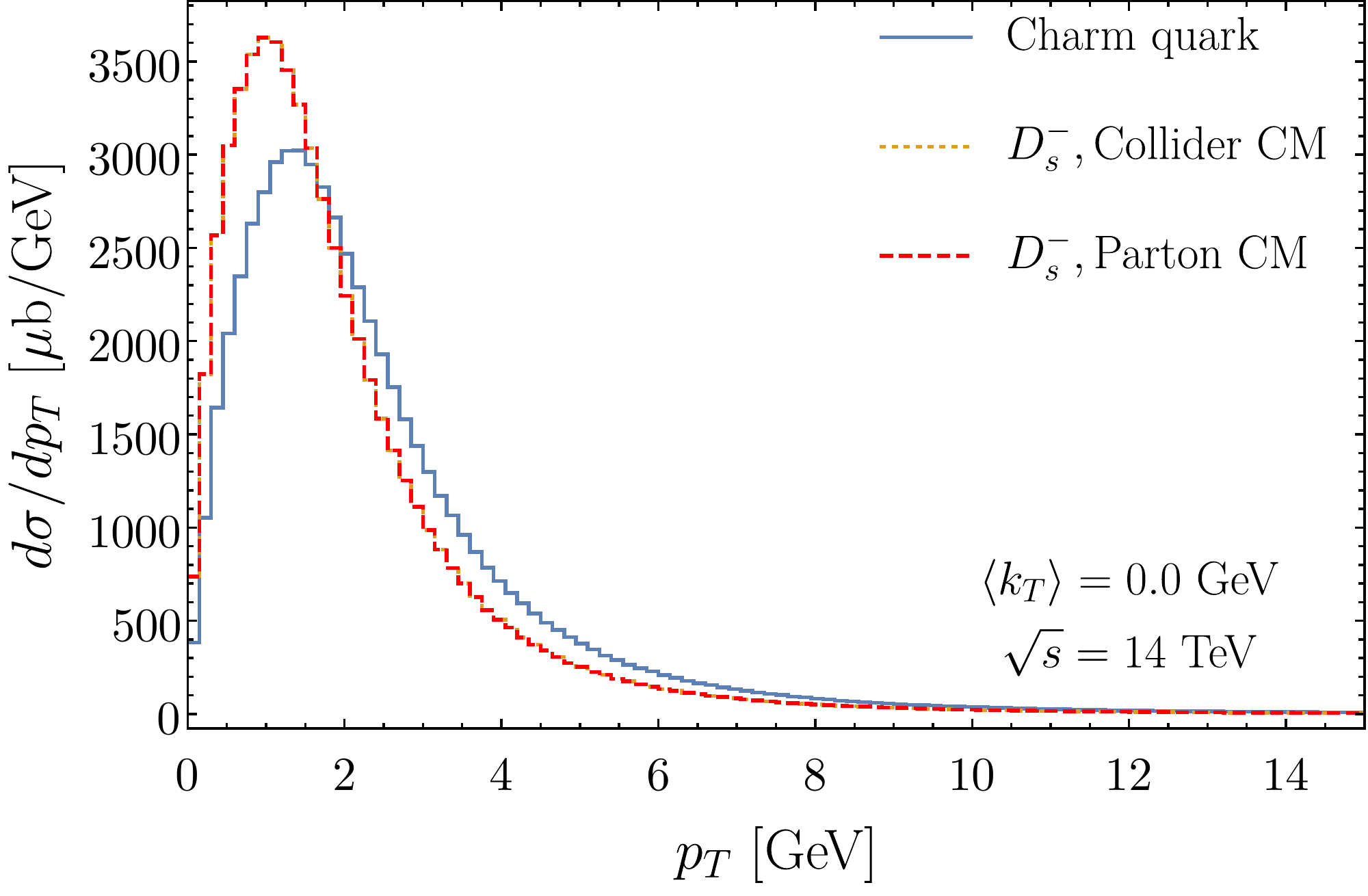}
\includegraphics[width=0.49\textwidth]{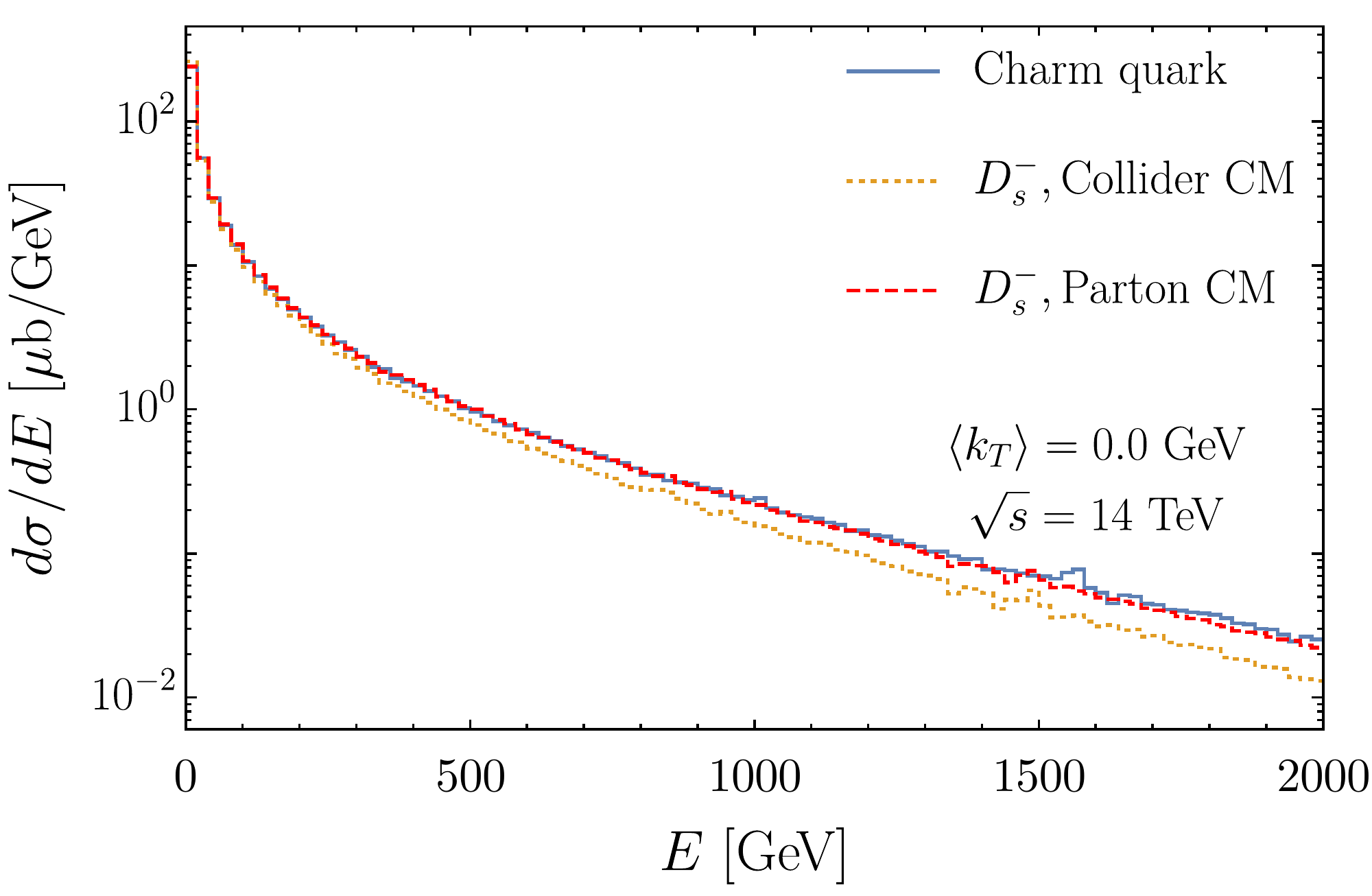}
\par\end{centering}
\caption{Effect of two different implementations of the fragmentation of a $c$-quark into a $D_s^-$ meson, in the $D_s^-$ rapidity $y$ (upper left), transverse momentum $p_{T}$ (upper right) and energy $E$ distributions (lower) at a $pp$ collider with $\sqrt{s}=14$ TeV, for $\left\langle k_{T}\right\rangle = 0$ GeV. 
For facilitating the comparison with the $c$-quark distributions, the fragmentation fraction of charm quark to $D_s^-$ is set to 1. In the second plot, the collider and parton CM frame $p_T$ distributions are identical. 
See text for more detail on the fragmentation options in the collider CM frame and parton CM frame.
\label{fig:old-new-frag}}
\end{figure}

We use the Peterson fragmentation functions for the $c$-quark to charm meson transition~\cite{Peterson:1982ak}. One approach to implement the fragmentation is to perform the reduction
of the charm quark 3-momentum $\vec{p}_{Q}$ (by $\vec{p}_{H}=z\vec{p}_{Q}$)
in the collider center of mass  (CM) frame. Such an implementation generates
a significant peak at $y_{H}=0$ in the hadron rapidity distribution,
as can be seen in Fig.~\ref{fig:old-new-frag}. This is caused by
the larger mass of the heavy hadron compared to that of the heavy
quark, $m_{H}>m_{Q}$, and also by the fragmentation which requires
$\left\langle 1/z\right\rangle >1$. 
However, such a significant peak
at $y_{H}=0$ is not found in the experimental observations. 
Alternatively, the fragmentation
can be performed with the heavy hadron receiving fraction $z$ of $\vec{p}_
Q$
in the CM frame of the initial state partons and then a boost is used to bring the heavy hadron to the collider CM frame. In such an
implementation, the unwanted peak at $y_{H}=0$ is highly suppressed due to the further integral  over $y_{\textrm{cm}}$,  the rapidity of the system of the two initial-state partons. The transverse momentum distributions are identical under these two implementations since the two reference frames are related by a longitudinal boost, and the transverse momentum does not change with such a boost, i.e., $\vec{p}_{HT}= z \vec{p}_{QT}$ in both implementations.  

A comparison of the energy distributions obtained using the two fragmentation implementations
is similar to that of the rapidity distributions, but under
the second implementation (in the parton CM frame), the energy of the hadron is a function of not only the energy of the heavy quark and $z$, but also the heavy quark rapidity and $y_{\textrm{cm}}$. 
This makes the energy distribution of the hadron in the parton CM implementation of fragmentation closer to that of the heavy quark energy distribution than in the case of collider CM  implementation of fragmentation.
Of course, the distributions obtained in the two implementations are all identical in the massless limit. In 
this work, we use the implementation of fragmentation in the parton CM frame. Implementation in the parton CM frame more closely resembles the frame where fragmentation functions are extracted in $e^+e^-$ collisions.

The parton level differential cross sections at NLO are convoluted with PDFs for $pp$ collisions.
In this work, the default set of NLO PDFs, in a decoupling scheme with three active flavours consistent with the matrix elements we use, is provided by the PROSA collaboration \cite{Zenaiev:2019ktw} and implemented through the LHAPDF interface \cite{Buckley:2014ana}. The PROSA19 PDFs (in the following we will rename PROSA19 simply as ``PROSA") result from a fit, including, among others, data on open heavy flavour production from HERA, LHCb and ALICE, as already mentioned in the Introduction.
Forty PDF eigenvectors in addition to the best fit account for uncertainties 
associated to the PDFs. They account for i) experimental uncertainties (``fit'' uncertainties), ii) model assumptions in performing the fit, including, among others, $\mu_R$ and $\mu_F$ variations, as well as variation of the parameters entering fragmentation functions and fractions and of the $\alpha_s(M_Z)$ value, iii) parameterization uncertainties related to the functional form of the PDFs at
the starting scale for evolution and the parameters appearing there. 
The prescription for combining the PROSA PDF eigenvectors in order to produce uncertainty bands in kinematic distributions is described in Appendix~\ref{sec:errorbands}. For comparison, we also show the central predictions obtained with other 3-flavour NLO PDFs: the CT14nlo~\cite{Dulat:2015mca}, 
ABMP16~\cite{Alekhin:2018pai} and
NNPDF3.1~\cite{Ball:2017nwa} 3-flavour sets, 
which have already been used for computing heavy-flavour production in previous works (see e.g. Ref.~\cite{Benzke:2017yjn, Accardi:2016ndt}).
Differently from the PROSA PDF sets, these fits do not account for any flavour hadroproduction data. 

For computations with each PDF set, we use as input the associated $\alpha_s(M_Z)$ value, and $\alpha_s$ evolution at two-loops.
The pole mass $m_c$, also input to the NLO calculation, is related to the $\overline{\rm MS}$ mass at the mass scale $\mu=\overline{m}_c(\mu)$ through four loops by the relation \cite{Chetyrkin:1999qi,Marquard:2015qpa,Ball:2016qeg}
\begin{equation}
m_c=\overline{m}_c(1 + 0.424\alpha_s+ 1.046\alpha^2_s+ 3.76\alpha^3_s+ 17.5\alpha^4_s+{\cal O}(\alpha^5_s))\ ,
\end{equation}
where $\alpha_s \equiv \alpha_s(\mu)$.
While converting heavy-quark masses from one mass renormalization scheme to the other, 
we include the first correction in our NLO evaluation. In particular, for the PROSA PDFs, the $\overline{\rm MS}$ charm mass is 1.242 GeV, a value that comes from the simultaneous fit of the $\overline{\rm MS}$ heavy-quark masses and the PDFs performed by the PROSA collaboration in Ref.~\cite{Zenaiev:2019ktw}. This translates into a pole mass of 1.442 GeV. Table \ref{tab:charm-masses} shows the charm quark mass values for the various PDFs used in this work.

\begin{table} 
    \centering
    \begin{tabular}{|l|c|c|}
    \hline
    PDF Set & $\overline{{\rm MS}}$ mass [GeV] & Pole mass [GeV]\\
    \hline
    \hline
     PROSA\textunderscore 2019\textunderscore FFNS \cite{Zenaiev:2019ktw}   & 1.242 & 1.442 \\
     \hline
     CT14nlo\textunderscore NF3 \cite{Dulat:2015mca}  &  -  & 1.3\\
     \hline
     ABMP16\textunderscore 3\textunderscore nlo 
     \cite{Alekhin:2018pai} & 1.175 & 1.376 \\
     \hline
     NNPDF3.1\textunderscore nlo\textunderscore pch\textunderscore as\textunderscore 0118\textunderscore nf\textunderscore 3
     \cite{Ball:2017nwa} & - & 1.51 \\
     \hline
    \end{tabular}
    \caption{The charm quark pole mass, and, if relevant, the $\overline{{\rm MS}}$  mass for different PDF sets. Conversions from the  $\overline{{\rm MS}}$ to the on-shell renormalization scheme are performed at 1-loop.}
    \label{tab:charm-masses}
\end{table}

For the renormalization scale $\mu_R$ and factorization scale $\mu_F$, we adopt two different functional forms, the scale used for the PROSA 2019 fit to heavy flavour production
\begin{equation}
\label{eq:mt2def}
    m_{T,2}^2\equiv \Bigl(p_T^2 + (2m_c)^2\Bigr)\, ,
\end{equation}
and the transverse mass
\begin{equation}
\label{eq:mt1def}
    m_{T,1}^2\equiv m_T ^2= \Bigl(p_T^2 + m_c^2\Bigr)\, ,
\end{equation}
which is more widely used as central $\mu_R$ and $\mu_F$ in many computations of heavy-quark pair production. The use of the modified scale $m_{T,2}^2$ finds its justification in the fact that it reduces the NNLO/NLO $K$-factor in the bulk of the phase-space for $c\bar{c}$ production, with respect to the other option. 
As we will show in the following Section, we verify that NLO scale uncertainties based on $m_{T,2}^2$ are also reduced in the kinematic region relevant for very forward heavy-flavour production 
compared to those obtained with $m_{T,1}^2$,
and that the uncertainty band is more symmetric around the predictions from the central scale.
In our characterization of scale uncertainties, we consider variations around the central scale $(\mu_R,\mu_F)= (1,1)m_{T,2}$, related to scale combinations 
$\{(0.5,1),\ (1,2),\ (1, 0.5),\ (2,1),$ $(0.5,0.5)$ and $(2,2)\}\,m_{T,2}$, by building the envelope of the corresponding predictions \cite{Cacciari:2012ny}. 
On the other hand, the PROSA PDF uncertainties from the 40 sets are determined by fixing $(\mu_R,\mu_F)= (1,1)\,m_{T,2}$, assuming the independence of PDF and scale uncertainties, that can then be added in quadrature, as also suggested by the PROSA collaboration in Ref.~\cite{Zenaiev:2019ktw, PROSA:2015yid}. 
Details of combining uncertainties associated with the 40 PDF sets are included in Appendix \ref{sec:errorbands}. These studies are all performed with $\langle k_T\rangle=0.7$~GeV.
For comparison, we also show predictions with the scale choice $(\mu_R,\mu_F)=(1,2)\,m_T$ and $\langle k_T\rangle=1.2$~GeV, parameters that lead to predictions which compare well with the LHCb data at $\sqrt{s}=$~13~TeV.

The evaluation of $D^+_s\to \nu_\tau$ decays 
is straightforward as the relevant process is the two-body decay $D_s^+\to \tau^+\nu_\tau$, with branching fraction $B=0.0548$~\cite{Tanabashi:2018oca}. The neutrino in this two-body decay is denoted as ``direct'' neutrino. The $\tau$ lepton also decays. Its neutrino is called the ``chain'' neutrino. Details of the implementation of $D_s^+\to \tau^+ \nu_\tau$ and $\tau$ decays appear in, for example, Ref. \cite{Bai:2020ukz}.

\begin{figure}
   \centering
\includegraphics[width=0.70\textwidth]{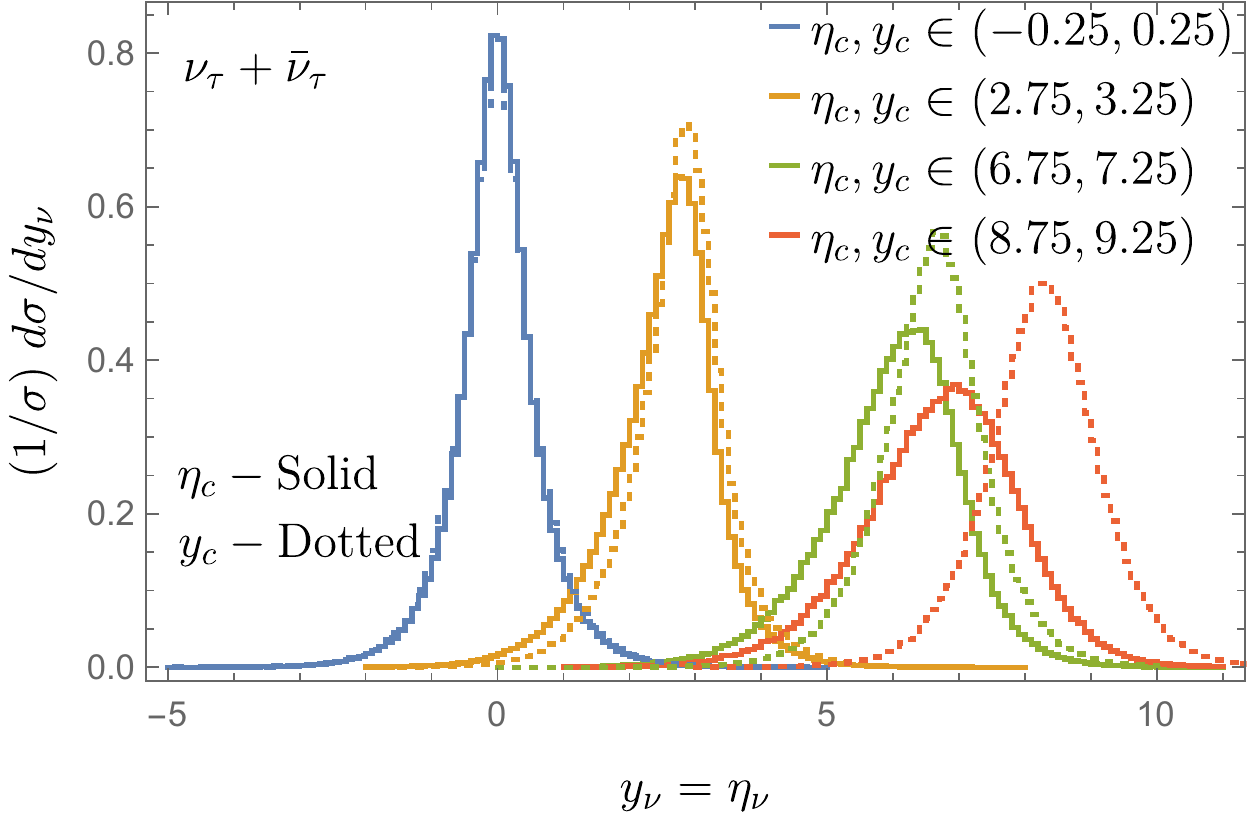}
\includegraphics[width=0.70\textwidth, trim=-1cm -0.8cm 0 0]{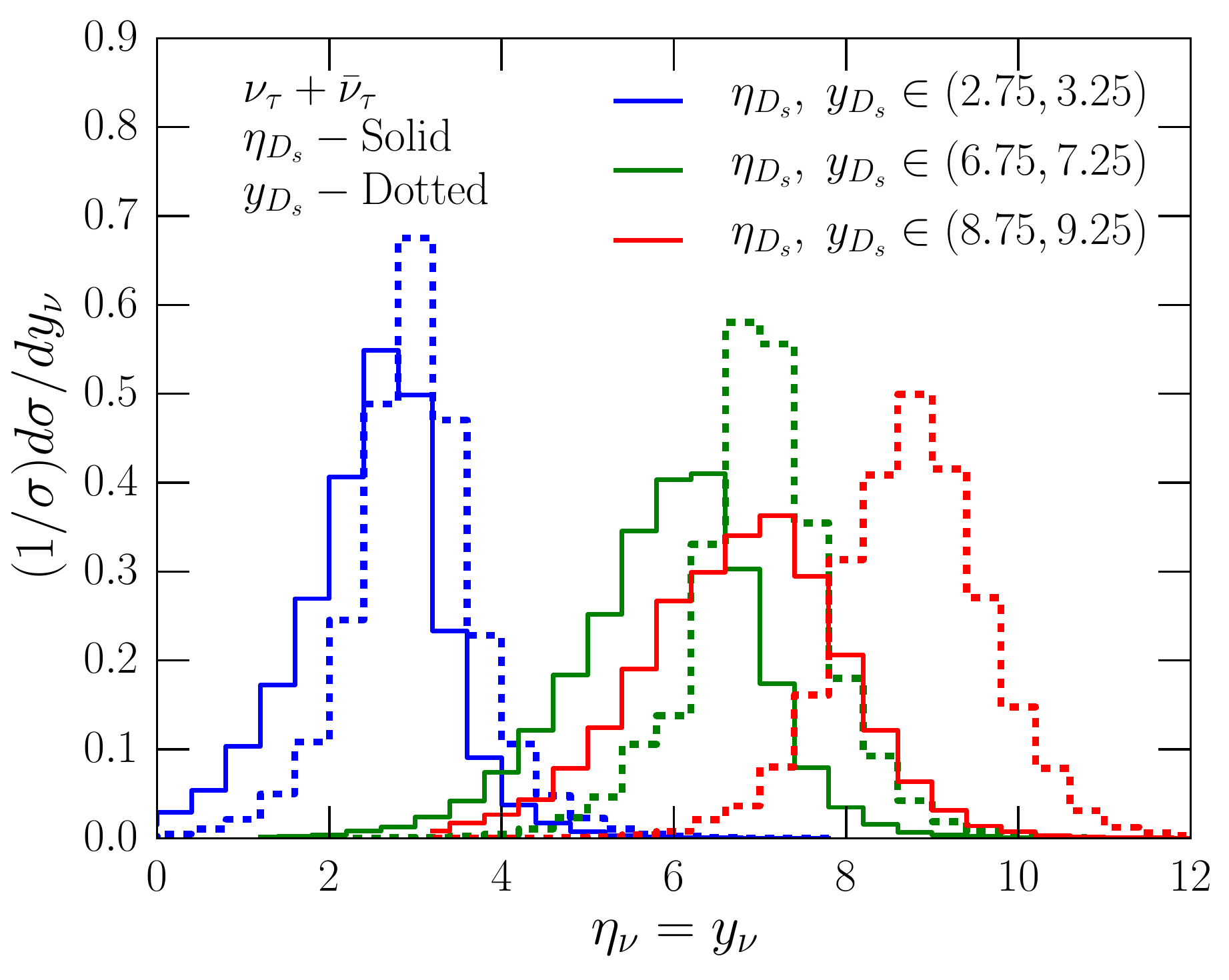} 
   \caption{Upper: the distribution of $y_\nu=\eta_\nu$ for $\nu_\tau+\bar{\nu}_\tau$ that come from $D_s$ decays that originate from charm quark rapidities (dotted histograms) and charm quark pseudo-rapidities (solid histograms) in the ranges $[-0.25, 0.25]$, $[2.75, 3.25]$, $[6.75, 7.25]$ and $[8.75, 9.25]$. 
   Lower: the distribution of $y_\nu=\eta_\nu$ for $\nu_\tau+\bar{\nu}_\tau$ that come from $D_s^\pm$ decays where the $D_s^\pm$ rapidities $y_{D_s^\pm}$ (dotted histograms) and $D_s^\pm$ pseudo-rapidities $\eta_{D_s}$ (solid histograms) lie in the ranges $[2.75, 3.25]$, $[6.75, 7.25]$ and $[8.75, 9.25]$.
   The distributions have unit normalization in each $D_s^\pm$ (pseudo)rapidity interval. They refer to $D_s^\pm$ produced in $pp$ collisions at $\sqrt{s}=14$ TeV, using as input $(\mu_R,\mu_F)=(1,1)m_{T,2}$ and $\langle k_T\rangle = 0.7 {~\rm GeV}$. 
  }
  \label{fig:rapidity-pseudo-compare}
\end{figure}

Detector size and positioning along the beam axis (the $z$-axis) lead to very small angles $\theta$ which translate to neutrino pseudorapidities $\eta_\nu=y_\nu=-\ln\tan(\theta/2)$. Since the neutrino rapidity and pseudorapidity are identical, we use the terms interchangeably for neutrinos. The distinction between rapidity and pseudorapidity is important for charm quark and $D_s^\pm$ distributions, as shown in Fig.~\ref{fig:rapidity-pseudo-compare}.
The upper panel of Fig.~\ref{fig:rapidity-pseudo-compare}, for charm quarks, shows that it is the charm rapidity $y_c$ rather than its pseudo-rapidity $\eta_c$ that correlates more closely with the neutrino $\eta_\nu$ for the large $\eta_\nu$ values most relevant to the forward LHC neutrino experiments. 
For example, for a charm rapidity window in the range $y_c=6.75-7.25$, the peak of the resulting neutrino distribution (green dotted line) is at $\eta_\nu=6.7$.  By contrast, for a charm pseudorapidity window in the range $\eta_c=6.75-7.25$, the corresponding $\eta_\nu$ distribution (green solid line) peaks at the lower value $\eta_\nu$=6.3. 

The displacements in the location of the $\eta_\nu$ peaks with respect to the  center of the charm pseudorapidity windows are even larger for angles closer to the beamline. 
For $\eta_c=8.75-9.25$, the peak of the resulting neutrino distribution is $\eta_\nu=7.0$, while
for $y_c=8.75-9.25$, the peak of the neutrino distribution is closer to the charm rapidity window, at $\eta_\nu=8.3$.
The same conclusions apply for 
the correlations between neutrino rapidities and $D_s$ meson rapidity and pseudorapidity windows, as shown in the lower panel of Fig.~\ref{fig:rapidity-pseudo-compare}. 
The neutrino rapidity is better correlated with the meson rapidity than the meson pseudorapidity.

On the other hand, neutrinos with a fixed (pseudo)rapidity receive contributions from decays of charmed mesons 
with a range of rapidities,  including even rapidities $y_{D_s} < \eta_\nu$. For example, in the lower panel of Fig.~\ref{fig:rapidity-pseudo-compare}, the red dotted histogram of neutrino $\eta_\nu$ distribution extends to larger values than the interval of rapidities  
$y_{D_s}$ of the parent meson, 8.75--9.25. These kinematics considerations have profound implications for the design of the far-forward neutrino experiments and the possibilities of constraining unknown QCD aspects through their data. In any case, we always use the full kinematics of charm production, fragmentation and decay to direct and chain neutrinos in our evaluation of the neutrino energy distributions for several $\eta_\nu$ ranges.

\subsection{Charm hadron production}

\begin{figure}
    \centering
    \includegraphics[scale=0.5]{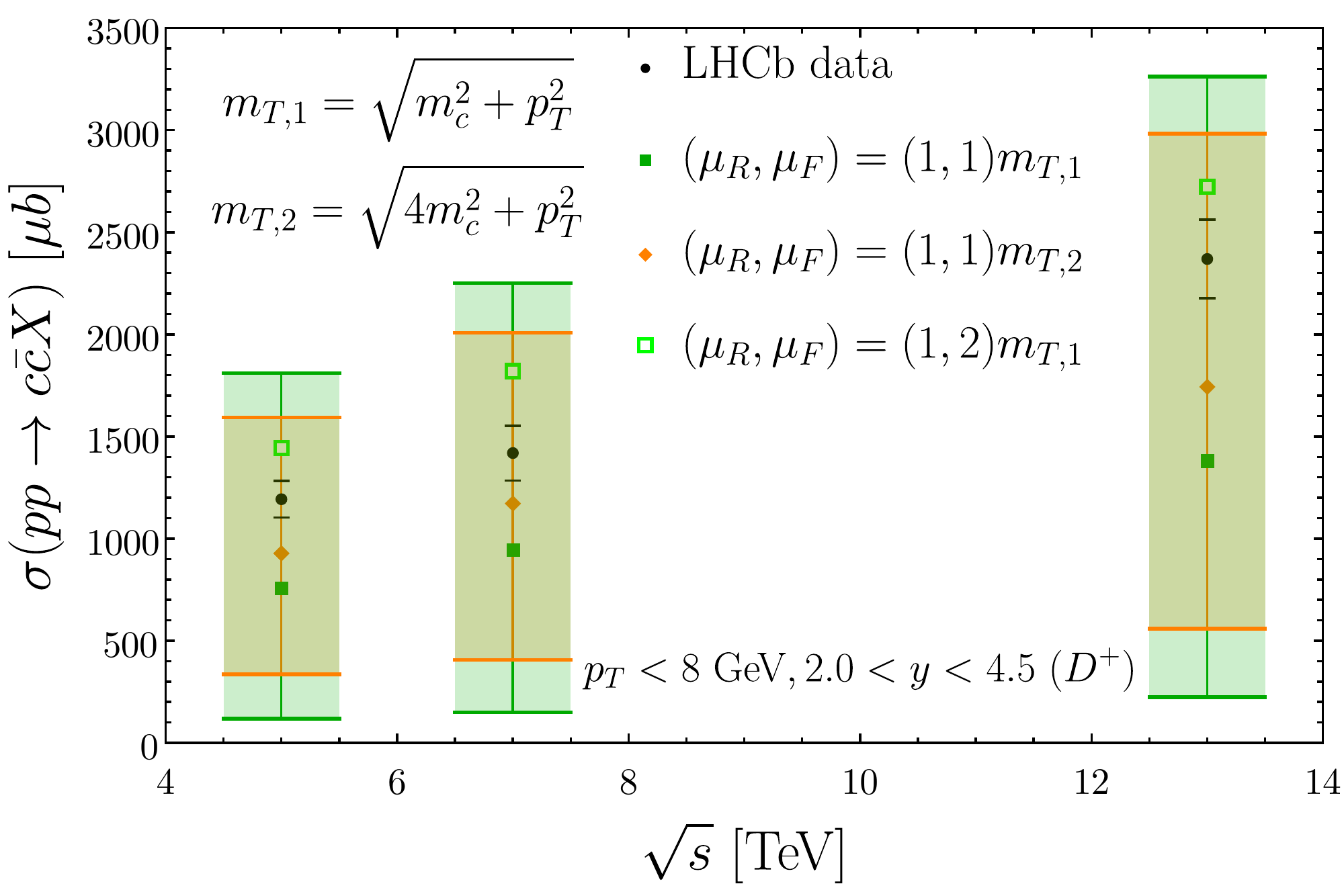}
    \caption{NLO QCD predictions for the equivalent charm-quark pair production fiducial cross sections $\sigma(pp\to c\bar{c}X)$ from charm meson ($D^+$) production in a fiducial region with  $p_T<8$ GeV and $2.0<y<4.5$ for $\sqrt{s}=5,$ 7 and 13 TeV, compared to the cross sections determined by the LHCb experiment \cite{Aaij:2016jht, LHCb:2013xam, Aaij:2015bpa} in the same fiducial region. 
    The orange (green) error bands represent the scale dependence uncertainties for the theoretical predictions, for renormalization and factorization scales proportional to $m_{T,2}$ ($m_{T,1}=m_T$). The green open marker represents central predictions with $(\mu_R, \mu_F) = (1, 2)m_{T,1}$.
}
    \label{fig:LHCb-ccbar}
\end{figure}

We begin with the $c\bar{c}$ production cross section. 
Measurements of this cross section have been extracted from LHCb data for $D^0$, $D^+$ and their charge conjugate mesons produced with $2.0<y<4.5$ for $\sqrt{s}=5$, 7 and 13 TeV. Accounting for the respective fragmentation fractions for $D^0$ and $D^+$, the LHCb collaboration extracted the corresponding $\sigma(pp\to c\bar{c} X)$ in this restricted kinematic regime \cite{Aaij:2016jht, LHCb:2013xam, Aaij:2015bpa},
shown in Fig. \ref{fig:LHCb-ccbar} by the black dots with error bars. Our NLO QCD  evaluation of the equivalent quantity is performed with charm production and fragmentation to $D^+$, with the $D^+$ momenta satisfying the LHC analysis cuts. Normalized by the inverse of the fragmentation fraction to $D^+$, our predictions are shown in Fig. \ref{fig:LHCb-ccbar} for scales $\mu_R,\mu_F\propto m_{T,2}$ (orange marker and error bars) and for $\mu_R, \mu_F\propto m_{T,1}=m_T$ (filled green marker and error bar).  The figure illustrates the fact that the scale uncertainties are somewhat smaller for $\mu_R,\mu_F\propto m_{T,2}$ compared to  $\mu_R, \mu_F\propto m_{T,1}=m_T$. 
The central predictions for both central scale choices lie below the black data points from the LHCb data from charm mesons in the same transverse momentum and rapidity range. For reference, we also show with the green open marker the corresponding cross sections for $(\mu_R,\mu_F)=(1,2)m_T$, which lie above the measured cross sections.

\begin{figure}
\centering
    \includegraphics[width=0.70\textwidth]{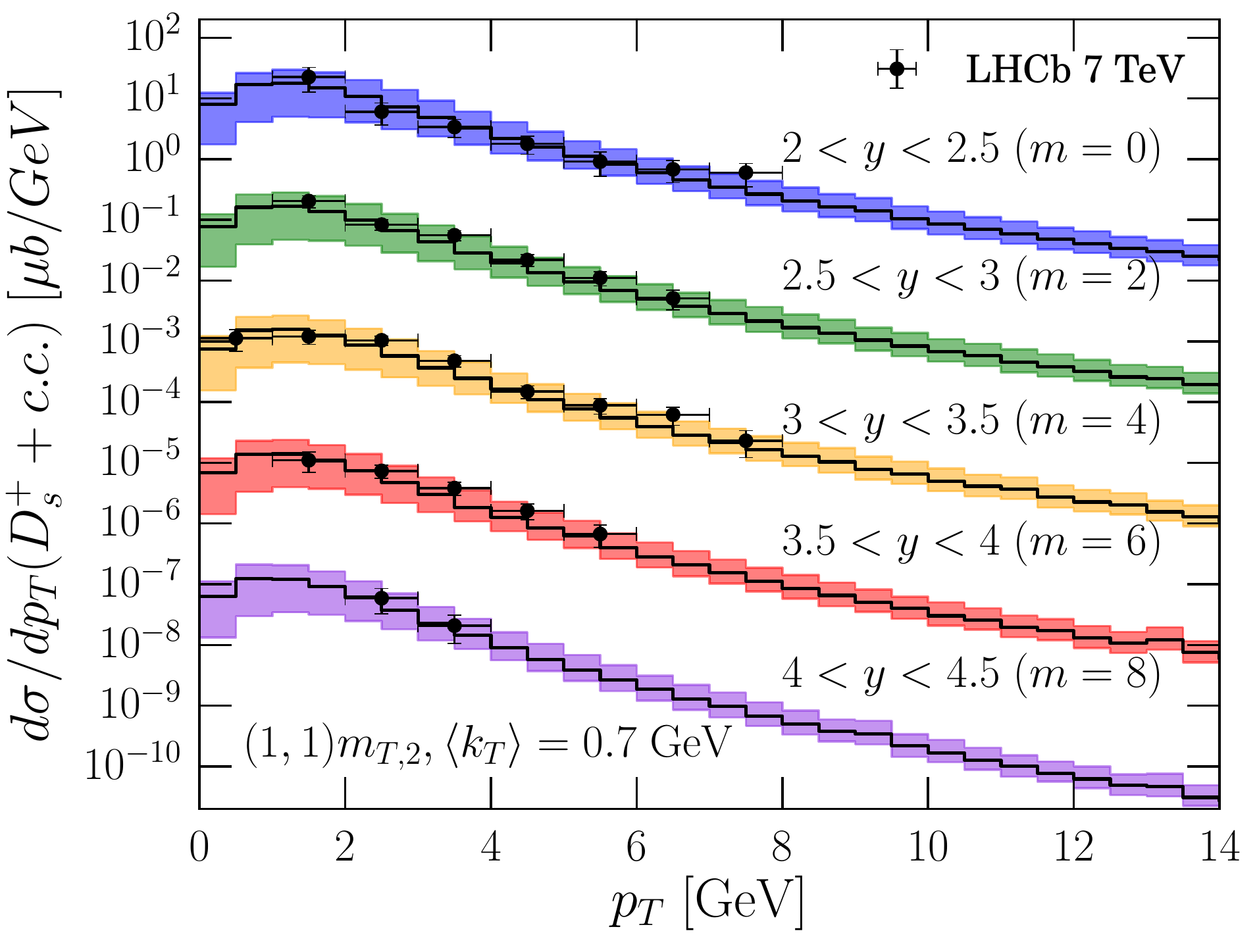}
    \includegraphics[width=0.70\textwidth]{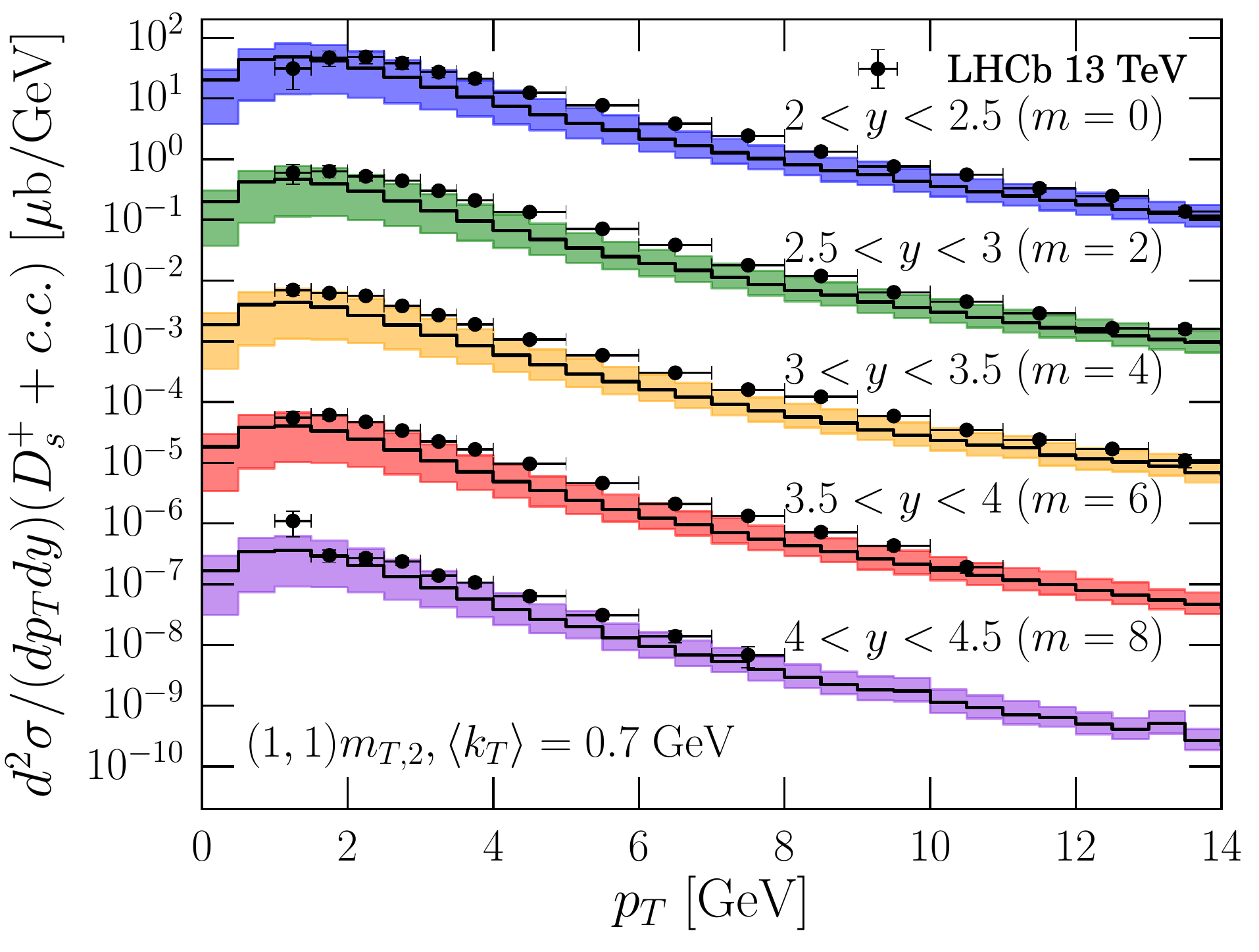}   
    \caption{
    The transverse momentum distribution in the $D_s^\pm$ rapidity ranges indicated in the figure, for production of  $D_s^+$ or its charge-conjugate meson in $pp$ collisions compared with LHCb data for  $d\sigma(D_s^+)  /dp_T+d\sigma(D_s^-)  /dp_T$  at $\sqrt{s}=7$ TeV \cite{LHCb:2013xam} (upper) and for $d^2\sigma(D_s^+)/dp_Tdy +d^2\sigma(D_s^-)/dp_Tdy$ at $13$ TeV   \cite{Aaij:2015bpa} (lower). 
    The $\Delta y$ bins are shifted by $10^{-m}$ where $m=0$, 2, 4, 6 and 8.
    The    central scale is set to $(\mu_R,\mu_F)= (1,1)m_{T,2}$ in theory predictions
    and $\langle k_T\rangle=0.7$ GeV.
The colored bands show the 7-point scale variation uncertainty around the solid black histograms for the central scale. 
}
    \label{fig:LHCb-Ds}
\end{figure}

\begin{figure}
\centering
   \includegraphics[width=0.70\textwidth]{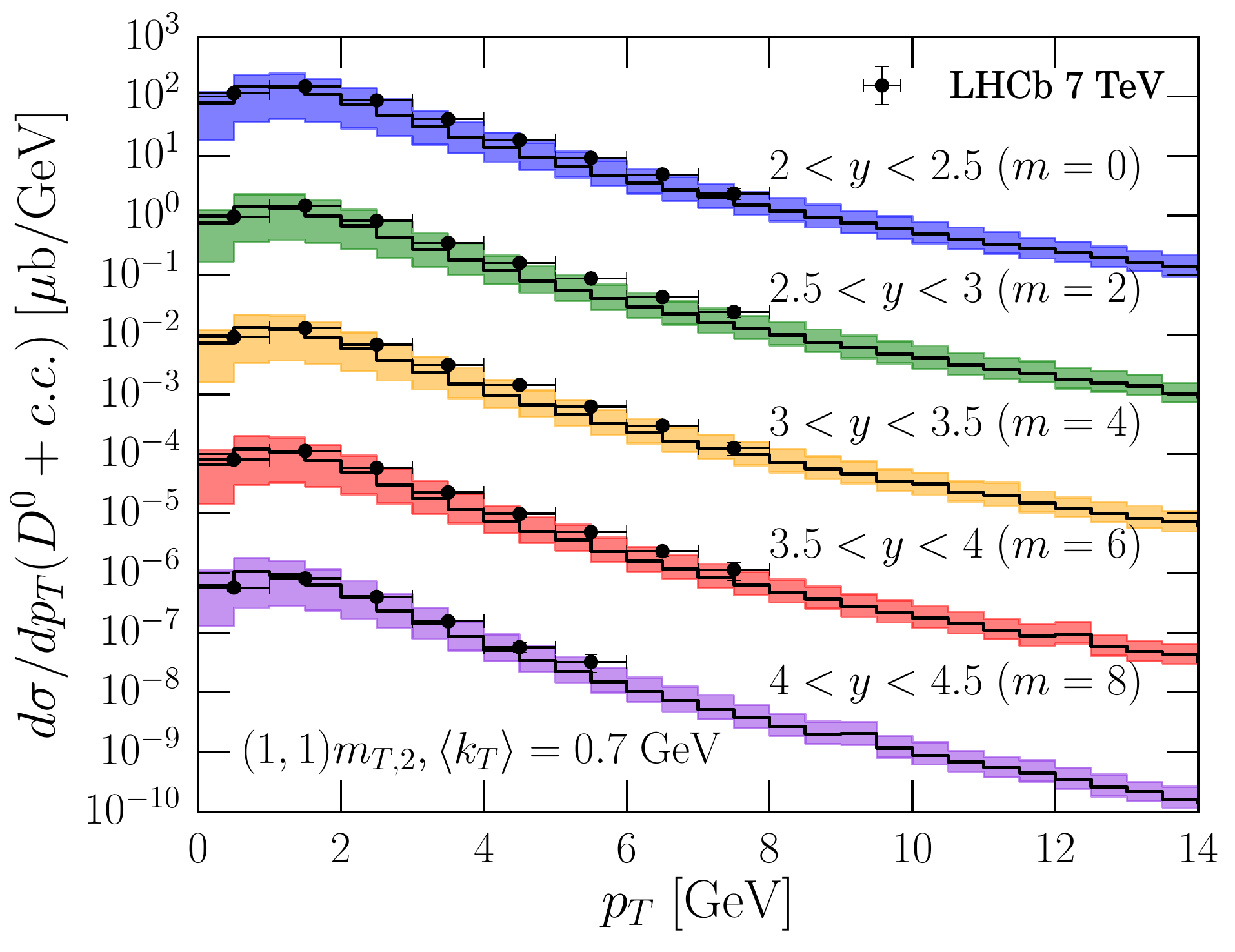}
    \includegraphics[width=0.70\textwidth]{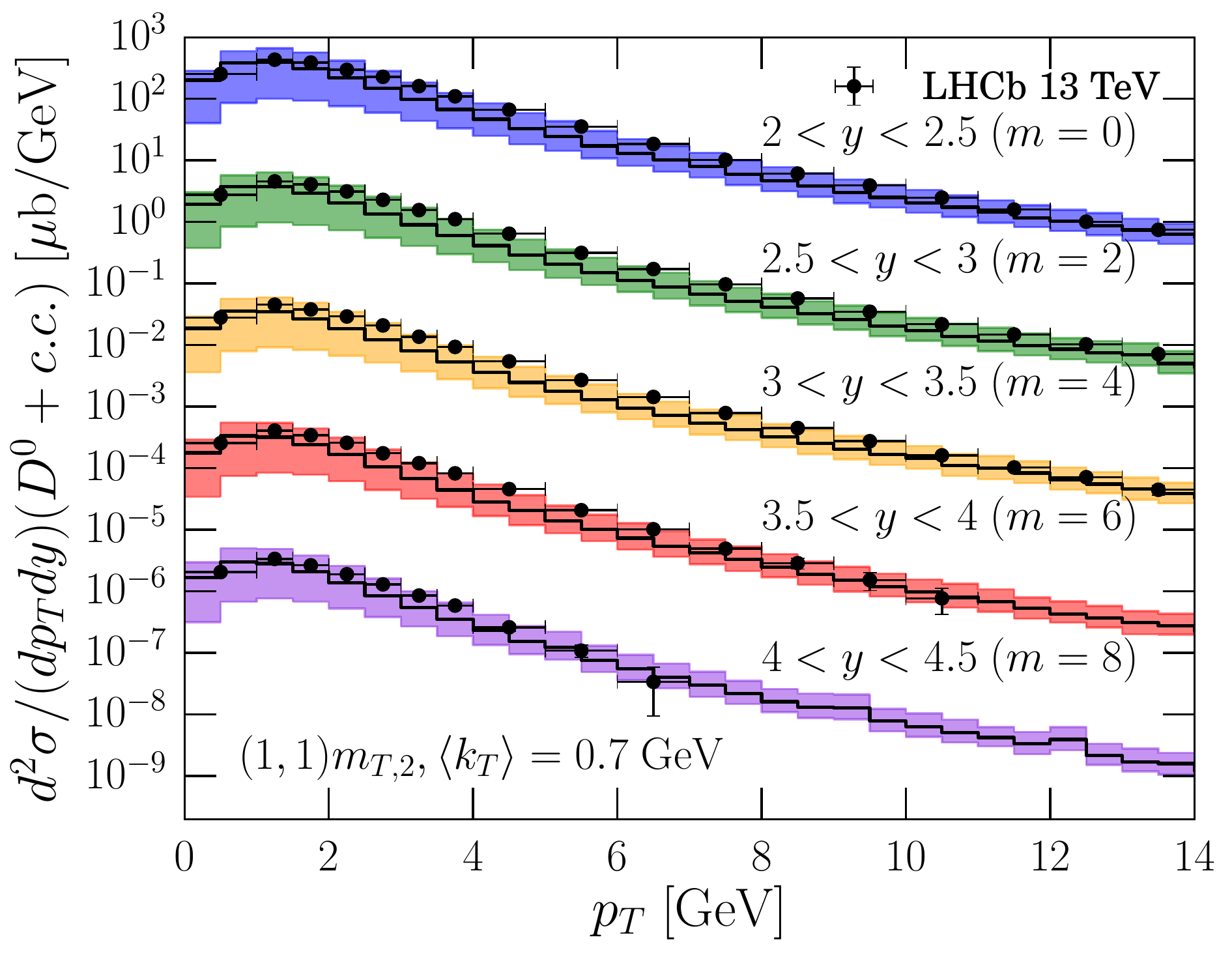}     
    \caption{
    Same as in Fig.~\ref{fig:LHCb-Ds}, but for production of  $D^0$ and its charge conjugate, compared with LHCb data for $\sqrt{s}=7$ TeV \cite{LHCb:2013xam} (upper) and\ $13$ TeV \cite{Aaij:2015bpa} (lower). }
    \label{fig:LHCb-D0}
\end{figure}

The feature that central theory predictions with our default central scale choice lie somewhat below the data, although theoretical predictions and experimental data are still compatible within the uncertainties, is also apparent in the LHCb transverse momentum distributions.
Figs. \ref{fig:LHCb-Ds} and \ref{fig:LHCb-D0} compare our predictions on the double differential cross sections for $D_s^\pm$ and $D^0+\bar{D}^0$ production in $p_T$ and $y$  with the corresponding 
LHCb data~\cite{LHCb:2013xam,Aaij:2015bpa}. 
The colored bands in the histograms show the uncertainty in the prediction of $d^2\sigma/dp_T\, dy$ associated with the 7-point scale variation around $(1,1)m_{T,2}$. 
The LHCb data generally lie within the scale uncertainty band of theory predictions and have smaller experimental uncertainties than the latter.

\begin{figure}
    \centering
    \includegraphics[width=0.70\textwidth]{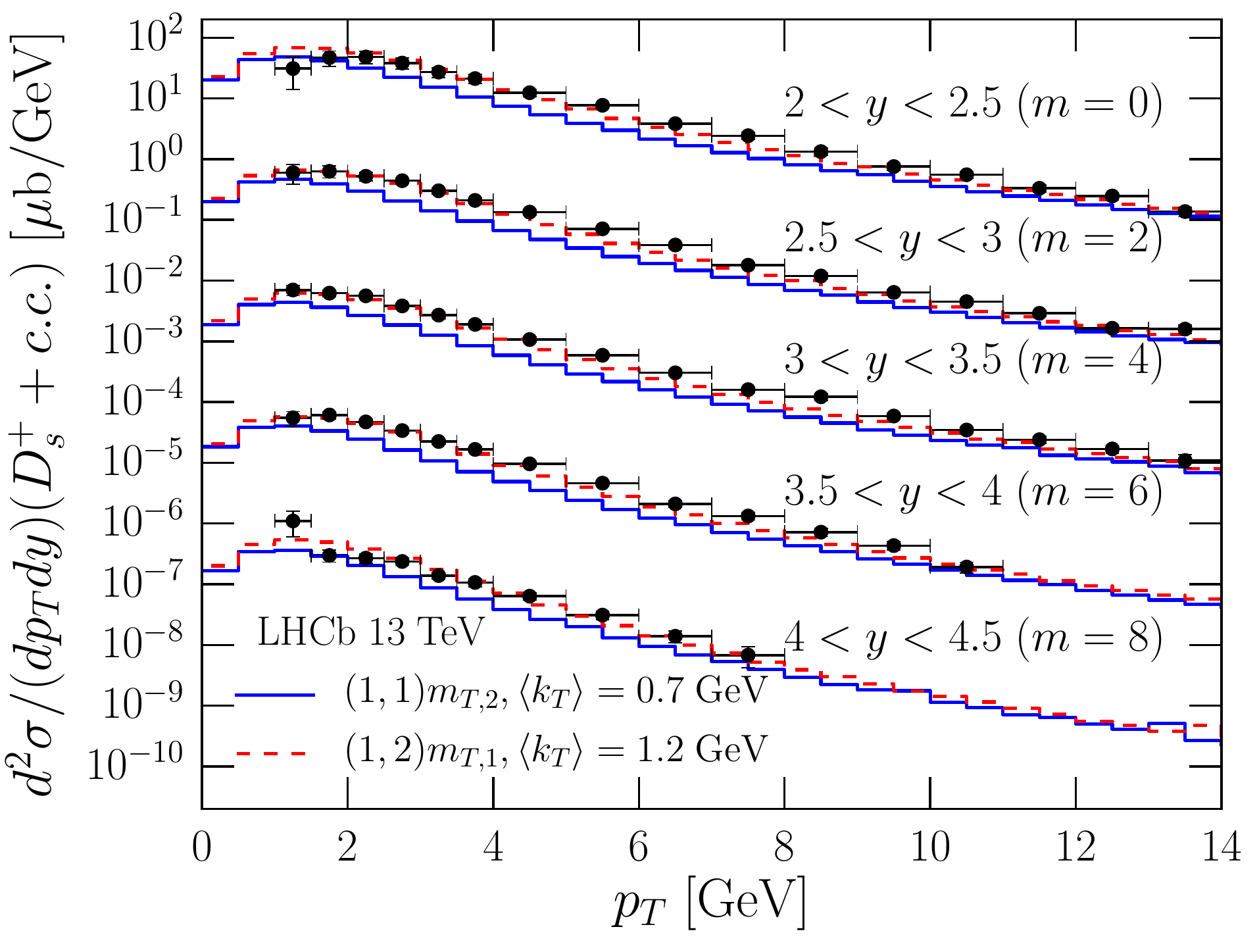}
    \includegraphics[width=0.70\textwidth]{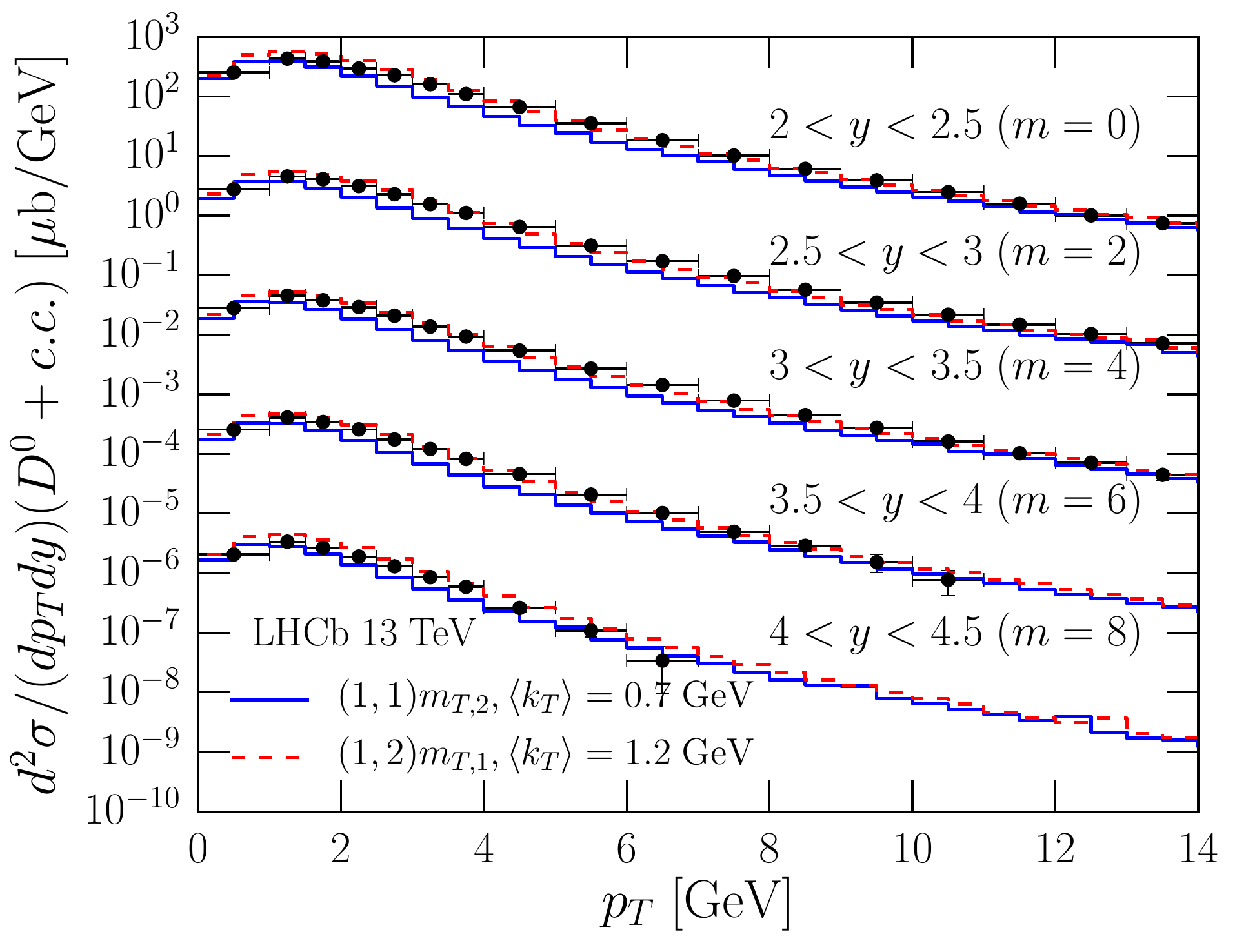}    
    \caption{The double-differential distribution $d^2\sigma/dp_T\, dy$ for  $D_s^++D_s^-$ (upper) 
    and $D^0+\bar{D}^0$ (lower) production 
    at $\sqrt{s}=13$ TeV  \cite{Aaij:2015bpa}, with input
    $(\mu_R,\mu_F)=(1,1)m_{T,2}$ and 
    $\langle k_T\rangle~=~0.7$~GeV (solid) and $(\mu_R,\mu_F)=(1,2)m_T$ and $\langle k_T\rangle = 1.2$ GeV (dashed), compared with LHCb data. The distributions are scaled as in Fig.~\ref{fig:LHCb-Ds}. }
    \label{fig:LHCb-Ds-alt}
\end{figure}

For reference in Fig.~\ref{fig:LHCb-Ds-alt}, in addition to predictions with the central scales $(\mu_R,\mu_F)= (1,1)m_{T,2}$ and $\langle k_T\rangle=0.7$ GeV (solid histogram), we show $d^2\sigma/dp_T\, dy$ with $(\mu_R,\mu_F)=(1,2)m_T$ and $\langle k_T\rangle = 1.2$ GeV (dashed histograms). These inputs yield histograms for the $D_s^\pm$ and $D^0+\bar{D}^0$ distributions that lie closer to the LHCb data than our default scale and $\langle k_T\rangle$ choices. 
Using $(\mu_R,\mu_F)=(1,2)m_T$ and $\langle k_T\rangle = 1.2$ GeV  as default input choices, however, would suffer from some theoretical drawbacks in the context of the current study of scale uncertainties. In particular, the choice of $(\mu_R,\mu_F)=(1,2)m_T$ as  central scale leads to the inclusion of scales $\mu_F/\mu_R=4$. This large ratio of scales leads to negative cross sections at very low $p_T$, indicating a breakdown of perturbative QCD at fixed relatively low (NLO) order, as considered here, in the standard  factorization scheme. Furthermore, the larger value of $\langle k_T\rangle = 1.2$ GeV is more difficult to interpret as an intrinsic $k_T$, considering that we expect the latter to be less than the scale of the proton mass. However, we note that large values of $\langle k_T\rangle$ are suggested in other processes \cite{Apanasevich:1998ki,Miu:1998ju,Balazs:2000sz} and may reflect the importance of higher-order QCD corrections, somehow mimicked by the use of high $\langle k_T\rangle$ values.

\begin{figure}
    \centering
   \includegraphics[width=0.70\textwidth]{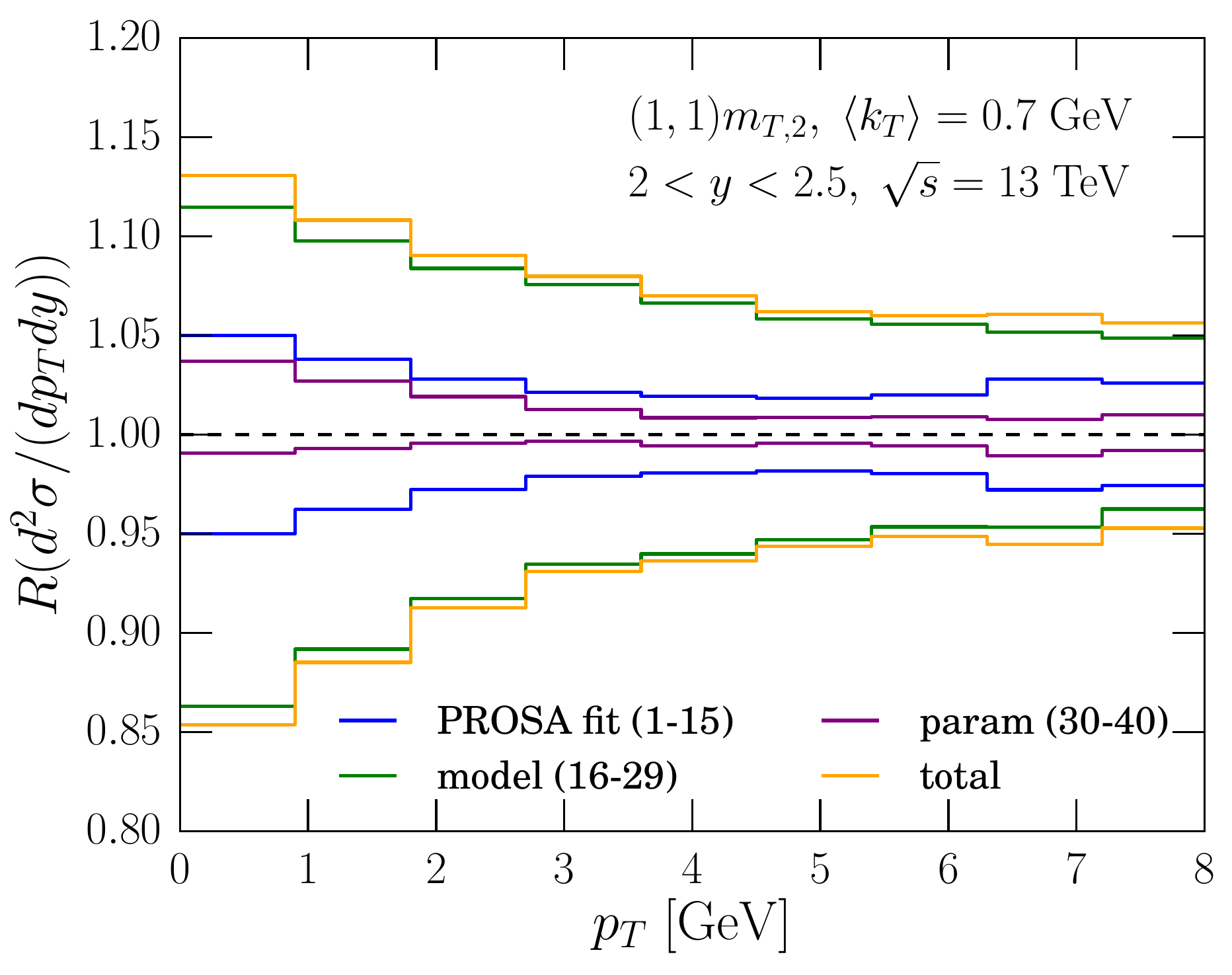}
      \includegraphics[width=0.70\textwidth]{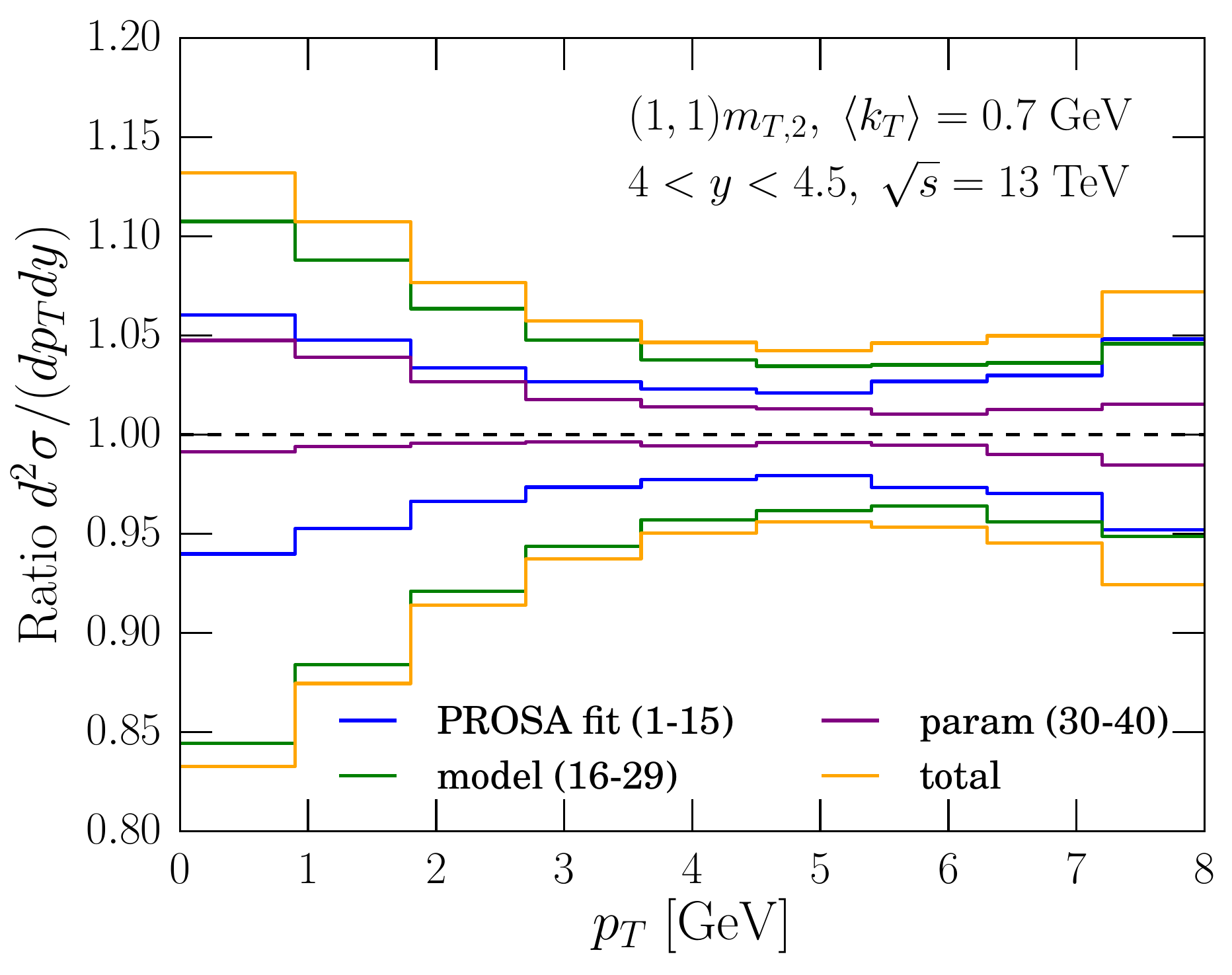}
      
    \caption{For the double-differential distribution $d^2\sigma/dp_T\, dy$ of  $D_s^++D_s^-$ production at $\sqrt{s}=13$ TeV with $(\mu_R,\mu_F)=(1,1)m_{T,2}$ and $\langle k_T\rangle = 0.7$ GeV, the ratio of the full PDF uncertainties and of their separate components due to fit, model and parameterization inputs, to the central PDF (best-fit), as a function of $p_T$, for two $D_s^\pm$ rapidity ranges: $2.0<y<2.5$ (upper) and $4.0<y<4.5$ (lower), considering the PROSA PDF fit.} 
    \label{fig:Ds-40PROSA}
\end{figure}

With our default scale and $\langle k_T\rangle$ values, Fig.~\ref{fig:Ds-40PROSA} shows the uncertainty band associated with the variation of the PROSA PDFs, on $p_T$ distributions for $D_s^\pm$ mesons
for the two LHCb more extreme rapidity ranges.  $2.0<y<2.5$ and $4.0<y<4.5$. 
The PROSA PDFs include 40 different sets representative of fit, model, and parameterization uncertainties, in addition to the central set, corresponding to the best-fit.
The 40 variations included in the PROSA PDF fit bring in uncertainties in the differential cross section of $D_s^\pm$ production within about $\pm$ 15\%, with the largest deviation in the low $p_T$ region, especially in the forward direction.  
The dominant contribution to the uncertainties come from the so-called PROSA ``model" uncertainties, which involve, among others, the uncertainty on various theory inputs for heavy quark production used in the fit (see Ref.~\cite{Zenaiev:2019ktw} for more detail).  

\begin{figure} [h]
    \centering
   \includegraphics[width=0.70\textwidth]{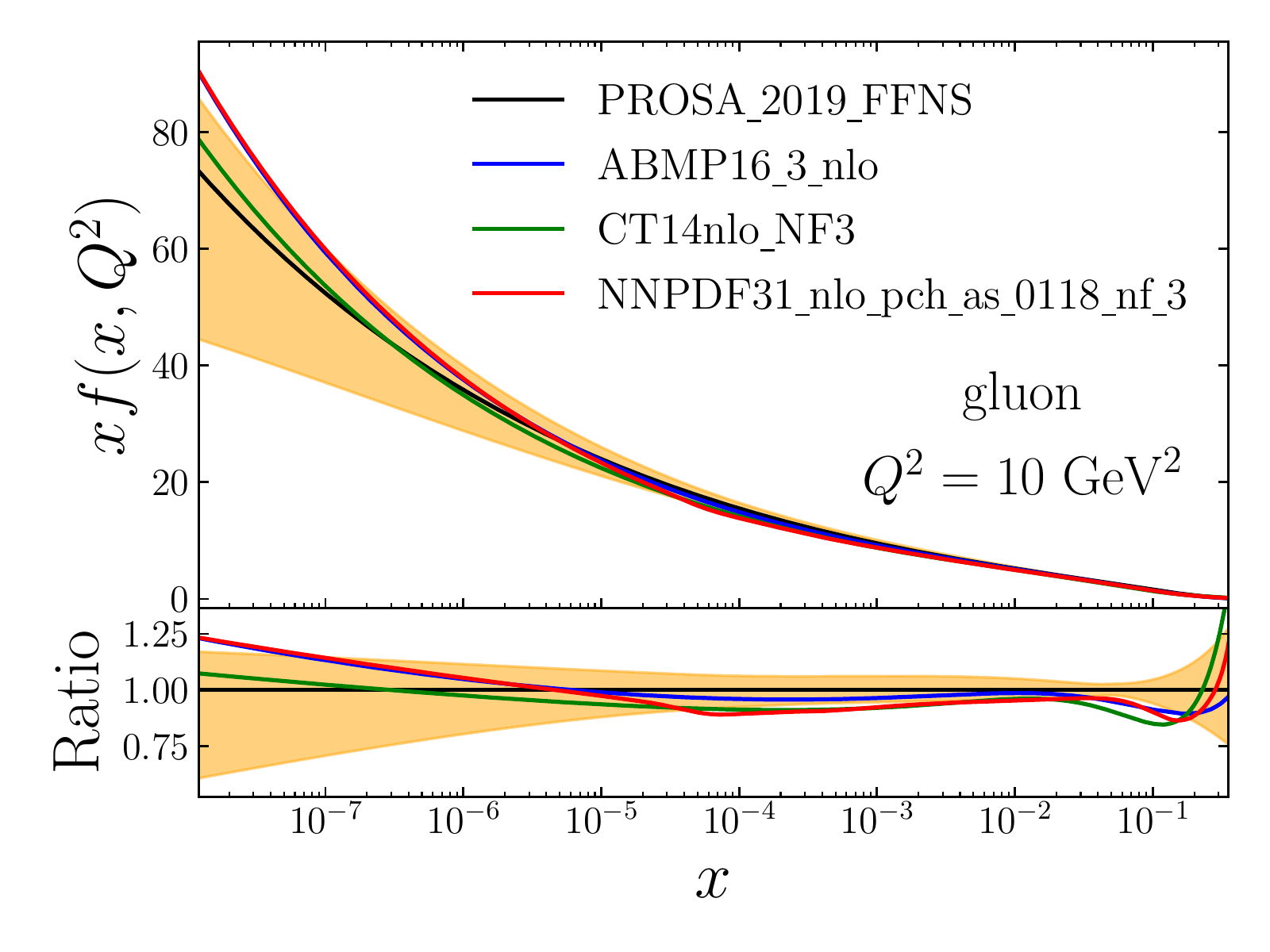}
      \includegraphics[width=0.70\textwidth]{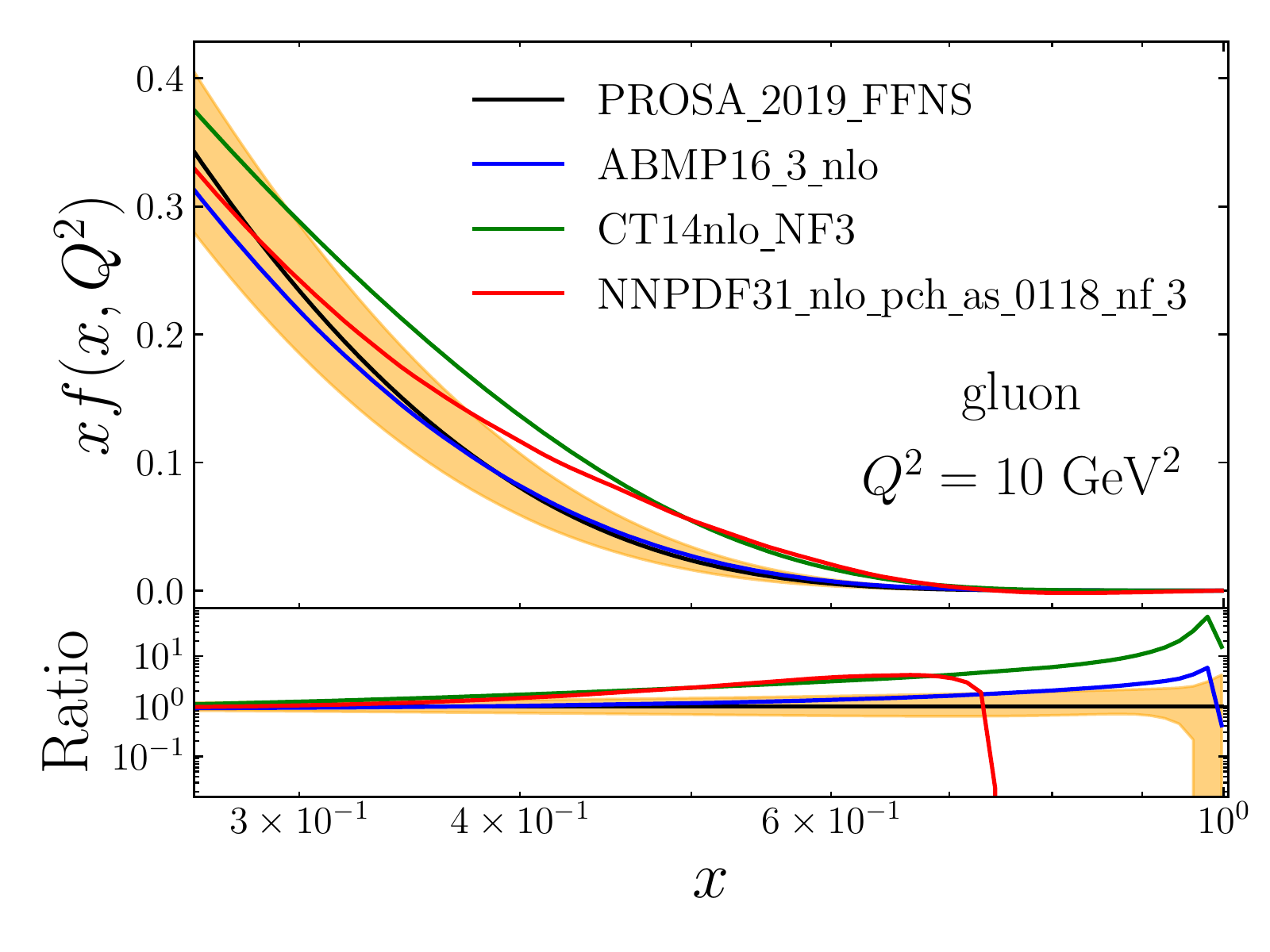}      
    \caption{Gluon PDFs $xf(x,Q^2)$ for $Q^2=10$ GeV$^2$ for $x<0.3$ (upper) and $x>0.3$ (lower) for the PROSA 2019 FFNS fit (black curve with orange uncertainty band). The  ABMP16, CT14 and NNPDF3.1 central gluon PDFs are also shown.  All PDFs are for 3 active flavours at NLO. The lower insets show ratios relative to the central PROSA PDF. Note the different scales in the two ratio plots.}
    \label{fig:PDFun}
\end{figure}

Other PDF fits in the 3-flavour scheme are available. We compare the PROSA PDF sets with the NLO sets of the CT14 \cite{Dulat:2015mca}, ABMP16 \cite{Alekhin:2018pai} and NNPDF3.1 \cite{Ball:2017nwa} collaborations. 
Particularly important for this work and heavy-flavour production in general, are gluon PDFs.
The small-$x$ and large-$x$ behavior of the gluon PDFs is shown in the upper and lower panels of Fig.~\ref{fig:PDFun}, respectively, where the PROSA PDF best-fit results (black curves) are plotted together with the related uncertainty bands (orange). Additionally the best-fit gluon distributions from the other aforementioned PDF fits are shown.
Beneath each panel ratios with respect to the central PROSA  PDF are also shown. The upper panel is limited to the $x$ range $10^{-8}<x<0.3$, whereas the lower panel focuses on $x>0.3$. 
Both ranges are interesting because far-forward production of charm quarks at large center-of-mass energies involves the product of small-$x$ and large-$x$ PDFs. The lower panel shows large deviations of the various PDF best-fits among each other for the largest $x$ values (note the different scale in the $y$ axis, when comparing the ratios of the upper and the lower panels).

\begin{figure} [h]
    \centering
      \includegraphics[scale=0.40]{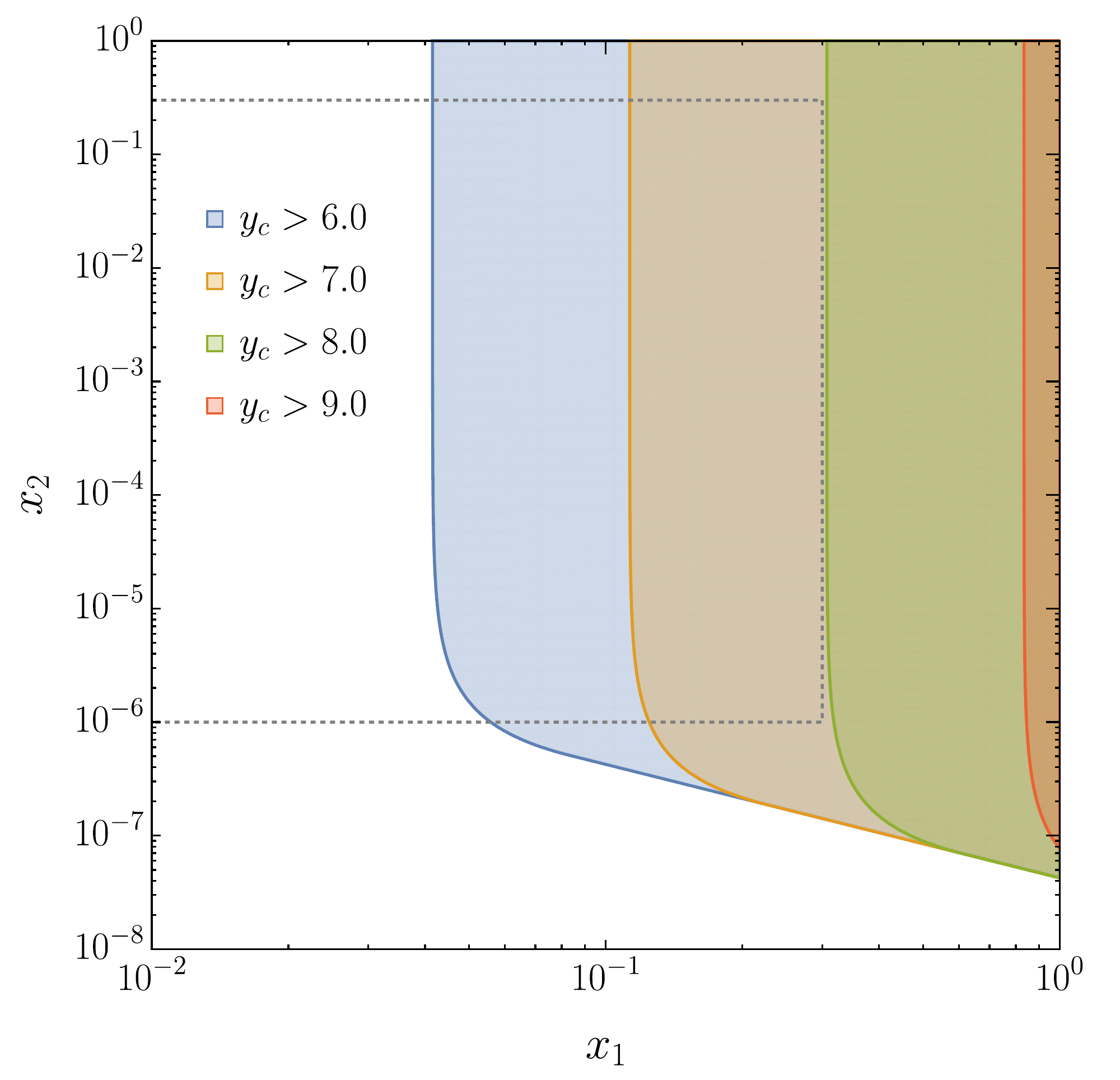}
   \includegraphics[scale=0.40]{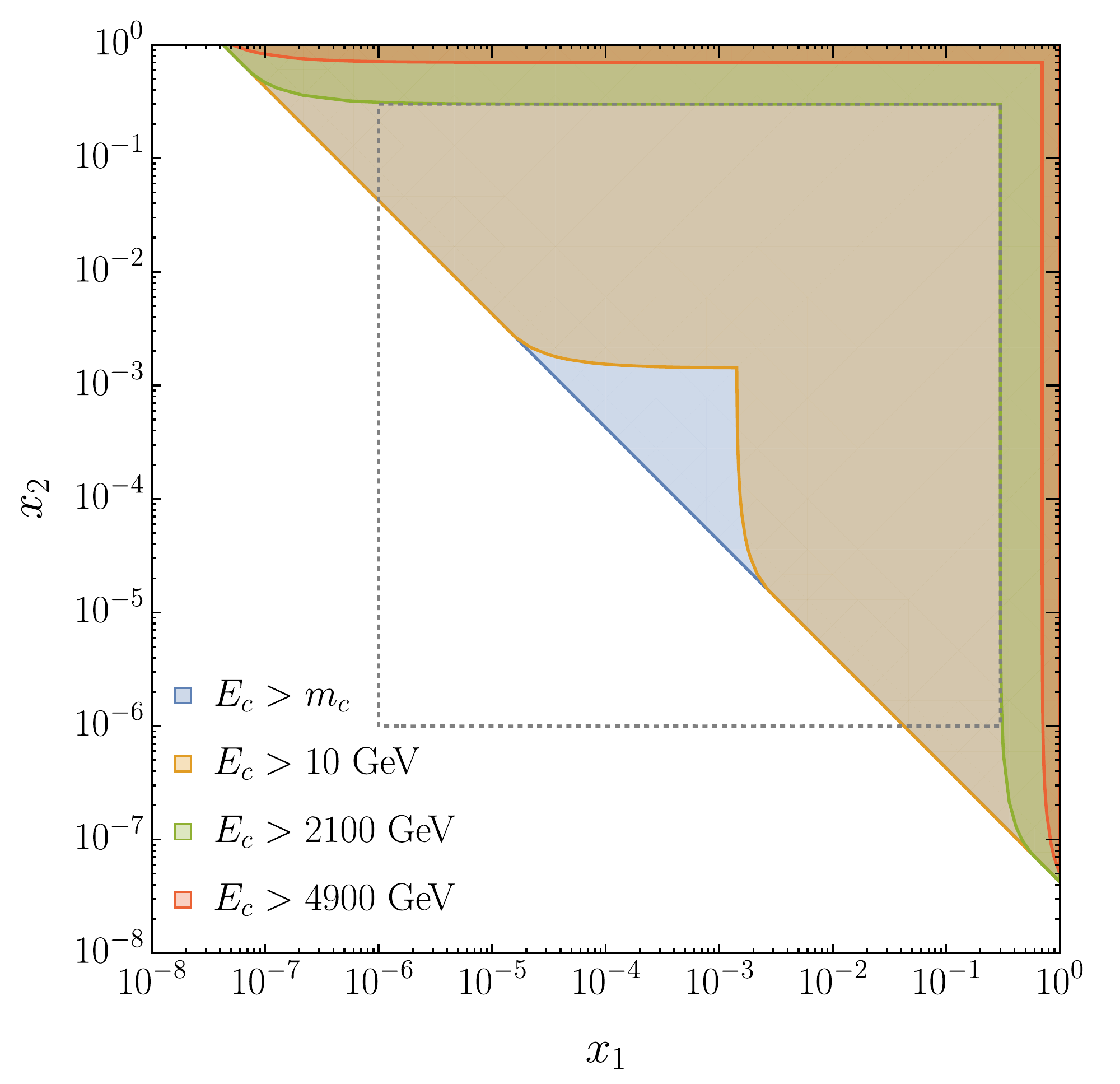}
   
    \caption{ For $pp \rightarrow c+X$ production for $\sqrt{s}=14$ TeV,
    the $(x_1,x_2)$ regions that contribute to charm quark 
    rapidity distributions (upper) for $y_c>6, 7, 8, 9$, respectively, where $x_1>x_2$ for $y_c>0$, and energy distributions (lower) for $E_c>m_c, 10$ GeV, 2.1 TeV and 4.9 TeV. See the text for further discussion.
    In both panels, the colored regions extend to large $x_1$, 
    overlapped by regions with further limitations on the $y_c$ or $E_c$ values. In the lower panel, the shaded regions for $E_c>m_c$ covers the full upper right triangle. The dotted lines show
     $x_i=10^{-6}$ and $x_i=0.3$. Between these values, the PDFs considered in this work are best constrained by data or differ less among each other.}
    \label{fig:xregions}
\end{figure}

To illustrate the regions of the partonic longitudinal momentum fractions  $(x_1,x_2)$ that contribute to charm production in $pp$ collisions,
the upper panel of Fig.~\ref{fig:xregions} shows the range of involved 
 $(x_1,x_2)$  values, depending on the rapidity of the produced charm quark in $pp \to cX$.  
 For $y_c>0$, the charm quark has a momentum component in the $+z$-axis direction, so $x_1>x_2$ where $x_1$ is the parton momentum fraction of the involved parton in the proton in the beam traveling in the $+z$-axis direction.  
The blue region in the upper panel and the colored regions to its right, show that combinations of partons with $x_1\gsim 0.04 $ and $x_2\gsim 4\times 10^{-8}$ contribute to the production of charm with $y_c>6$. The $(x_1,x_2)$ ranges 
are peaked on more extreme values when the minimum charm rapidity increases. For example, for $y_c>9$, $x_1\gsim 0.8$ and $x_2\gsim 10^{-7}$. 
On the other hand, as we saw in Fig.~\ref{fig:PDFun}, the PDF vary widely for large values~of~$x$.

The lower panel of Fig.~\ref{fig:xregions} shows the region of $(x_1,x_2)$ that contribute to energy distributions above a given minimum charm energy. Since there is not a rapidity restriction in the lower panel, there is not a condition on which parton momentum fraction is greater. The lower panel shows that when the forward region of positive $y_c$ is considered so $x_1>x_2$, the band of ($x_1$, $x_2$) combinations that contributes to the production of charm quarks with increasingly high energy relies on extreme parton $x$ values. 
In particular, the production of neutrinos with $E_\nu \sim \mathcal{O}({\rm TeV})$ and $x_1>x_2\ (y_c>0)$   
requires $x_1$ to be very large ($x_1 > 0.5$). 

The dotted boxes in both panels of
Fig.~\ref{fig:xregions} show the range $10^{-6}
 < x < 0.3$ 
in which the considered PDFs 
either are best constrained or better agree among each other.
At large $x$, the gluon PDF is  constrained by inclusive jet, dijet and $t\bar{t}$ distributions \cite{Amoroso:2022eow,H1:2021xxi,Czakon:2016olj,Ball:2017nwa,Bailey:2019yze,Czakon:2019yrx,Kadir:2020yml}. However, gluon PDF uncertainties remain large, in part due to tensions between different data sets. For example, for the   $t\bar{t}$ case, this issue  is discussed in, e.g., refs. \cite{Bailey:2019yze,Guzzi:2022dis}. 
In any case, using the PROSA19 PDFs, $y(t\bar{t})$ data from CMS at $\sqrt{s}=8$ TeV \cite{CMS:2017iqf} are reasonably well reproduced for a range of $t\bar{t}$ invariant masses $(\chi^2=23/15)$, slightly better than with HERAPDF2.0 ($\chi^2=24/15$) \cite{Reno:2022dis}. 
Different PDF sets will most diverge from each other for charm produced at large rapidities and/or at high energies.

\begin{figure}[h]
    \centering
         \includegraphics[width=0.60\textwidth]{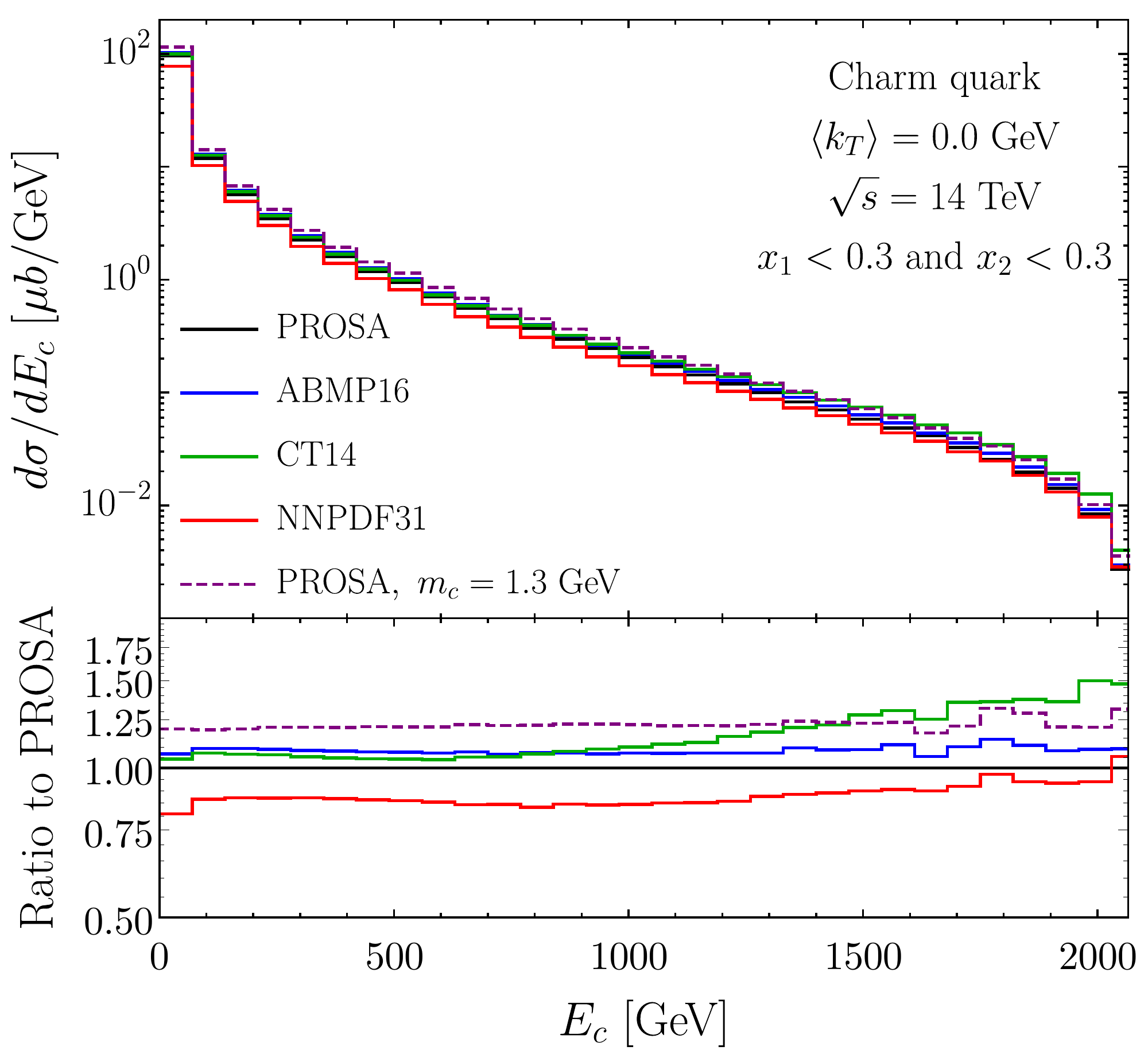}
   \includegraphics[width=0.60\textwidth]{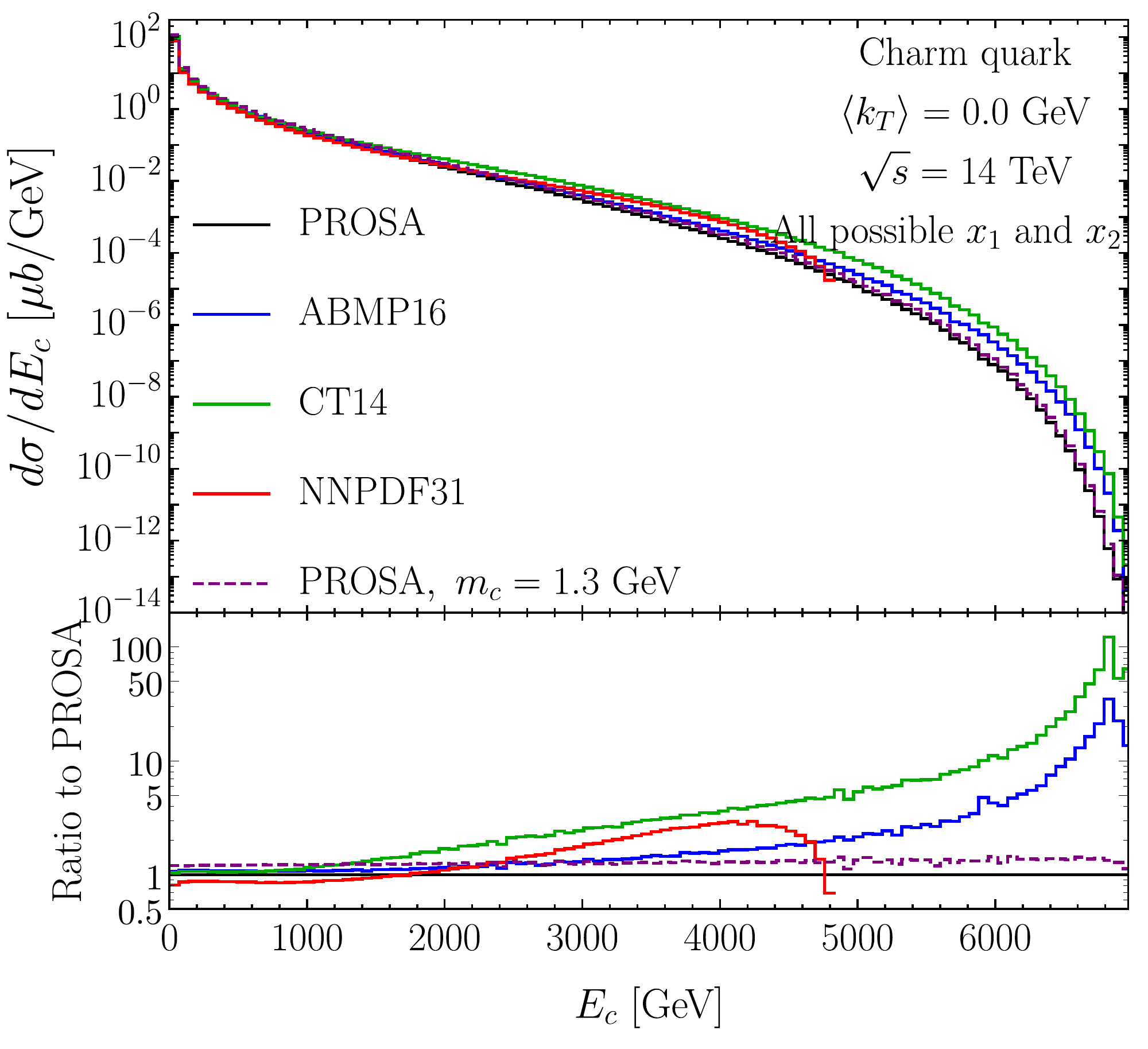}
    \caption{Charm energy distributions at NLO with $\langle k_T\rangle = 0$, with $x_1$ and $x_2$ integrations limited to the range $x_1,x_2<0.3$ (upper) and with full integration over all possible values of $x_1$ and $x_2$ (lower) for the  central predictions using as input the PROSA, ABMP16, CT14 and NNPDF3.1 PDFs. 
    The lower plots show the ratios of the charm energy distributions with respect to the prediction based on the central PROSA PDF, as a function of the charm energy.
    }
    \label{fig:PDF_Ec}
\end{figure}

To uncover the effect of large $x$ PDFs, we set $\langle k_T\rangle = 0$ and integrate over all rapidities.
The impacts of large-$x$ differences in the PROSA, ABMP16, CT14 and NNPDF3.1 PDFs on the evaluations of the charm quark energy distributions with these fits are shown in Fig.~\ref{fig:PDF_Ec}. We note that predictions for the various PDF sets are evaluated with the corresponding charm quark mass values from Table \ref{tab:charm-masses}. When $x_1,x_2<0.3$, the kinematic limit to the charm quark energy is $\sim$ 2.1 TeV. 
As shown in the upper panel of Fig.~\ref{fig:PDF_Ec}, the charm energy distributions from the four central PDF sets differ by at most 50\% at large $E_c$, and less at lower energies. 
The dashed purple histograms show the charm energy distribution using PROSA PDFs and $m_c=1.3$ GeV rather than $1.442$ GeV, showing an increase 
up to $\sim 25\%$ at low energy, and much lower at the highest energies, in the energy distribution.  
The distribution 
based on the NNPDF3.1 PDFs, computed using the charm mass value accompanying this PDF fit, i.e., $m_c = 1.51$ GeV, turns out to be the lowest. 

The lower panel of Fig.~\ref{fig:PDF_Ec} shows the charm energy distribution with integration over the full ranges of $x_1$ and $x_2$.  As anticipated, the energy distributions differ significantly for $E_c$
greater than a few TeV. The ratios of the deviations show a qualitatively similar behavior to the ratios of the large-$x$ gluon PDFs shown in Fig.~\ref{fig:PDFun}.
Figs. \ref{fig:xregions} and \ref{fig:PDF_Ec} show that predictions for very high charm quark rapidities ($y_c\gsim 8$) have significantly larger deviations derived from different PDF sets than predictions for lower charm rapidities. This is reflected in the wider range of predictions for FASER$\nu$ than for SND@LHC, for example, as discussed in Section~\ref{sec:evts}. We note that in Fig.~\ref{fig:PDF_Ec} the energy distribution of the charm quark is considered. 
When the charm quark is fragmented and the meson decays to tau neutrinos, the predictions from different PDFs start to differ already for lower neutrino energy values, below 2 TeV, as we show in the next section.

\section{Tau neutrino and antineutrino production from charm}
\label{sec:neutrinos}

\begin{figure}
\centering
    \includegraphics[width=0.49\textwidth, trim=2cm 0 0 0]{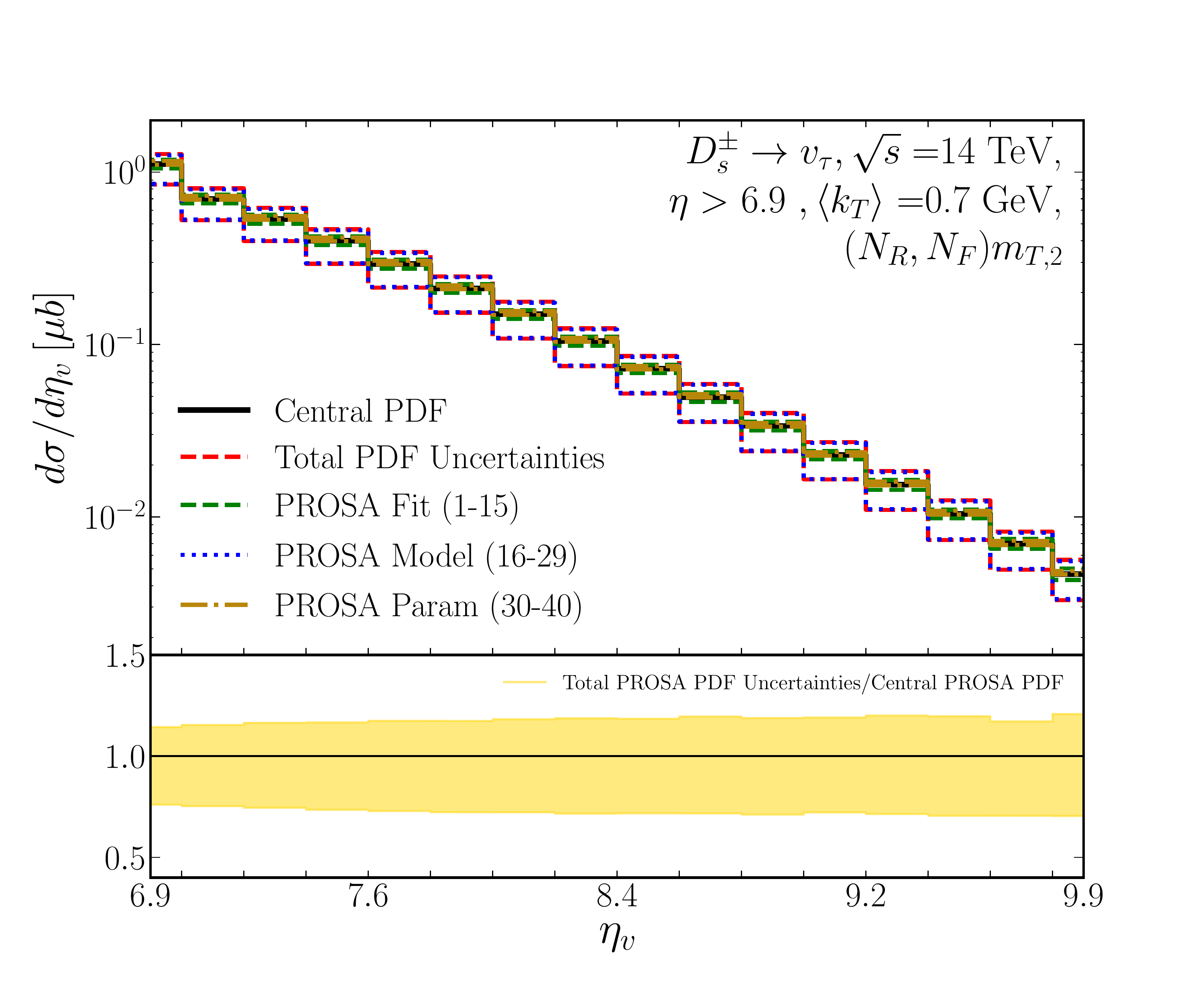}
    \includegraphics[width=0.49\textwidth, trim=2cm 0 0 0]{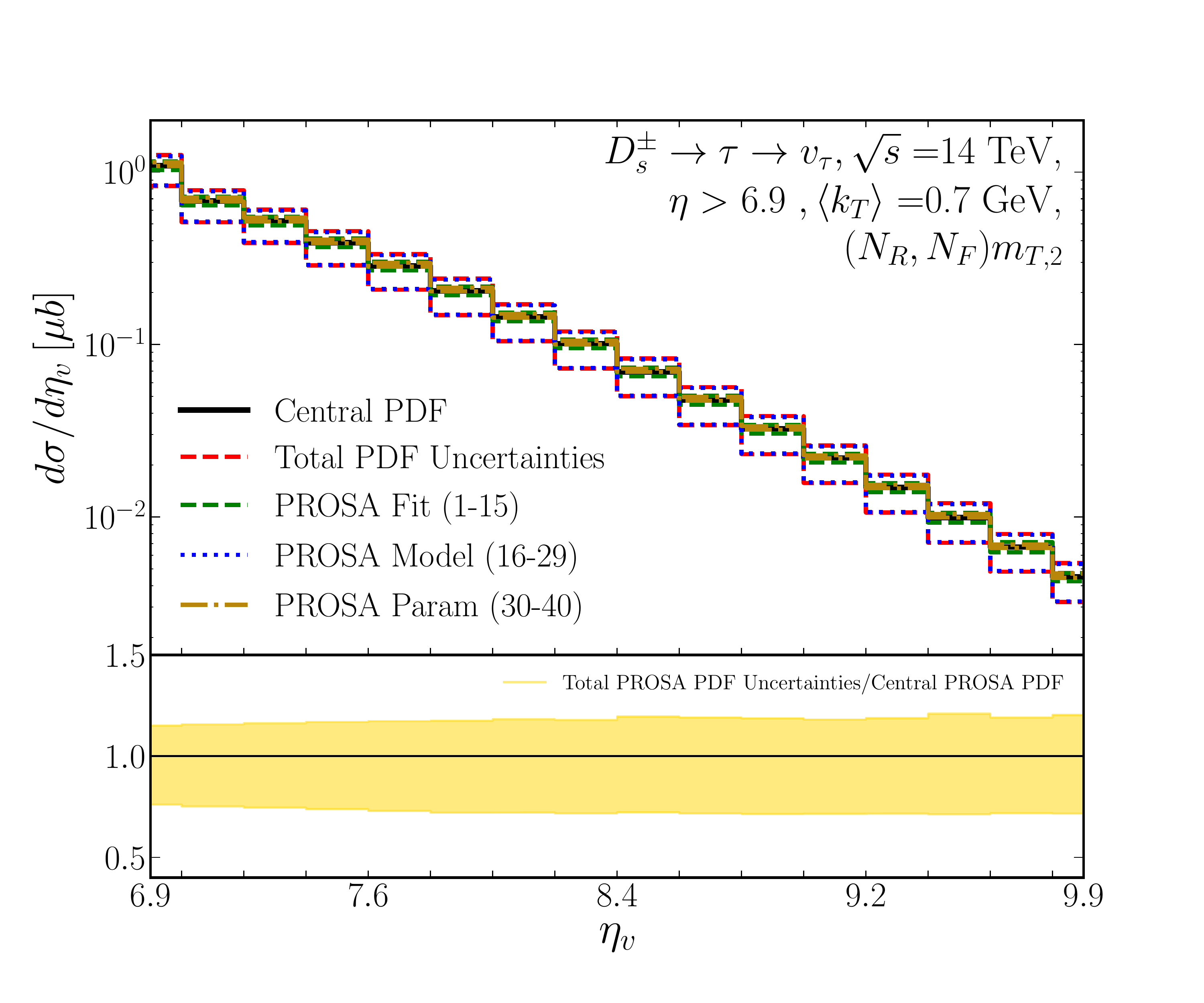}
  \hspace*{-0.93cm}\includegraphics[width=0.44\textwidth]{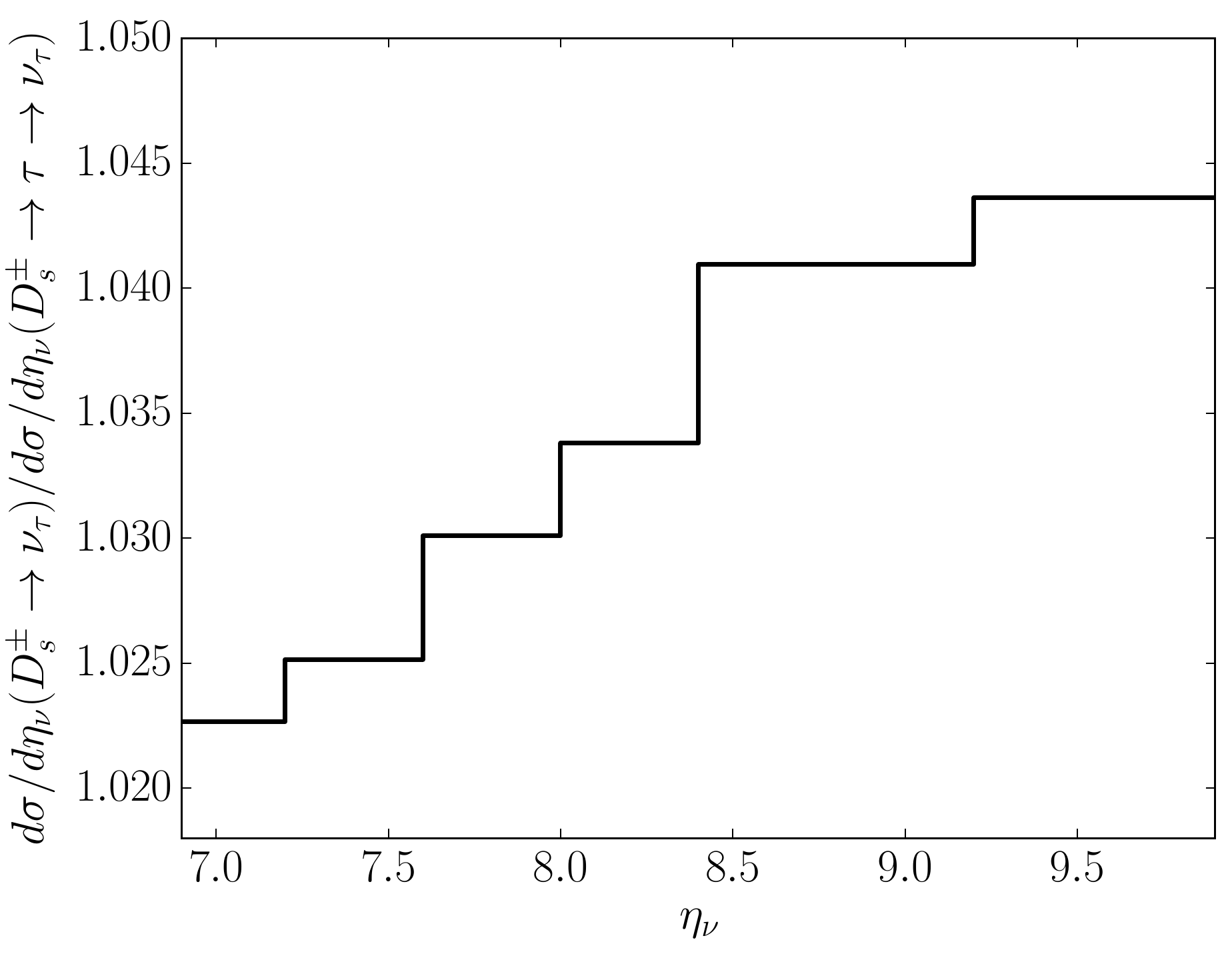}
    \hspace*{0.56cm} \includegraphics[width=0.45\textwidth]{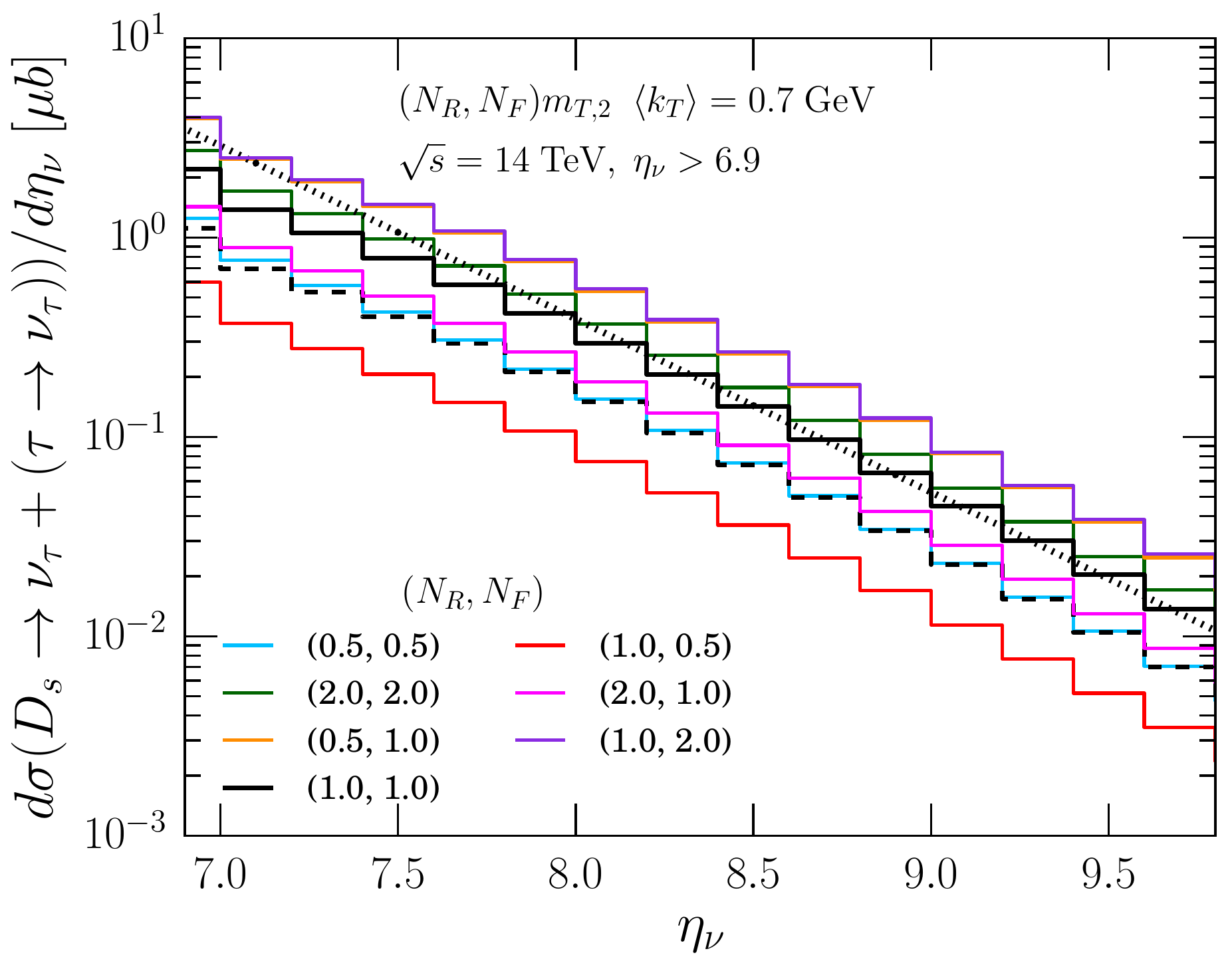}    
    \caption{
    The PROSA PDF uncertainties in the $\eta_\nu$ distribution  
    for $\sqrt{s}=14$ TeV. The upper left plot is for the direct $D_s\to \nu_\tau$ contribution and the upper right plot is for the chain decay $D_s\to\tau\to \nu_\tau$
    contribution. The lower left plot shows the ratio of the direct to chain decay distributions.  The lower right plot 
    shows the scale uncertainties in the $\eta_\nu$ distribution
    summing the direct and chain contributions, where the dashed histogram shows the direct contribution and the dotted line shows the approximation for $d\sigma/d \eta_\nu$ from eq. (\ref{eq:etaarea}).
}
    \label{fig:direct-chain-PROSAfits}
\end{figure}

We now turn to tau neutrino and antineutrino production from $D_s^\pm$ production and decay, including both the direct neutrinos $D_s\to \nu_\tau$ and chain decay neutrinos $D_s\to \tau\to \nu_\tau$. 
We observe that data on 
$\nu_\tau$ and $\bar{\nu}_\tau$ are not available in standard LHC experiments, for which all neutrinos produced in an event just contribute to the event as missing energy. We assume a  branching fraction  $B(D_s\to \tau\nu_\tau)=0.0548$ \cite{Tanabashi:2018oca}. Our perturbative evaluation produces equal numbers of $D_s^+$ and $D_s^-$. The energy distributions of left-handed $\nu_\tau$'s are equal to the energy distributions of right-handed $\bar\nu_\tau$'s. All of the predictions shown in this section are for the sum $\nu_\tau+\bar\nu_\tau$, and are generically referred to as ``neutrinos."

We begin with the neutrino rapidity distributions. 
Fig.~\ref{fig:direct-chain-PROSAfits} shows the PROSA PDF uncertainties for the $\eta_\nu$ distribution of the tau neutrinos from direct ($D_s\to \nu_\tau$, upper left panel) and chain decay ($D_s\to \tau\to \nu_\tau$, upper right panel). The two contributions to the neutrino rapidity distributions show similar PDF uncertainties, in the range of $\pm 20-30\%$ in the ratio to the central PDF. The PROSA model uncertainty dominates. The lower left plot of 
Fig.~\ref{fig:direct-chain-PROSAfits} shows the ratio of the $\eta_\nu$ distributions of the direct and chain neutrinos. Their contributions are nearly equal across the considered rapidity range $\eta_\nu > 6.9$, with a maximum difference of very few percent at the largest $\eta_\nu$. 

The lower right plot of Fig.~\ref{fig:direct-chain-PROSAfits} shows the $\eta_\nu$  distributions from the sum of direct plus chain contributions and the scale uncertainty. For reference, the dashed line is just the direct contribution. 
The scale uncertainties, again, are significantly larger than the total PROSA PDF uncertainties. The scale uncertainty envelope for the rapidity distributions range between the lower edge of the envelope that is $\sim 25\%$ of the central value (a correction of $\sim -75\%$ to the central value) to close to a factor of 2 times 
the central scale result for the upper edge of the scale uncertainty envelope, nearly independent of energy. 

At high rapidity, the lower right plot shows that the rapidity distributions for different scale combinations have a common behavior 
 (i.e., the same shape) as a function of rapidity, and they only differ among each other for the normalization. 
The dotted line 
corresponds to the analytical formula
\begin{equation}
\label{eq:etaarea}
    \frac{d\sigma}{d\eta_\nu}\simeq \Bigl( 0.214\ \mu {\rm b}\Bigr)\,e^{-2(\eta_\nu-8.3)}\,.
\end{equation}
This represents the central scale histogram values to within $\pm 5\%$ for  $\eta_\nu>8.3$ in Fig.~\ref{fig:direct-chain-PROSAfits}, but eq. (\ref{eq:etaarea}) lies above the histograms in the lower right panel  for $\eta_\nu<8.3$. The scaling behaviour at large $\eta_\nu$  appears to be a universal feature, independent of $N_R$ and $N_F$ for $(N_R,N_F)m_{T,2}$ scales when $\eta_\nu>8.3$.

We probe the functional behavior of $d\sigma/d\eta_\nu\sim e^{-2\eta_\nu}$ for $\eta_\nu>8.3$ with the following considerations. We consider a cylindrical detector at a distance $D_d$  from the interaction point, aligned and centered along the beam axis ($z$-axis), ``on-axis," for which the minimum detectable $\eta_\nu$ is labelled as $\eta_1$. Given the relation between the pseudorapidity and angle $\theta$ relative to the $z$-axis, the cross sectional area of the detector is 
\begin{eqnarray}
\label{eq:area}
    A_d(\eta_1)&\simeq& 4\pi D_d^2 e^{-2\eta_1}\\
    \label{eq:omega}
    &=& D_d^2\Omega_\nu(\eta_1)
    \,,
\end{eqnarray}
where $\Omega_\nu(\eta_1)\equiv4\pi e^{-2\eta_1}$
is the approximate solid angle enclosed by a circle around the $z$-axis for which the angle $\theta$ relative to the $z$-axis corresponds to $\eta_1$ and is a distance $D_d$ from the interaction point. The
functional form of $d\sigma/d\eta_\nu$  in eq.
(\ref{eq:etaarea}) is therefore related 
to $\Omega_\nu(\eta_\nu)$. 
We will come back to this point in our evaluation of the number of events per ton of target (see Section \ref{sec:evts}), which depends on the transverse area and the depth of the target. The scaling of $d\sigma/d\eta_\nu$  with $A_d(\eta_\nu)$  for large $\eta_\nu$ has already been noted in Ref.~\cite{Kling:2021gos}.

\begin{figure}[htbp]
    \centering
\includegraphics[width=0.49\textwidth, trim=2cm 0 2cm 0]{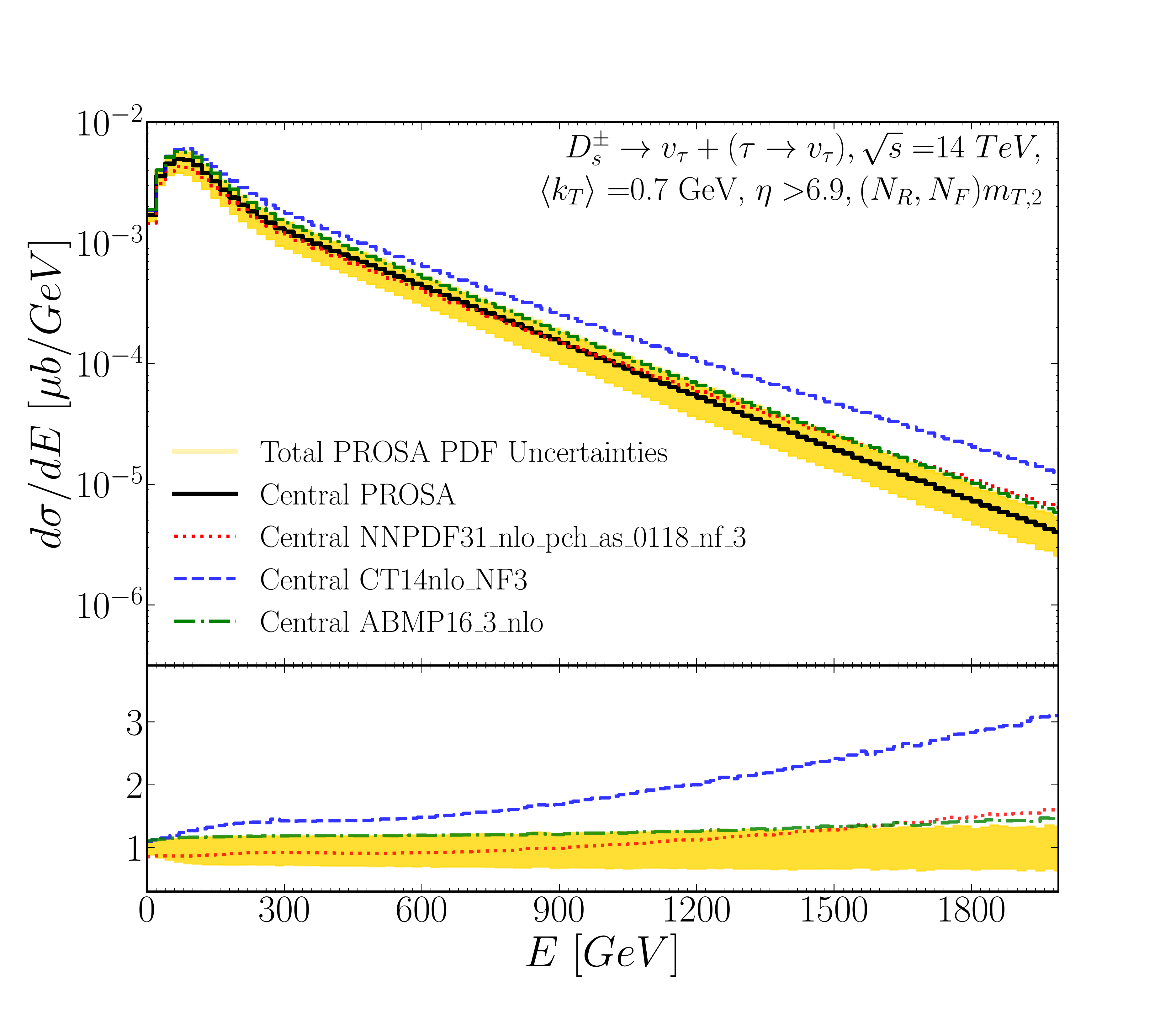}
\includegraphics[width=0.49\textwidth, trim= 2cm 0 2cm  0]{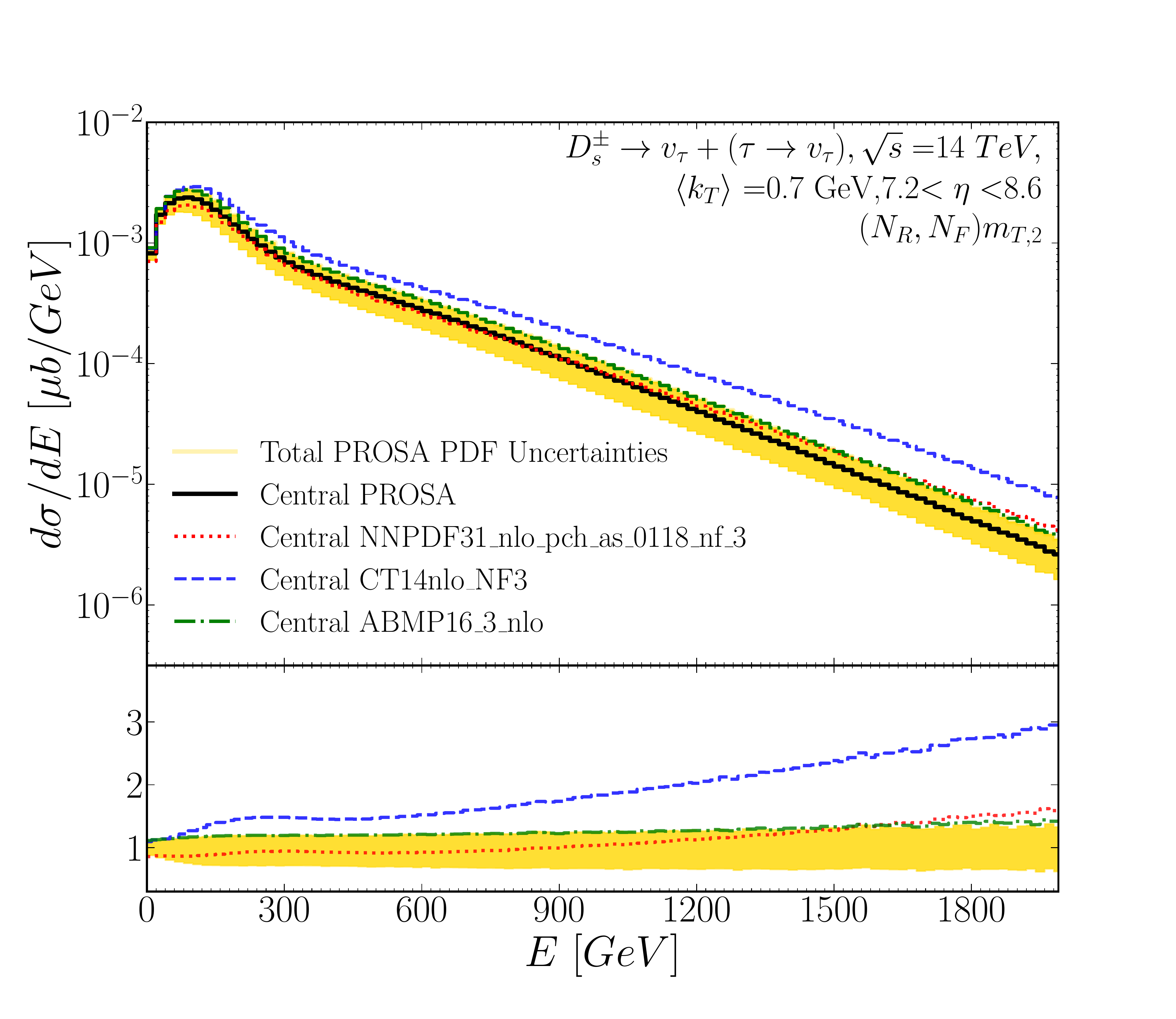}    
\includegraphics[width=0.49\textwidth, trim=2cm 0 2cm 0]{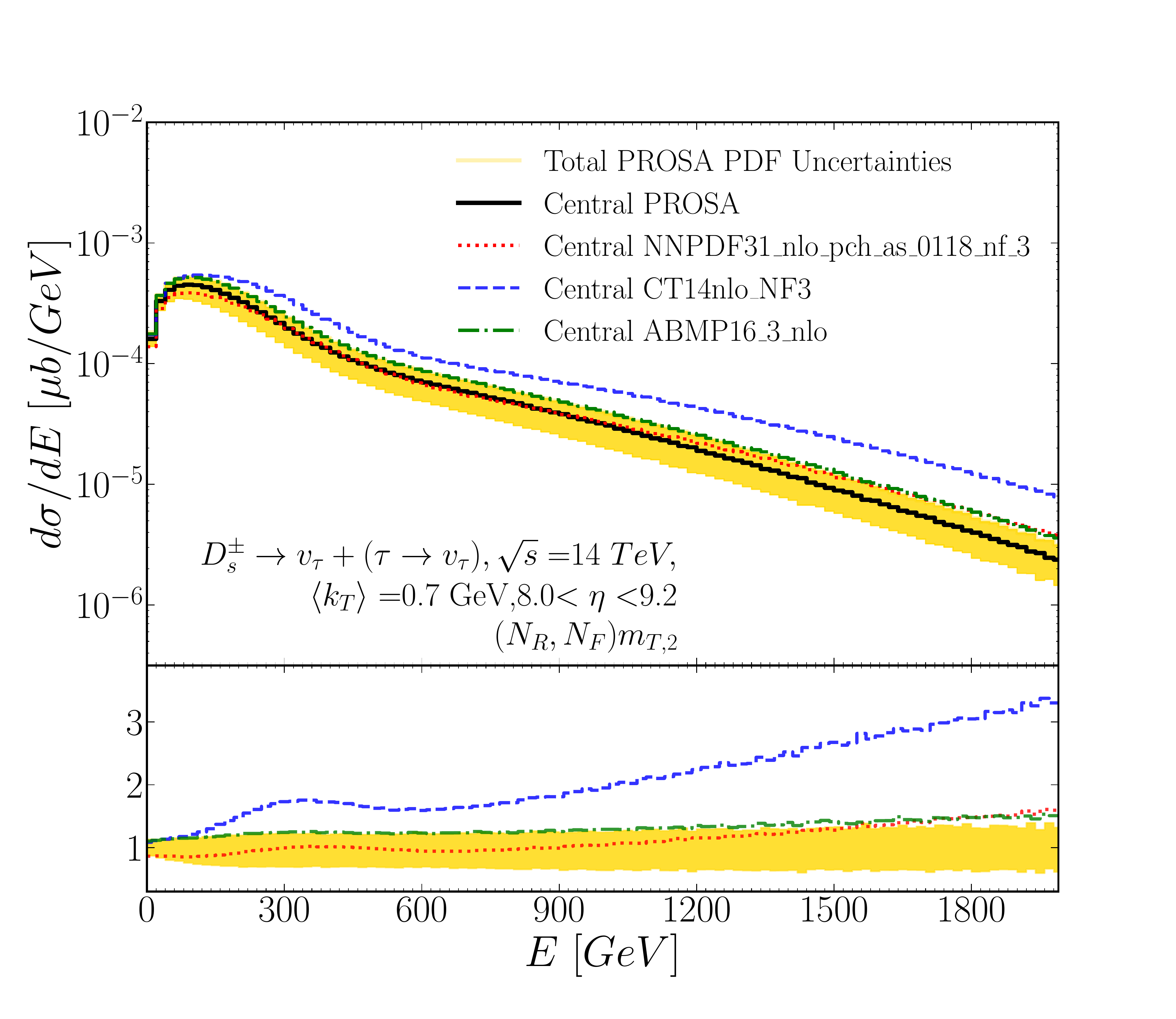} 
\includegraphics[width=0.49\textwidth, trim=2cm 0 2cm 0]{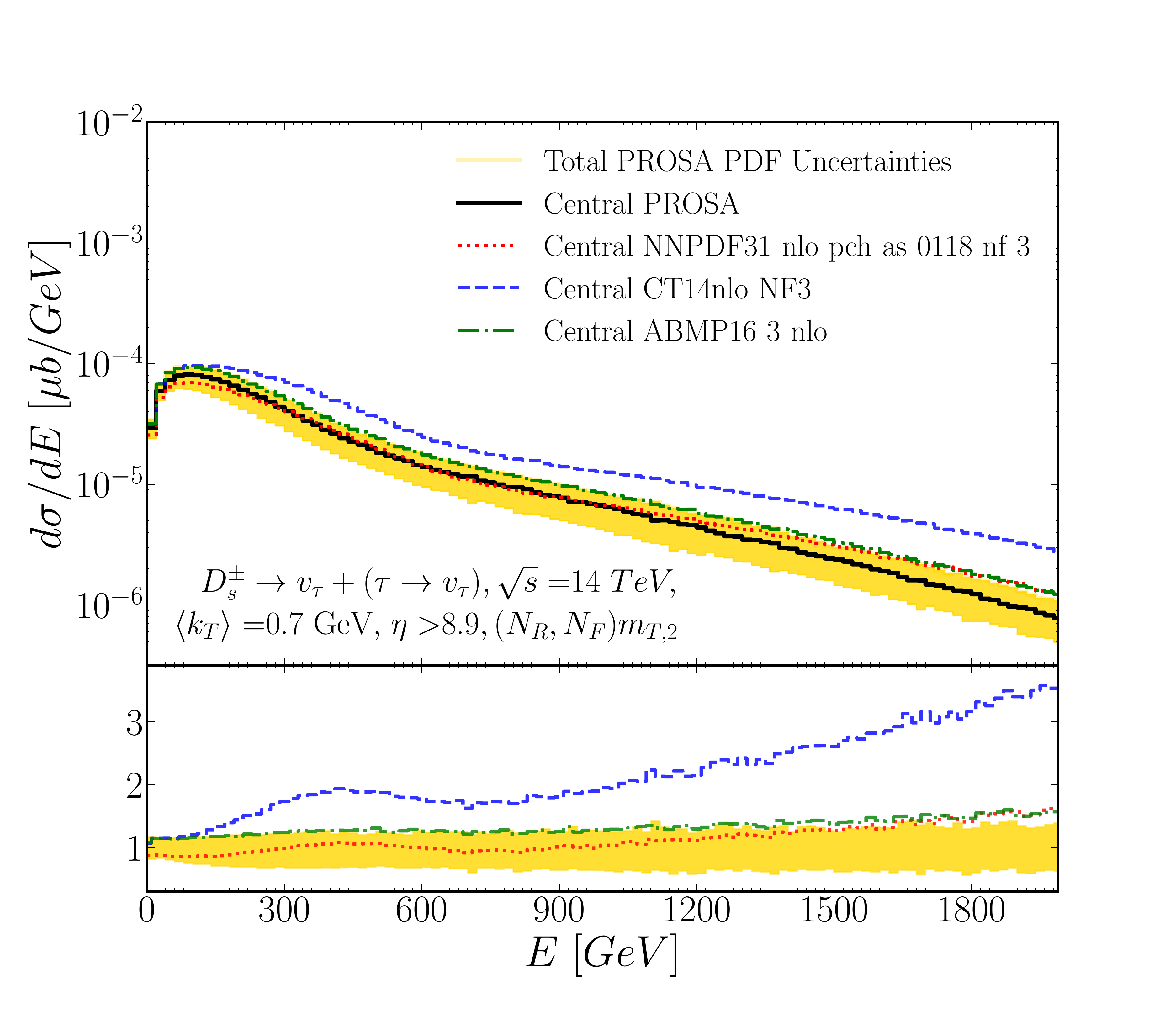}
     \caption{
      The PROSA PDF uncertainties in the NLO distribution of tau neutrino energy in the pseudorapidity range  $\eta _\nu> 6.9$ (upper left), 7.2 $< \eta _\nu<$ 8.6 (upper right), 
      $8.0 < \eta_\nu < 9.2 $ (lower left) 
      and $\eta_\nu > 8.9$ (lower right), respectively, for $pp$ collisions at $\sqrt{s} = 14$ TeV. The PROSA uncertainty envelope is shown in yellow in each panel. The red dotted, green dashed, blue dot-dashed curves correspond to the ratio of the NNPDF3.1, CT14 and ABMP16 NLO predictions to the central PROSA NLO prediction. 
     } \label{fig:energy-pdfuncertainties}
\end{figure}

\begin{figure}[h]
    \centering
    \includegraphics[width =0.49\textwidth, trim=2cm 0 2cm 0]{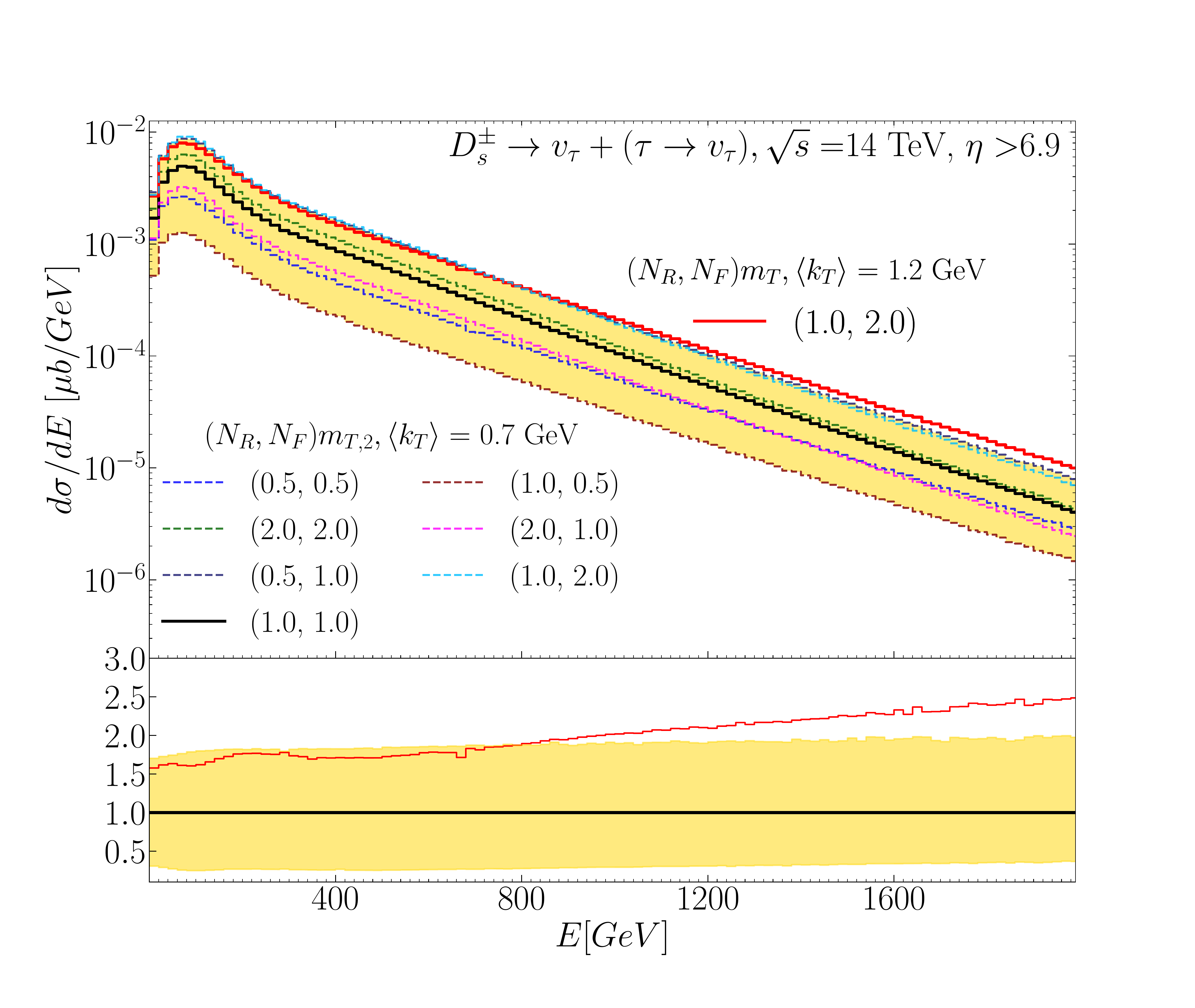}
    \includegraphics[width =0.49\textwidth, trim=2cm 0 2cm 0]{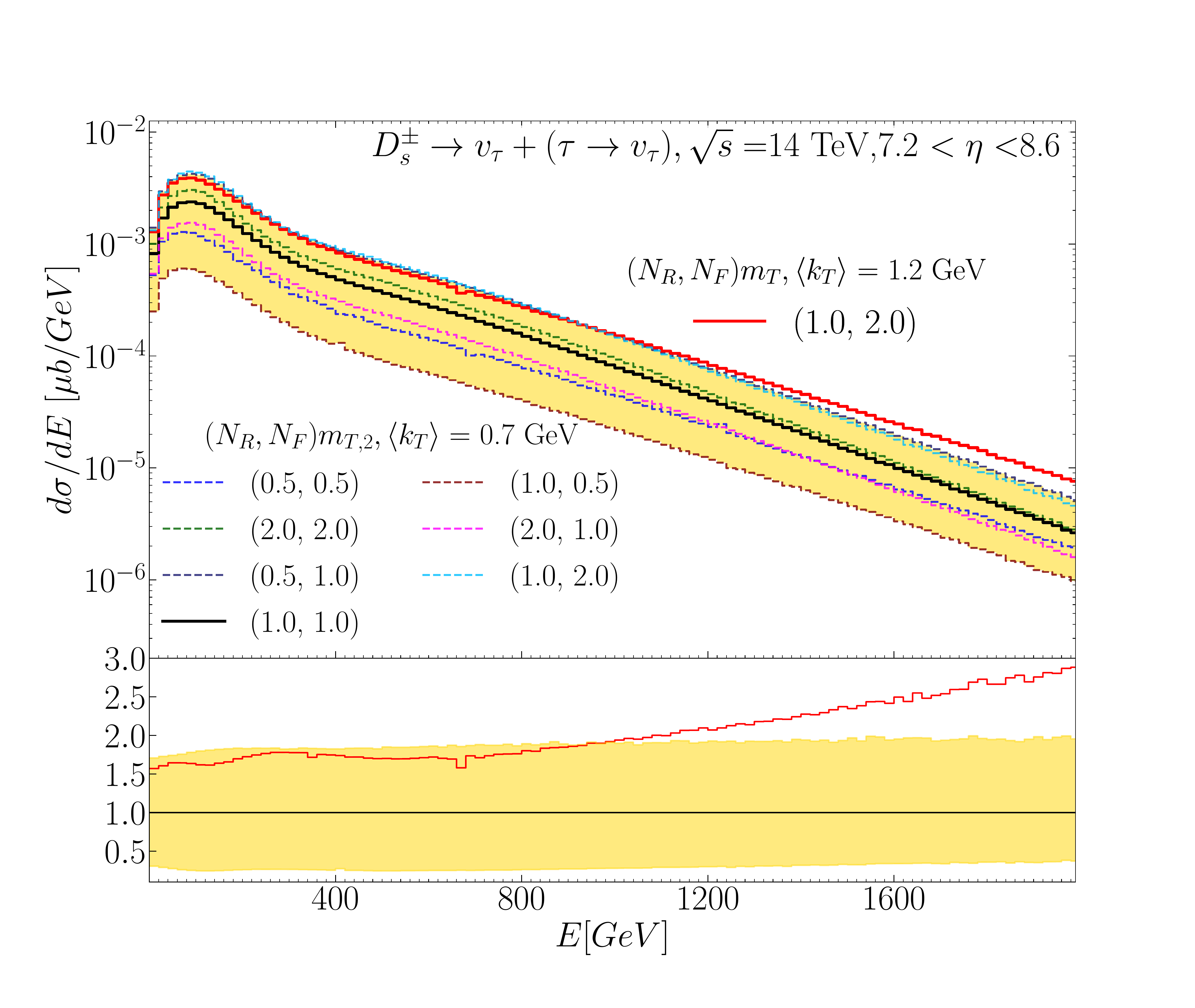}
    \includegraphics[width =0.49\textwidth, trim=2cm 0 2cm 0]{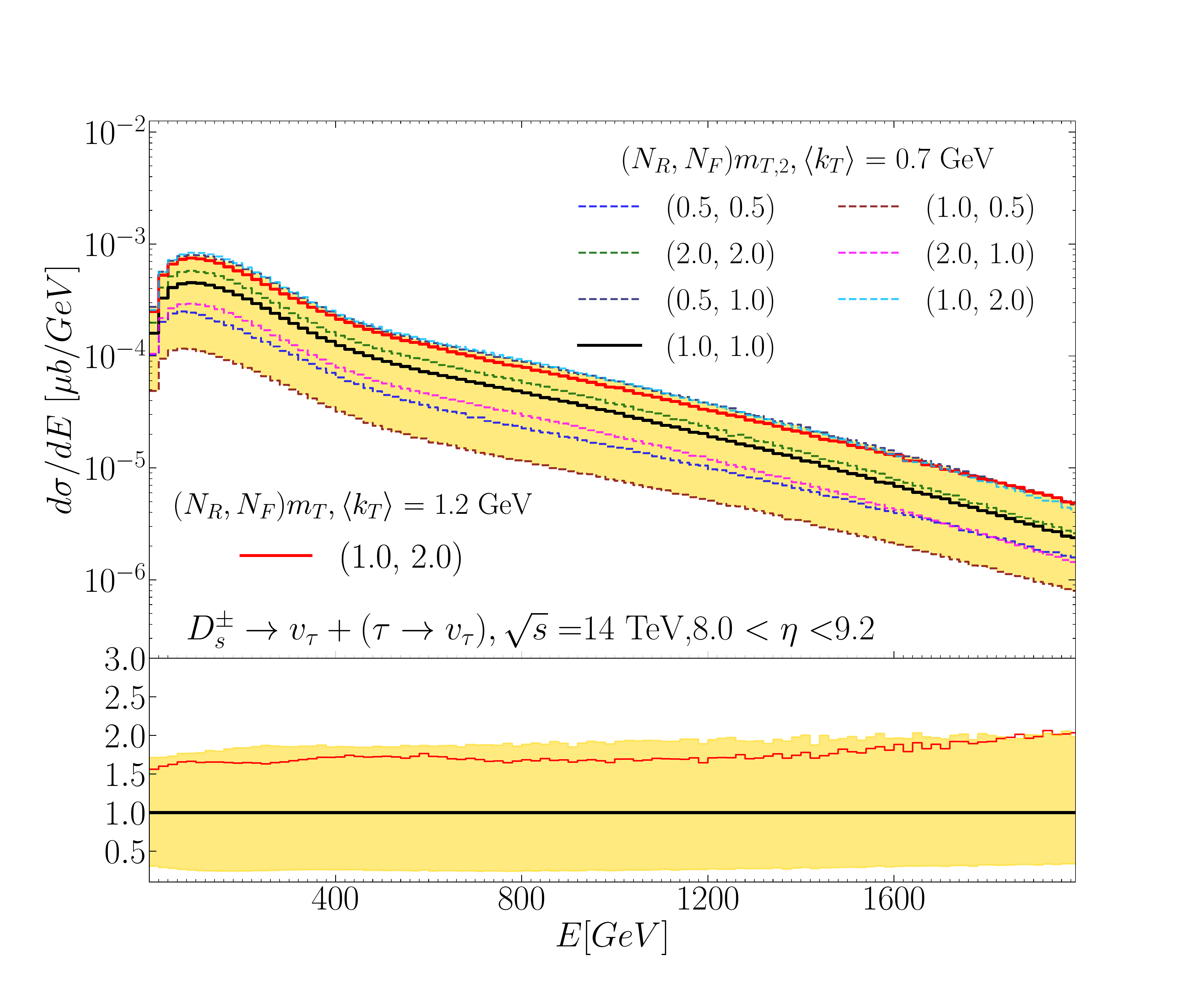}
   \includegraphics[width =0.49\textwidth, trim=2cm 0 2cm 0]{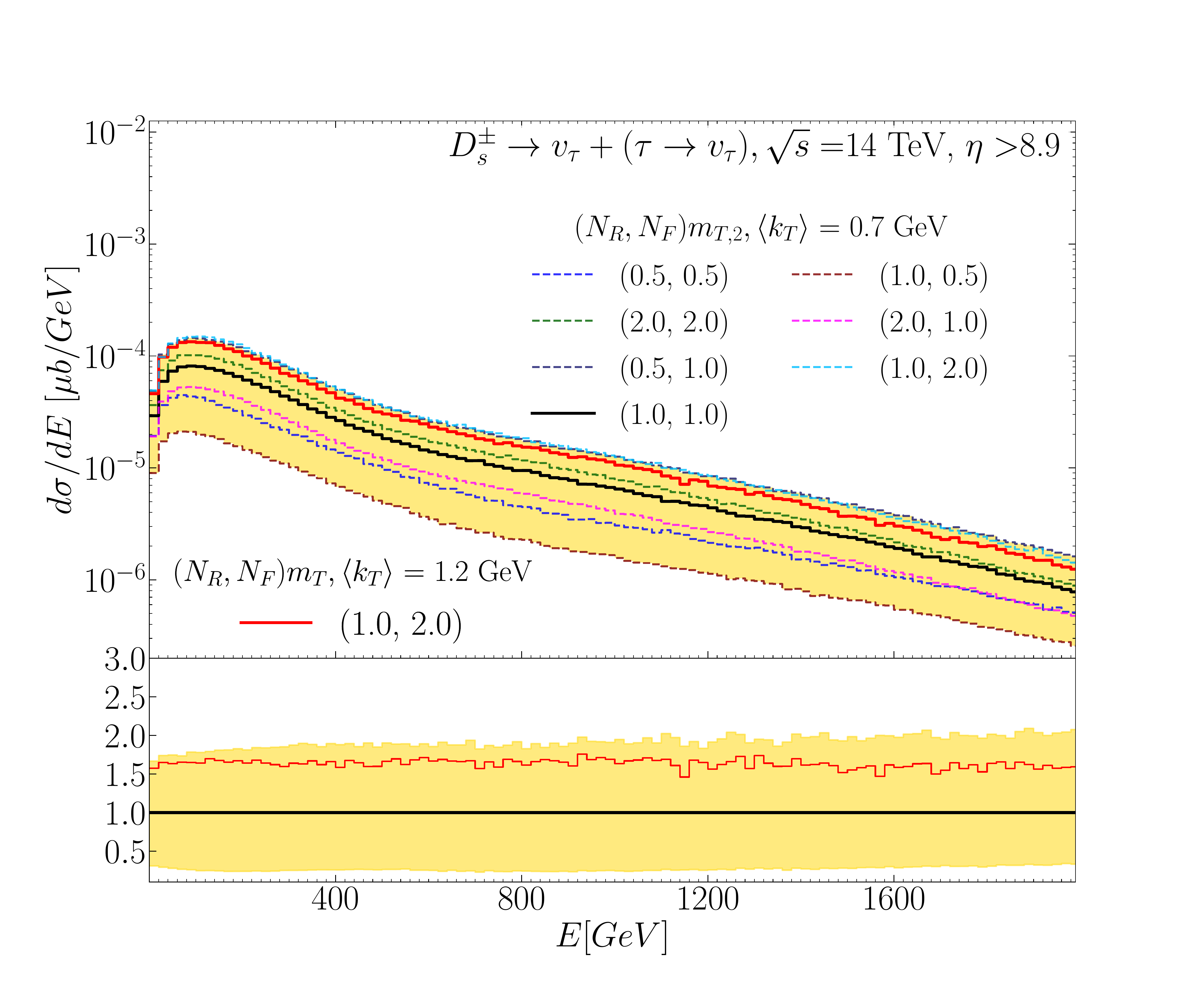}  
    \caption{
    Central predictions and scale uncertainties using as input  $(\mu_R,\mu_F)=(1,1)m_{T,2}$ and $\langle k_T\rangle = 0.7$ GeV (black histograms and bands) for the $E_\nu$ distributions for $\eta_\nu>6.9$ (upper left), $7.2<\eta_\nu<8.6$ (upper right),
    $8.0<\eta_\nu<9.2$ (lower left) and $\eta_\nu>8.9$ (lower right) in $pp$ collisions with $\sqrt{s}=14$ TeV. 
    Also shown are energy distributions for scales $(1,2)m_T$ with $\langle k_T\rangle =1.2$ GeV (red solid). The scale dependence envelope relative to the central scale choice of $(\mu_R,\mu_F)=(1,1)m_{T,2}$ is shown in the ratio plot for each rapidity range. The ratio of the $(1,2)m_T$,  $\langle k_T\rangle =1.2$ GeV prediction to the central prediction with the default parameter set is also shown with the red line in each ratio plot. 
    }
    \label{fig:e-direct-chain-scale-tot}
\end{figure}

For a detector on-axis of radius 1 m at a distance of 480 m from the interaction point, $\eta_\nu > 6.87$. We show results for $\eta_\nu > 6.87$ and other higher neutrino rapidity ranges corresponding to SND@LHC ($7.2<\eta_\nu<8.6$), FASER$\nu$ ($\eta_\nu>8.9$) and the range of 8.0 -- 9.2.
For the energy distributions of tau neutrinos whose rapidity is restricted to be higher than $\eta_{\rm min}=6.87\simeq 6.9$, 
the PROSA PDF uncertainties gradually increase 
from low tau neutrino  energies 
up to the 2 TeV energy range, as illustrated in Fig.~\ref{fig:energy-pdfuncertainties}. 
The black histogram and yellow band reflect the central PROSA PDF predictions and PDF uncertainty bands, respectively. 
Relative to the central result, the PDF uncertainty band has approximately the same shape as a function of energy for all  the $\eta_\nu$ ranges shown. 
As also in case of the rapidity distributions, the ABMP16 and NNPDF3.1 predictions lie within the PROSA PDF uncertainty band. However, in the $E_\nu$ range of 1-2 TeV, they tend towards the upper edge of the PDF uncertainty band. Again, the CT14 predictions are higher. The energy distributions obtained with the CT14 PDFs are larger by almost a factor of $\sim 4$ for $\eta_\nu>8.9$ at high energy. 
Only the PROSA PDF uncertainty bands are shown in Fig.~\ref{fig:energy-pdfuncertainties}. One expects that the uncertainty bands for the other PDFs will 
partly overlap with the PROSA bands. 
While the considered PDFs are compatible for a range of partonic $x$ values,
Fig.~ \ref{fig:energy-pdfuncertainties}  illustrates that
for tau neutrino energy distributions at high rapidities, predictions from different PDFs may vary widely.

Fig. \ref{fig:e-direct-chain-scale-tot} shows that the scale uncertainties for the same rapidity ranges  as in Fig.~\ref{fig:energy-pdfuncertainties} are largely independent of energy and much larger than the PROSA PDF uncertainties. Consistent with the neutrino rapidity distribution scale uncertainties, neutrino energy distributions corresponding to different scale combinations as obtained in the scale variation procedure range between $\sim 25\%$ to  a factor of $\sim$ 2 compared to predictions with the central scale.

In addition to our default scale dependence on $m_{T,2}$, Fig.~\ref{fig:e-direct-chain-scale-tot} shows the neutrino energy predictions using as input $(1.0,2.0)m_{T,1}$ with $\langle k_T\rangle =1.2$ GeV, a choice that better matches the LHCb experimental charm meson transverse momentum distributions.
With these inputs, the upper two panels show that the 
corresponding histogram lies at the upper edge of the scale uncertainty band below $\sim 1$ TeV, and, for rapidities below $\sim$ 8, is above the scale uncertainty band for tau neutrino energies above 1 TeV, as we discuss below. 

Tables \ref{tab:6.9}--\ref{tab:8.9} in Appendix \ref{sec:tables} list, for various rapidity ranges, the central values and uncertainties of the predictions for energy distribution obtained with input $(1,1)m_{T,2}$, $\langle k_T\rangle$ =0.7 GeV and the PROSA PDFs,  corresponding to the plots of Fig.~\ref{fig:energy-pdfuncertainties} and \ref{fig:e-direct-chain-scale-tot}. Tables \ref{tab:6.9a}--\ref{tab:8.9a} list the predictions for scales $(1,2)m_T$ with $\langle k_T\rangle =1.2$ GeV and the PROSA PDFs, and the NNPDF3.1, CT14 and ABMP16 predictions with $(1,1)m_{T,2}$ and $\langle k_T\rangle =0.7$ GeV. The Tables are available as ancillary files on the \href{https://arxiv.org/src/2112.11605v1/anc}{arXiv link} to this paper.

\begin{figure}
    \centering
     \includegraphics[width=0.525\textwidth]{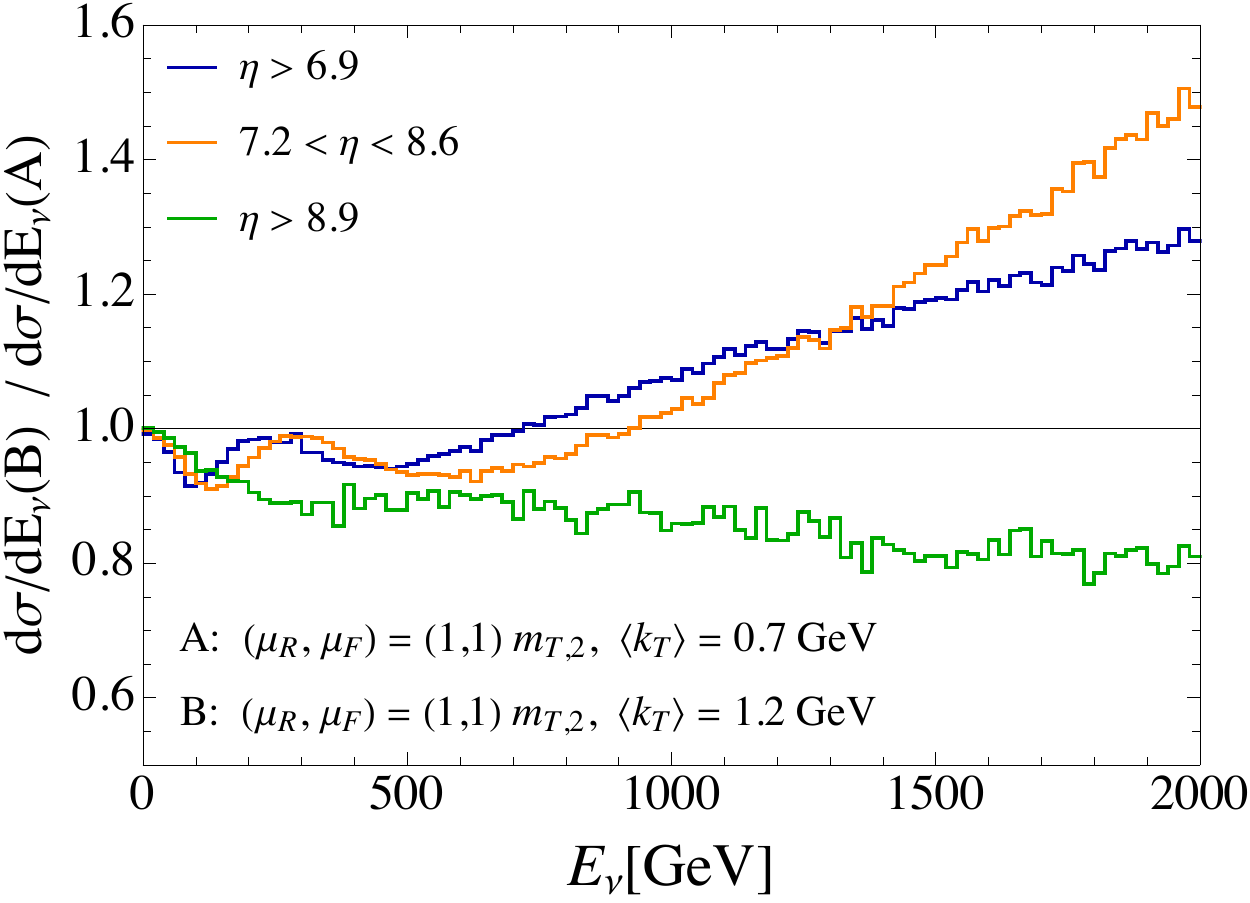}    
    \includegraphics[width=0.525\textwidth]{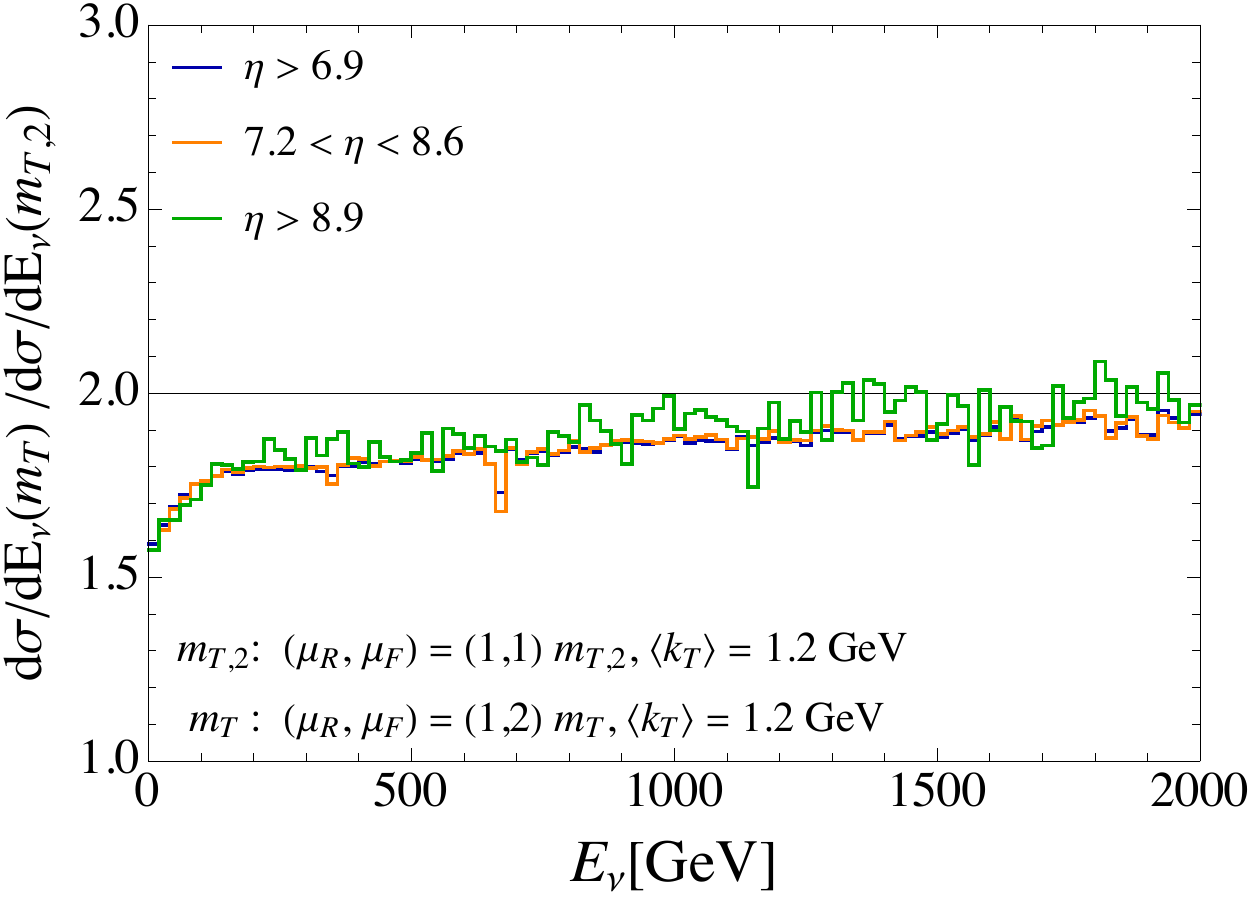} 
    \includegraphics[width=0.525\textwidth]{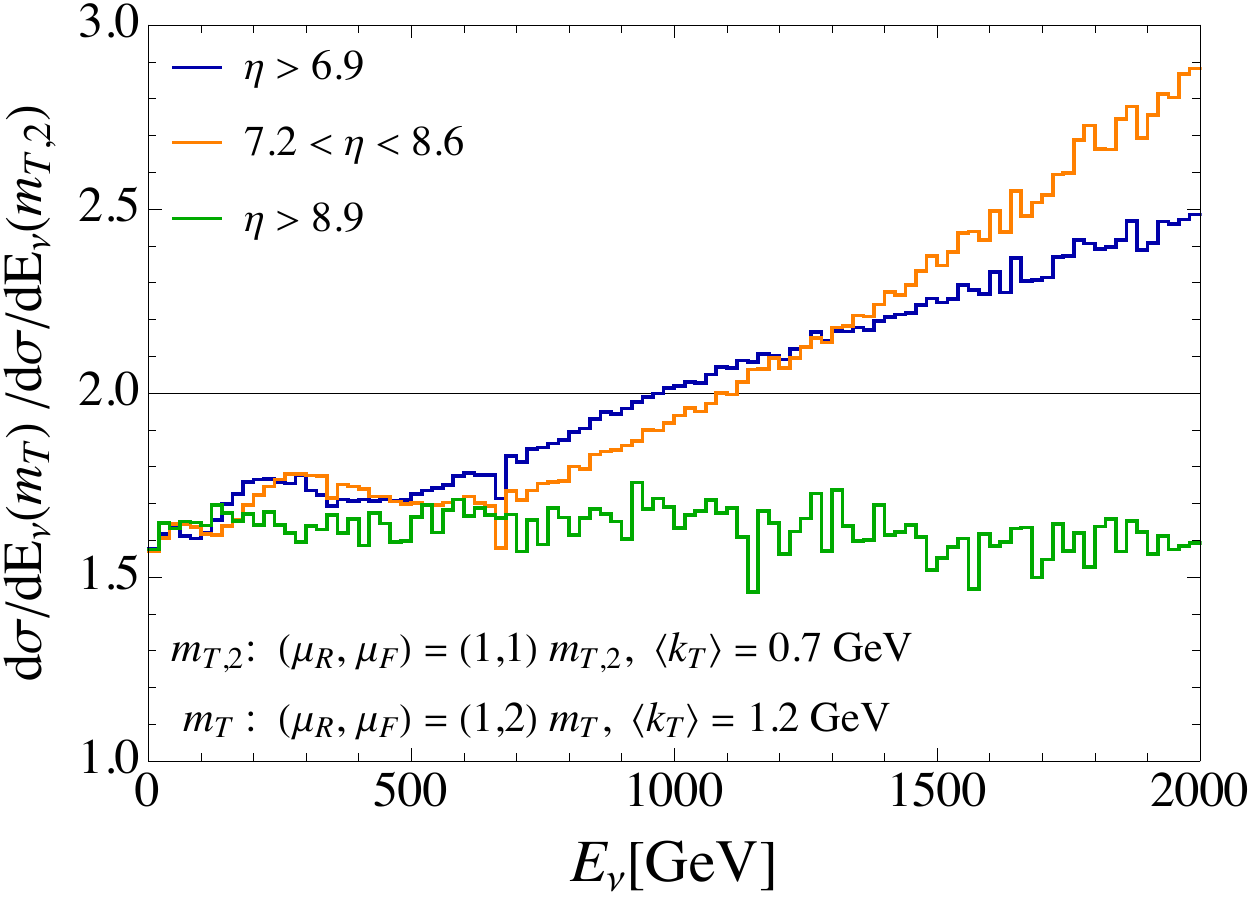}     
    \caption{Ratio of the tau neutrino plus antineutrino distribution $d \sigma/d E_\nu$  for $\eta_\nu>6.9$, $7.2<\eta_\nu<8.6$ and $\eta_\nu>8.9$. The upper plot has the ratio with $m_{T,2}$ scale dependence and $\langle k_T\rangle=1.2$ GeV (numerator) and 0.7 GeV (denominator). The middle plot has $\langle k_T\rangle=1.2$ GeV fixed with scale dependence $(1,2)m_T$ (numerator) and $(1,1)m_{T,2}$ (denominator).
    The lower plot has both $\langle k_T\rangle$ and scale differences in the ratio.}
    \label{fig:dsigmade-ratios}
\end{figure}

Fig.~\ref{fig:dsigmade-ratios} shows detailed comparisons of the 
ratio of energy distributions obtained with
the two scale choices, $m_{T,1}$ and $m_{T,2}$ defined in eqs. (\ref{eq:mt2def}-\ref{eq:mt1def}) and $\langle k_T\rangle$ = 0.7 and 1.2 GeV. With the aim of disentangling $\langle k_T\rangle$ and scale dependence effects and investigating each of them separately, we first fix the renormalization and factorization scales and vary $\langle k_T\rangle$. We then fix $\langle k_T\rangle$ and change the scales. Finally we vary the scales and $\langle k_T\rangle$.

The upper  panel of Fig.~\ref{fig:dsigmade-ratios} shows the ratio of energy distributions for $\langle k_T\rangle=1.2$ GeV to $\langle k_T\rangle=0.7$ GeV, for the scales fixed at $(\mu_F,\mu_R)=(1,1)m_{T,2}$. The $\langle k_T\rangle$ smearing moves very low $p_T$ mesons to higher $p_T$, thereby increasing the distribution in 
the high-$p_T$ tails 
up to 10 GeV and larger at LHCb, as shown in Ref. \cite{Bai:2020ukz}. 
At high energies,
\begin{eqnarray}
p_{T,\nu}\lsim 0.27\ {\rm GeV}\Biggl(\frac{E_\nu}{\rm TeV}\Biggr),\quad \eta>8.9\\
p_{T,\nu}\lsim 2\ {\rm GeV}\Biggl(\frac{E_\nu}{\rm TeV}\Biggr),\quad \eta>6.9.
\end{eqnarray}
At very forward rapidity, the larger of the two $\langle k_T\rangle$ values considered here, namely $\langle k_T\rangle=1.2$ GeV, 
pushes a fraction of the differential cross section outside of the allowed rapidity region, thereby decreasing the neutrino energy distribution relative to the distribution evaluated with a smaller $\langle k_T\rangle$. For $\eta_\nu>8.9$, the ratio of the energy distributions $d\sigma(\langle kT\rangle = 1.2$ GeV$)/dE_\nu$ to $d\sigma(\langle kT\rangle = 0.7$ GeV$)/dE_\nu$ shown in the upper panel of Fig.~\ref{fig:dsigmade-ratios}
is about 0.75 
for $E_\nu=2$ TeV. 
For $\eta_\nu > 6.9$ and $7.2 < \eta_\nu < 8.6 $, neutrinos with higher $p_T$ are detectable. The ratio of energy distributions increases by $\sim 25\%$ for $E_\nu=2$ TeV for $\langle k_T\rangle=1.2$ GeV compared to $\langle k_T\rangle=0.7$ GeV.

A larger effect on the neutrino energy distribution 
is generated by the difference in scale choice, for a fixed $\langle k_T\rangle$. 
The middle panel of Fig.~\ref{fig:dsigmade-ratios} shows the ratio of the predictions for the neutrino energy distribution obtained with the $(1,2)m_{T,1}$ and $(1,1)m_{T,2}$ scale choices, for a fixed $\langle  k_T \rangle$ = 1.2 GeV. The ratio of $d\sigma(\mu_R=m_T,\mu_F=2 m_T)/dE_\nu$ to $d\sigma(\mu_R=\mu_F=m_{T,2})/dE_\nu$ in the three different rapidity ranges largely overlap.
This means that the effect of changing the renormalization scales is largely rapidity independent. 
At low energy, $d\sigma / dE_\nu$ evaluated with $(1,2)m_{T,1}$ is $\sim 1.6$ time larger than with $(1,1)m_{T,2}$. The ratio between the energy distributions in the middle panel of Fig. \ref{fig:dsigmade-ratios} 
increases to $\sim 1.9$ as $E_\nu\to 2$ TeV. Specifically for
the two scale choices: $(1,2)m_{T,1}$ and $(1,1)m_{T,2}$,
for the charm transverse momentum near $p_T\to 0$, $m_{T,2}\simeq 2 m_{T,1}$, so the factorization scales are approximately equal and the renormalization scales differ by a factor of 2. For
higher $p_T$ the renormalization scales 
are nearly the same, whereas the PDFs are evaluated at factorization scales that differ by a factor of $\sim 2$.

The lower panel of Fig.~\ref{fig:dsigmade-ratios} shows the ratio of the neutrino energy distributions for different rapidity ranges obtained with $(1,2)m_{T,1}$ and $(1,1)m_{T,2}$, in association with $\langle k_T\rangle=1.2$ GeV and 0.7 GeV, respectively. The trends of the ratios follow the dependence on the ratio at fixed scale for different $\langle k_T\rangle$, scaled by the roughly energy independent ratio of differential cross sections evaluated at two scales and fixed $\langle k_T\rangle$ value.
For $\eta_\nu>6.9$ and $7.2 < \eta_\nu < 8.6$, the ratio increases with neutrino energy, whereas for $\eta_\nu>8.9$, the ratio is nearly energy independent.

\section{Tau neutrino and antineutrino charged current events}
\label{sec:events}

\subsection{Interaction cross sections}
\label{subsec:sigmanu}

PDF uncertainties also apply to the tau neutrino and antineutrino charged current (CC) cross sections. 
However, their size is much smaller than in case of charm production due to the fact that in the characteristic ranges of $x$ and $Q^2\gg m_c^2$ values relevant for the 
calculation of the CC cross sections, the PDF fits are very well constrained by the already available experimental data.
\begin{figure}
    \centering
    \includegraphics[width = 0.7\textwidth]{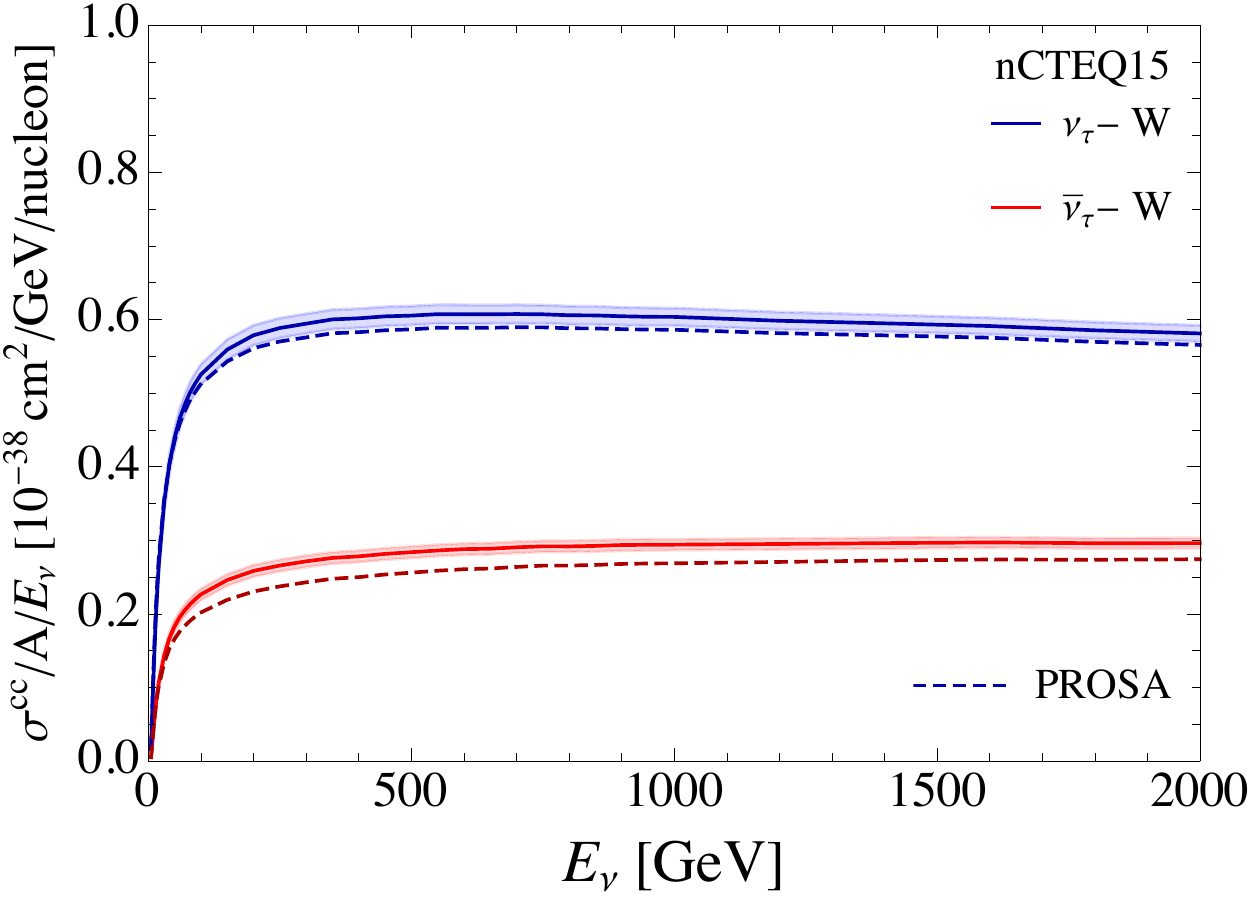}  
    \caption{The charged current cross sections per nucleon  for deep-inelastic scattering 
    of tau neutrinos and antineutrinos
    on tungsten (W) for different PDF sets: nCTEQ15-W (i.e. nCTEQ15 for tungsten) and PROSA (VFNS). For the latter, we assume isospin symmetry in order to build neutron PDFs. The predictions are scaled by (anti-)neutrino energy and the bands reflects uncertainty due to the 32 sets of nCTEQ15-W.   }
    \label{fig:sigma-nuc}
\end{figure}
Fig.~\ref{fig:sigma-nuc} shows the tau neutrino and antineutrino cross sections per nucleon for charged current interaction in a tungsten target, scaled by incident (anti-)neutrino energy. 
For the interaction cross sections, we use the nCTEQ15 \cite{Kovarik:2015cma} as a default PDF set as in Ref.~\cite{Bai:2020ukz}, here for tungsten, and evaluate the deep inelastic scattering (DIS) cross sections at NLO in QCD including target mass and tau lepton mass corrections \cite{Kretzer:2002fr,Kretzer:2004wk,Jeong:2010nt}. For $Q^2<2$ GeV$^2$, we extrapolate the neutrino and antineutrino structure functions with a prescription outlined in Ref. \cite{Reno:2006hj} that is based on the Capella et al. parameterization \cite{Capella:1994cr}. This prescription has low-$Q^2$ behavior that is similar to the Bodek-Yang PDF dependent prescription \cite{Bodek:2002ps,Bodek:2002vp,Bodek:2004pc,Bodek:2005de,Bodek:2021bde}.

The target mass corrections are small, even for tau antineutrino scattering, for energies above 25 GeV \cite{Kretzer:2004wk}. The tau mass correction accounts for a $\sim 25\%$ suppression of the tau neutrino cross section relative to the muon neutrino cross section for 100 GeV incident energies, reducing to $\sim 5\%$ tau mass effect at 1 TeV \cite{Jeong:2010nt}.
The tau neutrino and antineutrino CC cross section  for incident neutrino energies below 10 GeV do not make significant contributions to the total number of events over all energy spectrum.
We therefore neglect quasi-elastic and resonant scattering contributions, which are smaller than the DIS contribution for $E_\nu \gsim 10$ GeV.  
For neutrino energies above 100 GeV, the resonance region accounts for only a few percent of the total cross section \cite{Feng:2022inv,Jeong:2022tba}. We find that the contribution from $Q^2<1$ GeV$^2$ to the cross section at 100 GeV is less than 3\% (5\%) for $\nu_\tau$ ($\bar\nu_\tau$) scattering with our low-$Q^2$ extrapolation of the structure functions \cite{Feng:2022inv,Jeong:2022tba}. The fact that NLO QCD corrections to the neutrino interaction cross section for most of the energy range of interest are small \cite{Jeong:2010nt} means that our neglect of factorization and renormalization scale dependence of the neutrino cross section is reasonable. 
The bands in the figure display the uncertainty due to the 32 different PDF sets of nCTEQ15 for tungsten, which are obtained according to eq. (\ref{eq:ncteqerror}).
Deviation of the charged current cross sections due to the 
 nCTEQ15 PDF uncertainty from the one with the central PDF set is about 2--3 \% for $E_\nu \gsim 100 \, {\rm GeV}$ for both tau neutrinos and antineutrinos. Table \ref{tab:sigcc} in Appendix \ref{sec:tables} lists the $\nu_\tau$ and $\bar\nu_\tau$ CC cross sections per nucleon, including PDF uncertainties from the nCTEQ15 PDFs, for scattering with a tungsten target.

\begin{figure}
    \centering
    \includegraphics[width=0.7\textwidth]{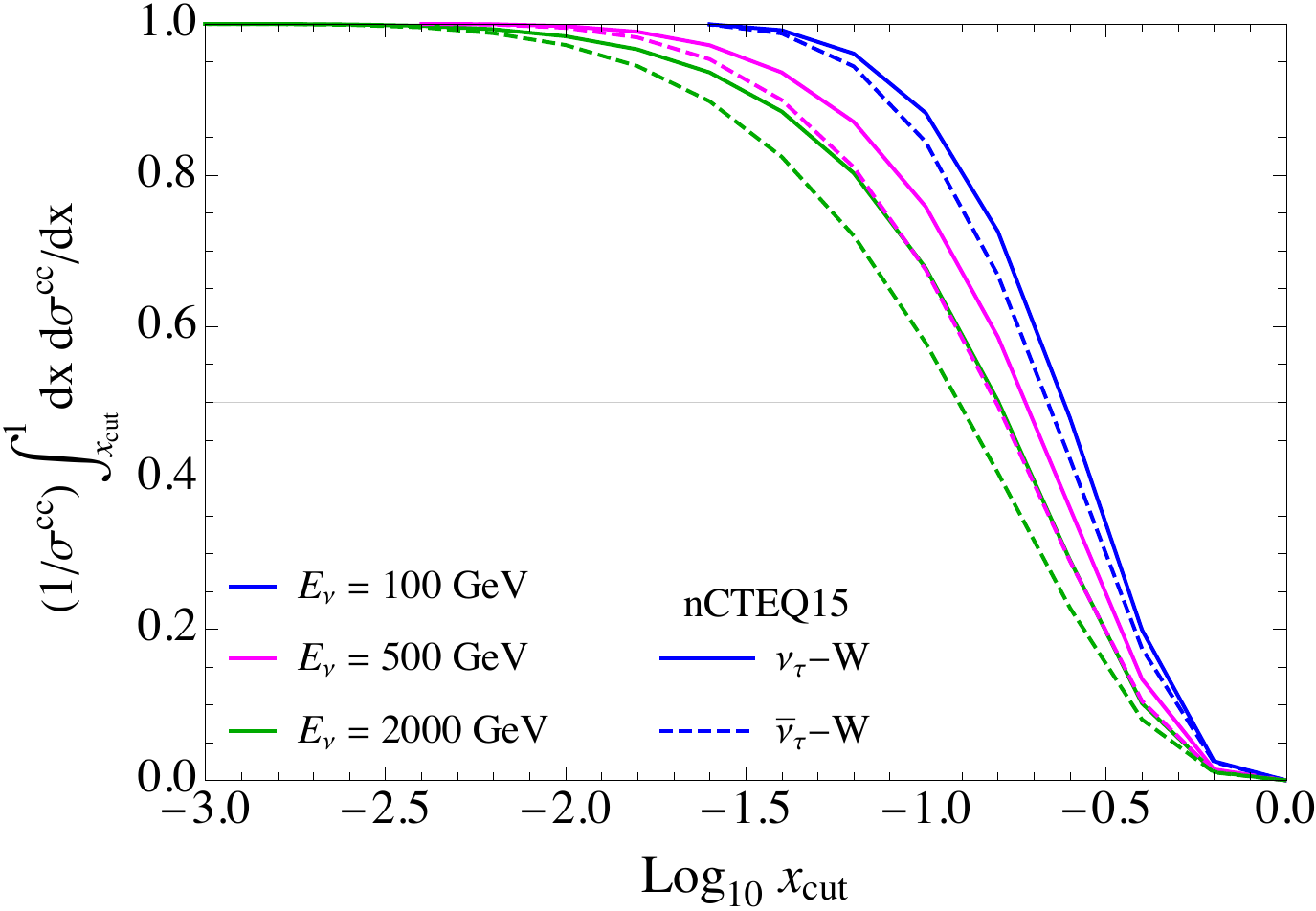}    
    \caption{For selected tau neutrino energies, the fractions of the charged current neutrino-tungsten cross section as a function of $x_{\rm min}=x_{\rm cut}$ are presented for tau neutrinos and antineutrinos.}
    \label{fig:sigma-frac}
\end{figure}

For comparison, we also present the cross sections evaluated with the PROSA PDF set in the variable flavour number scheme (VFNS) \cite{Zenaiev:2019ktw}. We use the VFNS version to be consistent with the nCTEQ15 PDFs, extracted in the same scheme, and considering the kinematical range of $Q^2$ values we are mostly interested in. For example, for $E_\nu=100$ GeV, the $\nu_\tau$ charged current DIS cross section on nucleons has $\langle Q^2\rangle=23$ GeV$^2$, while for $\bar{\nu}_\tau$ at the same incident energy, $\langle Q^2\rangle=13$ GeV$^2$ \cite{Reno:2021xx}. Most of the CC events come from higher energies where $\langle Q^2\rangle$ is even higher.
The PROSA collaboration so far only fitted proton PDFs, and not yet nuclear PDFs. In that case, we obtained the neutron PDFs according to isospin symmetry.
Only shown, with the dashed curves, are the predictions 
for the tau neutrino and antineutrino CC cross sections per nucleon for interactions with tungsten with the best fit of the PDF since other PROSA PDF sets necessary for evaluating  uncertainty are not provided in the VFNS. Table \ref{tab:sigcc-prosa} in Appendix \ref{sec:tables} lists the $\nu_\tau$ and $\bar\nu_\tau$ CC cross sections per nucleon for the PROSA VFNS PDFs.
The resulting cross sections with the PROSA VFNS PDFs are smaller than the central results with nCTEQ15 by $\sim$~3\% for tau neutrinos and $\sim$~7--11\% for antineutrinos in the energy range of interest, $100 \lsim E_{\nu_\tau (\bar{\nu}_\tau)} / {\rm GeV} \lsim 2000$, hence not in the uncertainty band of the nCTEQ15, although one could expect that, in case an uncertainty band would accompany the PROSA VFNS PDF fit as well, this would overlap with the nCTEQ15 uncertainty band.
{The discrepancy 
between the central predictions with the two PDF best-fits
is due to the combined effect of the differences between the PDFs and the way nuclear effects are incorporated.  
The largest impact comes from} the difference between the nCTEQ15 and PROSA PDFs in the large $x$ region, where the PDFs are not well constrained by the data. 
In Fig.~\ref{fig:sigma-frac}, we present the fractions of the CC cross section
according to the minimum values of parton-$x$ in the evaluation. 
Fig.~\ref{fig:sigma-frac} illustrates that the
$x$ range extends to lower $x$ values as the neutrino energy increases.
We find that  
for $E_\nu=2$ TeV, 68 (58)\% of the contribution to the CC tau neutrino (antineutrino) cross sections 
comes from 
phase-space configurations characterized by
$x > 0.1$, and 98 (96)\% is from $x>0.01$.

\subsection{Charged-current event distributions in the forward detectors}
\label{sec:evts}

In this section, we investigate the expected number of the tau neutrinos and antineutrinos events for the next stages of the LHC, Run 3 and High Luminosity LHC (HL--LHC). 
In our evaluation here, we consider contributions only from $D_s^\pm$ meson production and decays, which yield most of the tau neutrino and antineutrino events.
The contribution from $B$ mesons was investigated in Ref. \cite{Bai:2020ukz}; these  amount to $\sim$ 4 \% of total events.

As in Ref. \cite{Bai:2020ukz}, we calculate the number of neutrinos that can be detected at the LHC using
\begin{equation}
    \frac{dN}{dE_\nu} = \frac{d \sigma}{dE_\nu} (pp \to \nu_\tau  X + \bar\nu_\tau  X ) \times {\cal L} \times {\cal P_{\rm int}} \ 
\end{equation}
with the integrated luminosity ${\cal L} = 150 {\, \rm fb^{-1}}$ and $3000 {\, \rm fb^{-1}}$ for Run 3 and HL--LHC, respectively. All of our results presented above are integrated over $2\pi$ azimuthal angle.
Here, we use the interaction probability in the detector given by
\begin{equation}
    {\cal P_{\rm int}} = 
    (\rho_{W} \times  L_{\rm d} \times N_{\rm avo}) 
    \frac{\sigma_{\nu W}}{A_W} \ ,
\end{equation}
assuming that detectors are made of tungsten, thus the mass number is $A_W = 184$, 
and the density $\rho_W = 19.3 {\, \rm g / cm^3}$.  

We show results for a cylindrical detector aligned with the beam axis and capable of detecting neutrinos with $\eta_\nu > \eta_1$, constituted by a fixed mass of tungsten equal to 1 ton ($M_0=10^6$ g).
The mass can be written in terms of the solid angle $\Omega_\nu(\eta)$ defined in Eq.~(\ref{eq:area}),  the distance $D_d$ from the interaction point and the length $ L_d$ of the cylinder, according to the following relation, 
\begin{equation}
M_0 = L_d(\eta_1)\, D_d^2\,
\Omega_\nu(\eta_1)\rho_W\,.
\end{equation}
$L_d$ depends on $\eta_1$, since $\eta_1$ is related to the radius of the detector placed at a distance $D_d$ from the IP.
When looking in the rapidity range $\eta_1<\eta_\nu<\eta_2$, the relation is:
\begin{equation}
M_0 = L_d (\eta_1,\eta_2)\, D_d^2\,
[\Omega_\nu(\eta_1)-\Omega(\eta_2)]\rho_W\,.
\end{equation}
Here, the length $L_d(\eta_1,\eta_2)$ is related to the inner and outer radii of the cylindrical detector through the corresponding detectable rapidity range.
For $D_d=480$ m and detection of $\eta_\nu>\eta_1=8.9$  in the full $2\pi$ azimuthal angle around the $z$-axis, $L_d=1.06$ m for one ton of tungsten. 
For $\eta_\nu>6.9$, the corresponding detector depth is $L_d=0.016$ m, whereas for $7.2<\eta_\nu<8.6$, $L_d=0.034$ m. For a fractional azimuthal coverage $\Delta \phi$, the detector depth for 1 ton increases,
$L_d\to L_d\cdot 2\pi/\Delta\phi$.

By including a factor $1/M_0$, the number of events per unit energy per detector mass at a distance $D_d$ from the interaction point, under the neutrino pseudorapidity cuts 
incorporated in $d\sigma/dE_\nu$ to select forward neutrinos, can be written as
\begin{equation}
    \frac{1}{M_0}\frac{dN}{dE_\nu} =
 \frac{d \sigma}{dE_\nu} (pp \to \nu_\tau X + \bar\nu_\tau  X) \times\frac{ {\cal L}\, N_{\rm avo}\, \sigma_{\nu W}/A_W  }{D_d^2 \,\Omega_\nu}\,,
\end{equation}
where we have suppressed the dependence of $d\sigma/dE_\nu$ and $\Omega_\nu$ on $\eta_1$ (or $(\eta_1,\eta_2)$ where relevant).
For large $\eta_1$ ($\eta_1\gtrsim 8.3$),
the approximate solid angle scaling of the
rapidity dependent differential cross section shown in eq. (\ref{eq:etaarea}) approximately cancels the rapidity dependence of the solid angle to yield similar numbers of events per unit detector mass for each rapidity range at large enough rapidity \cite{Kling:2021gos}.
The number of events per unit mass for $\eta_\nu>8.9$ is larger than the number of events per unit mass for $\eta_\nu>6.9$, since for $\eta_\nu\lesssim 8.3$, the rapidity dependent differential cross section lies below the rapidity scaling curve (the dotted line in the right panel of Fig. \ref{fig:direct-chain-PROSAfits}).

On the other hand, in the case of two detectors of equal thickness $L_d$  of the same material, looking respectively to pseudorapidities $\eta > \eta_1$ and $\eta > \eta_2$, 
with $\eta_1 \gg \eta_2$, the total number of events will be much smaller in detector 1 ($\eta>\eta_1$)  with respect to detector 2 ($\eta> \eta_2$)(see Table \ref{table:events1.5-m}) due to both the 
much smaller $d\sigma/dE_\nu$, e.g., as shown in Fig. \ref{fig:energy-pdfuncertainties}, and the smaller mass of detector 1. 
In designing a far-forward experiment, both the transverse and the longitudinal size of the spaces to host them need to be considered.  


\subsubsection{During the Run 3 stage}
The LHC is scheduled to operate Run 3 from 2022 to 2024. During this stage, two new experiments, FASER$\nu$ and SND@LHC will be conducted to detect neutrinos in the large rapidity region.
The FASER$\nu$ detector is made of tungsten and emulsion with a size of 25 cm $\times$ 25 cm $\times$ 1.3 m, and the target mass of 1.2 ton \cite{Ariga:2020pyj}. 
Considering that the baseline is 480 m, the FASER$\nu$ experiment can probe the pseudorapidity range of $\eta \gtrsim 8.9$.
SND@LHC is designed to cover the pseudorapidity range $7.2 \lesssim \eta \lesssim 8.6$ and
uses a tungsten, emulsion, and scintillating fiber  detector with target mass of 830 kg \cite{Ahdida:2020evc}. While the rapidity ranges are not precise due to the fact that the detector shapes  might differ from simple cylinders,
we use these representative ranges for all the results presented here.

\begin{figure}
    \centering
    \includegraphics[width=0.70\textwidth]{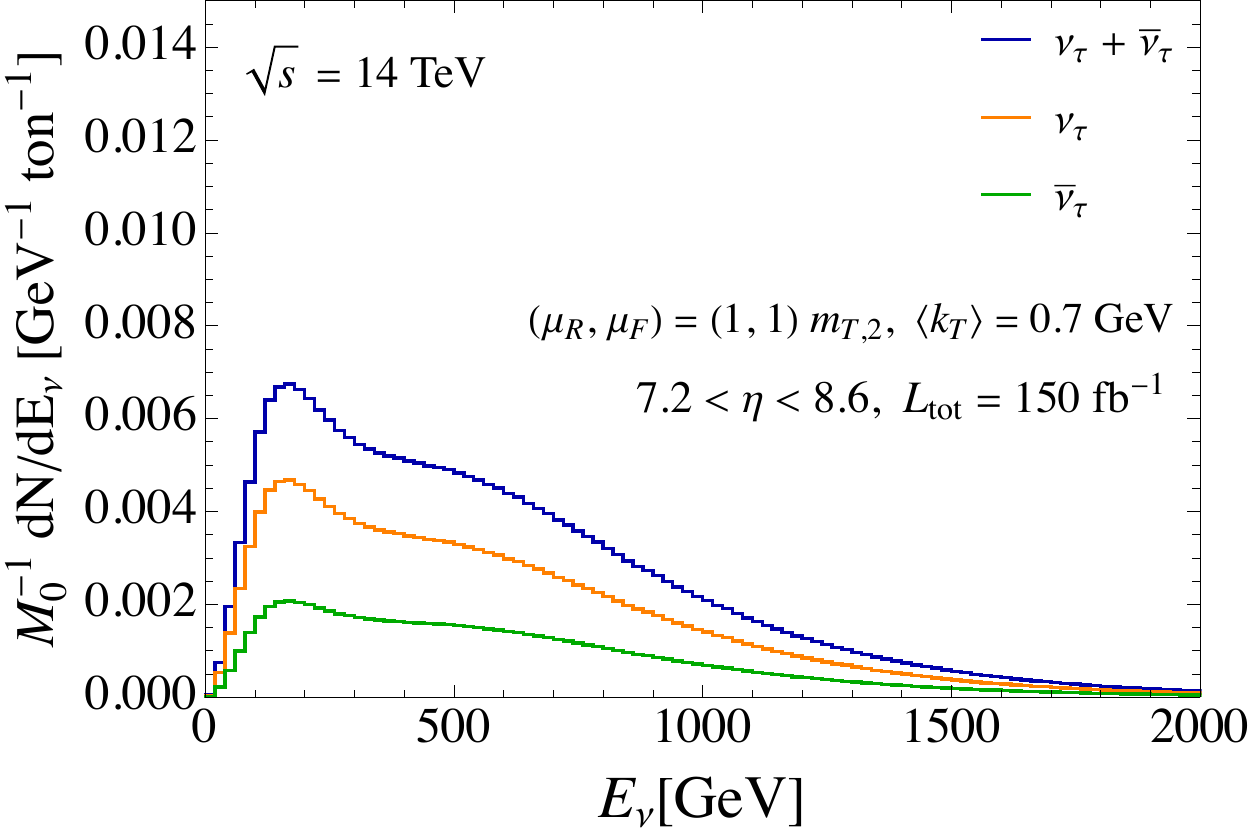} 
    \includegraphics[width=0.70\textwidth]{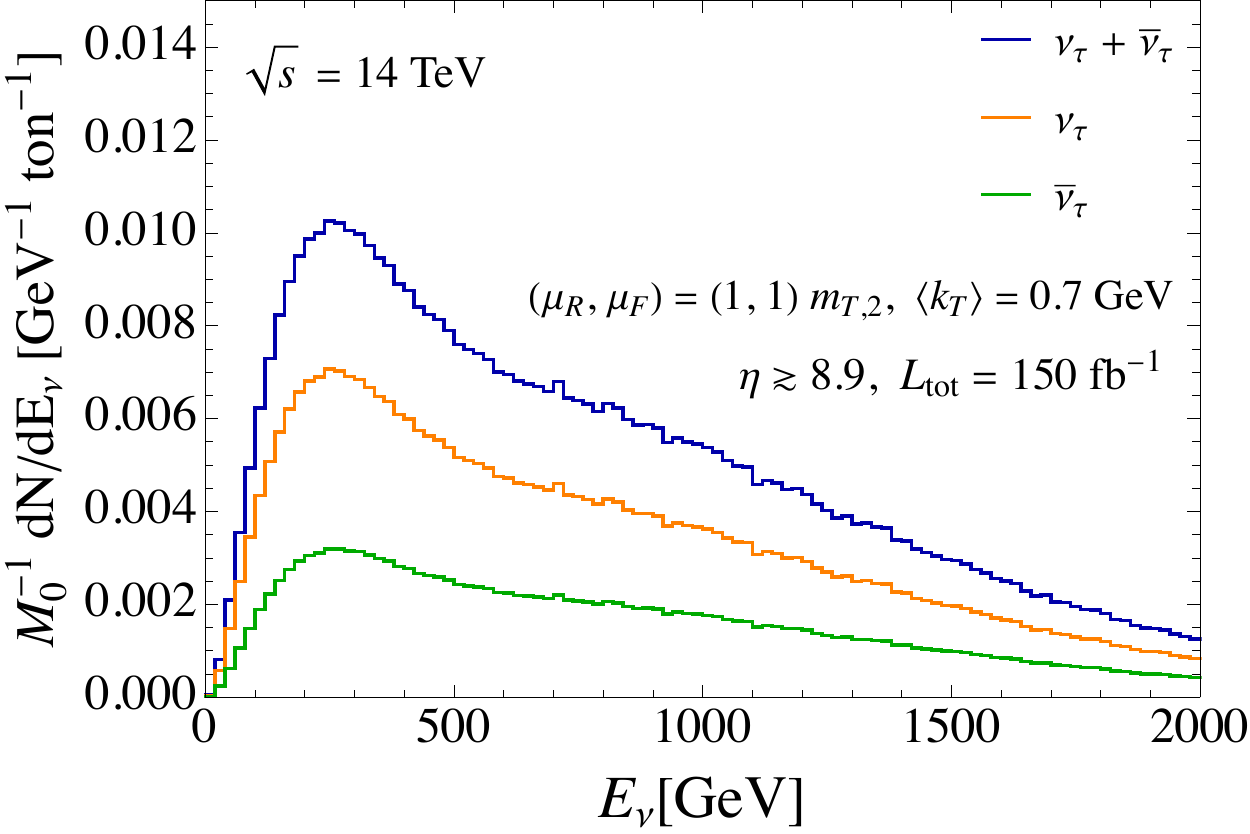}         
    \caption{The energy distribution of the number of tau neutrino and antineutrino charged-current events per neutrino energy per 1 ton of detector a distance $D_d=480$ m from the collider interaction region for the pseudorapidity ranges of $7.2 \lesssim \eta \lesssim 8.6$ (upper) and $\eta \gtrsim 8.9$ (lower) for the neutrinos produced in $pp$ collisions at $\sqrt{s} =$ 14 TeV.}
    \label{fig:dnde-run3}
\end{figure}

Fig.~\ref{fig:dnde-run3} shows the number of tau neutrino and antineutrino events per energy per ton of tungsten detector for the two 
aforementioned
rapidity cuts, considered for the plots in the upper and lower panel, respectively.
The presented predictions were obtained with our default scale choice, $(\mu_R, \mu_F) = (1, 1)\, m_{T,2}$ and $\langle k_T \rangle =$ 0.7 GeV.
The shape of the energy distribution of events reflects those of contributions from direct neutrinos 
at low energies and from chain decay neutrinos
at high energies. In our perturbative QCD evaluation of charm pair production, using a
fragmentation function 
$c\to D_s^+$ and $\bar{c}\to D_s^-$, the $\nu_\tau$ and $\bar\nu_\tau$ energy distributions are identical. However, the resulting number of events is larger for tau neutrinos than for tau antineutrinos, since the tau neutrino charged-current deep-inelastic-scattering (DIS) cross sections are about two times larger than the antineutrino cross sections \cite{Jeong:2010nt}. 
The fact that the number of events per unit energy per ton is larger for $\eta > 8.9$ than for $7.2 < \eta_\nu < 8.6$, as discussed above, follows from the deviation from solid angle scaling of the neutrino rapidity distribution for $\eta_\nu<8.3$. 
In the lower right panel of Fig. \ref{fig:direct-chain-PROSAfits}, for the bin centered at $\eta_\nu=7.3$, $d\sigma/d\eta_\nu$ is 
$\sim$~2/3 of the solid angle scaling result that applies to larger $\eta_\nu$,  the dotted line in the lower right panel of Fig.~\ref{fig:direct-chain-PROSAfits}. 
This factor of $2/3$ is the approximate ratio of the 
$M_0^{-1} d N / d E_\nu$ peaks in the upper  and lower panels in Fig. \ref{fig:dnde-run3}.

\begin{figure}
    \centering
    \includegraphics[width=0.60\textwidth]{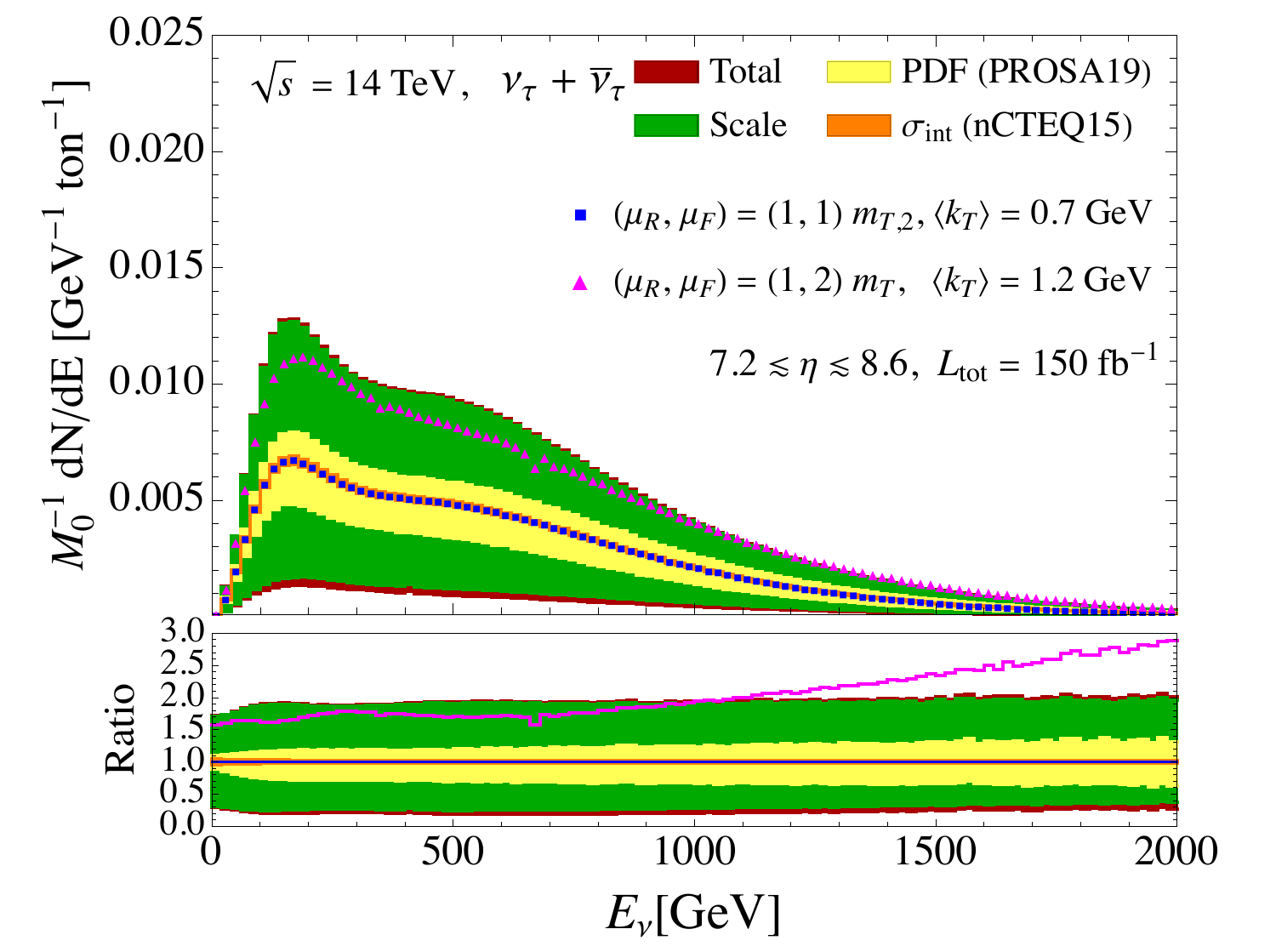}   
    \includegraphics[width=0.60\textwidth]{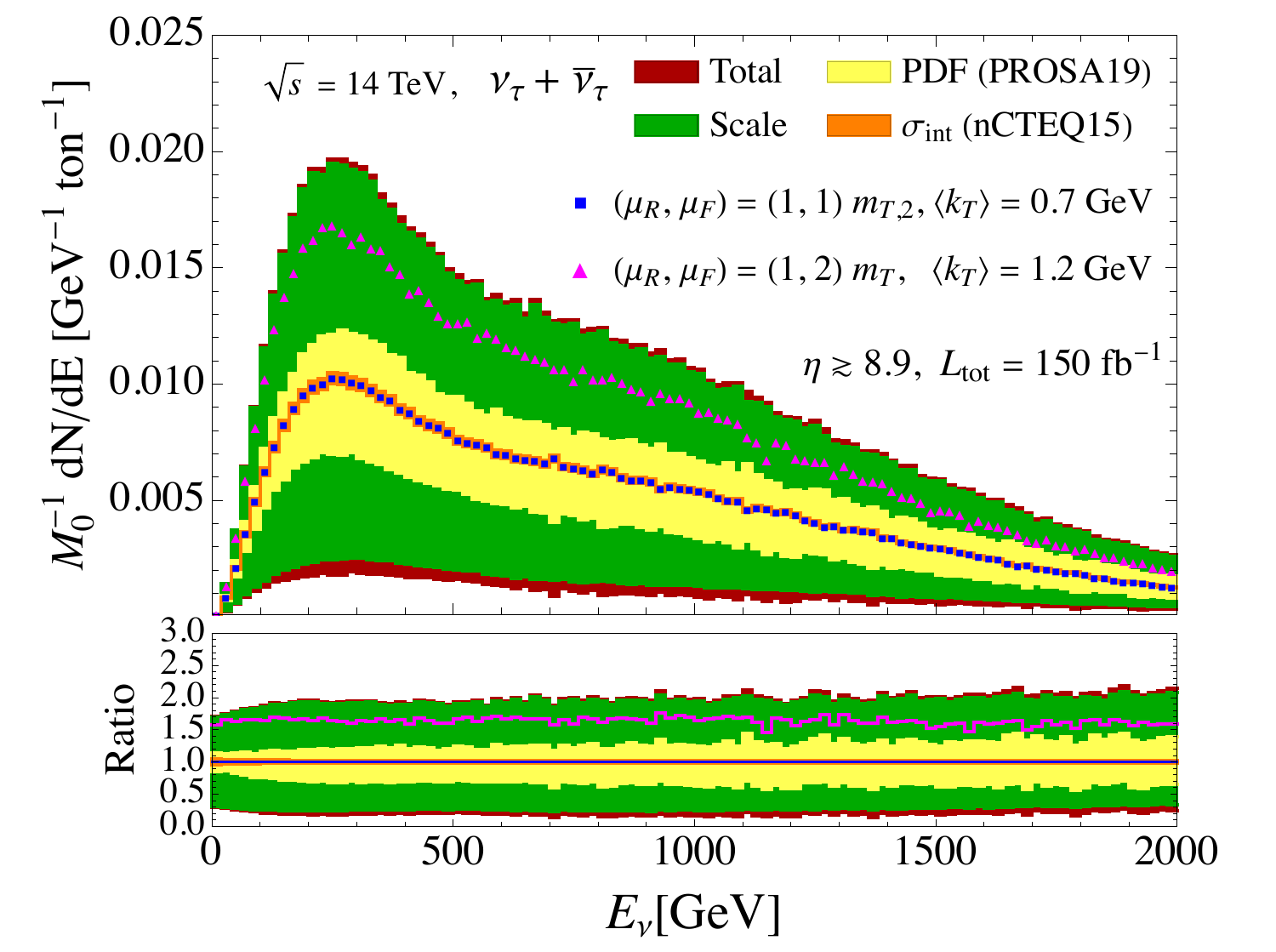}       
    \caption{Energy distribution of tau neutrino and antineutrino CC interaction events per ton of detector at a distance $D_d=480$ m from the collider interaction point, in the pseudorapidity ranges of $7.2 <\eta _\nu< 8.6$ (upper) and $\eta_\nu > 8.9$ (lower).
    Central predictions refer to  $(\mu_R, \mu_F) = (1, 1)\, m_{T,2}$ and $\langle k_T \rangle =$ 0.7 GeV. 
    The presented uncertainties are due to seven variations of the QCD scales (green) around the central value, the PDF uncertainty sets of the PROSA FFNS fit \cite{Zenaiev:2019ktw} (yellow) 
    and of the nCTEQ15 fit for tungsten \cite{Kovarik:2015cma} (orange)
    used for production and interaction, respectively, as well as the combination of the QCD and PDF uncertainties.
    Central predictions with $(\mu_R, \mu_F) = (1, 2)\, m_T$ and $\langle k_T \rangle =$ 1.2 GeV are also shown (magenta triangles) for comparison. The corresponding ratios to the central predictions of energy distributions of charged-current event numbers with error bands and with $(\mu_R, \mu_F) = (1, 2)\, m_T$ and $\langle k_T \rangle =$ 1.2 GeV are shown in the ratio plots.
    }
    \label{fig:dnde-uncert}
\end{figure}

In Fig.~\ref{fig:dnde-uncert}, we show the uncertainties in the energy distribution of $\nu_\tau + \bar{\nu}_\tau$ events due to the conventional QCD seven-point scale variation procedure (green) and the 40 different PROSA PDF sets (yellow). These uncertainties affect the charm production cross sections, which, in turn, yield the neutrino flux. We show as well the uncertainty band related to the 32 
nCTEQ15 PDF sets, affecting the cross section for neutrino CC DIS 
inside the detector. The total uncertainties from these three factors combined in quadrature are shown with red. 
The upper boundary of the total uncertainty band is about 2 times larger than the central predictions, whereas the lower boundary is about 80\% lower, as one can infer from the lower insets of the panels where ratios of the uncertainty bands to the central predictions for the energy distributions of the CC event numbers are shown. 
As shown in the figure, the largest contribution to the total uncertainty comes from the QCD scale variation, and the smallest contribution comes from the uncertainty in the interaction cross section arising from the variants of nCTEQ15 PDFs for tungsten, which amounts to about $\pm$2\% for $E_\nu \gsim 100$ GeV.
The different sets of the PROSA PDF fit, used in charm production, lead to a further uncertainty on the energy distribution of the events amounting to about  $+(20 - 30)$\% and $-(30 - 40$)\% of the central prediction, similar in the two rapidity ranges shown in Fig. \ref{fig:dnde-uncert}. 

We also present the predictions evaluated with our parameter set, $(\mu_R, \mu_F) = (1, 2)\, m_T$ and $\langle k_T \rangle =$ 1.2 GeV, (magenta triangles) for comparison. 
As discussed above, the predictions obtained with this set
are larger than those with the default parameter set.
As can be expected from Fig.~\ref{fig:dsigmade-ratios}, the alternative parameter set leads to about a 60 -- 70\% increase in the event rate with respect to our default parameter set for $E_\nu \lsim$ 500 GeV and 
$7.2 < \eta_\nu < 8.6$. The discrepancy increases with energy, so that the alternative parameter set prediction is a factor of  $\sim$ 2 (3) relative to the default set  at $E_\nu \sim$ 1000 (2000) GeV.
On the other hand, for $\eta > 8.9$, the discrepancy between the predictions evaluated with the two parameter sets 
is about 60\% for all the presented energies.

\begin{figure}
    \centering
    \includegraphics[width=0.60\textwidth]{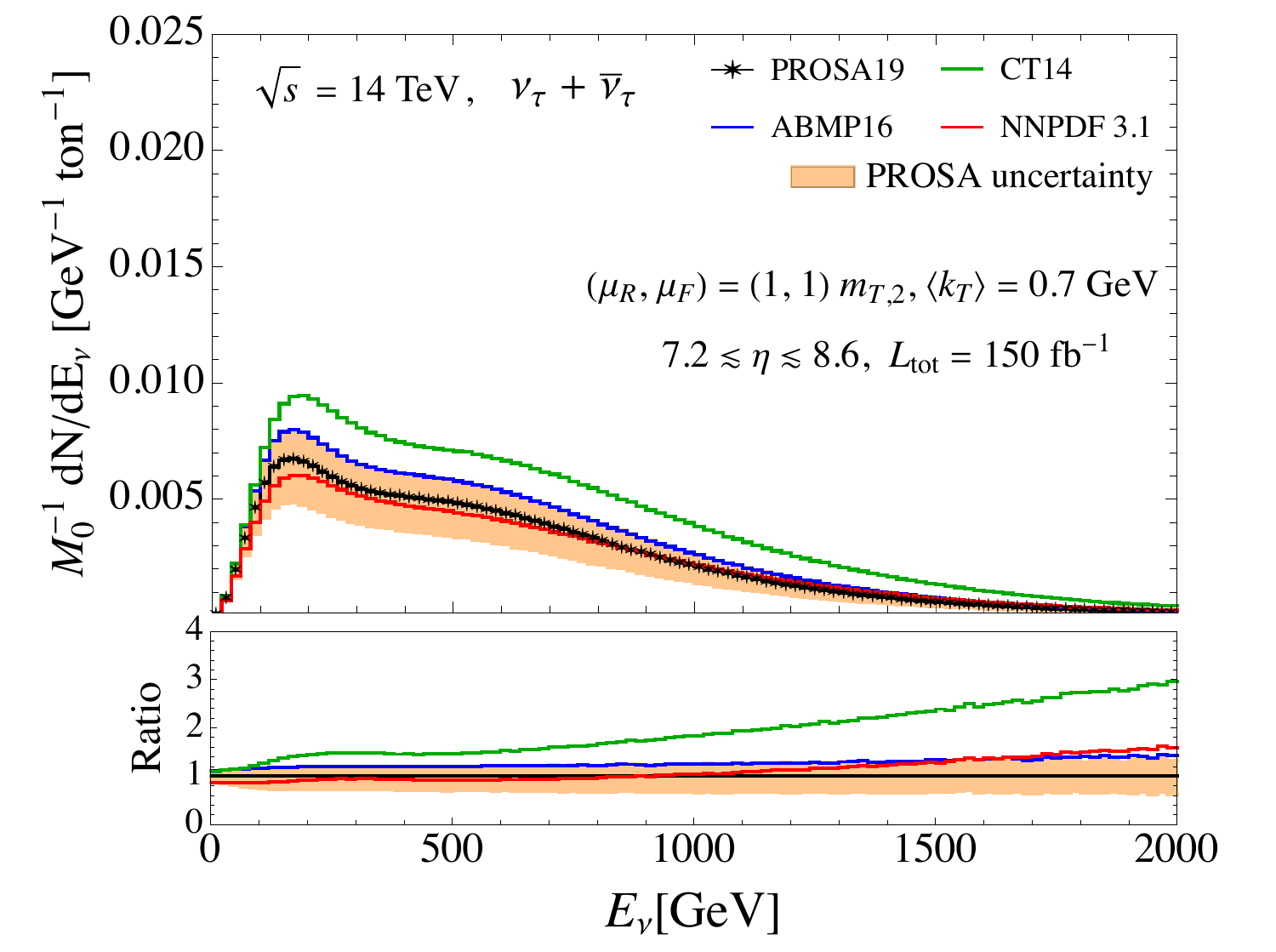}   
    \includegraphics[width=0.60\textwidth]{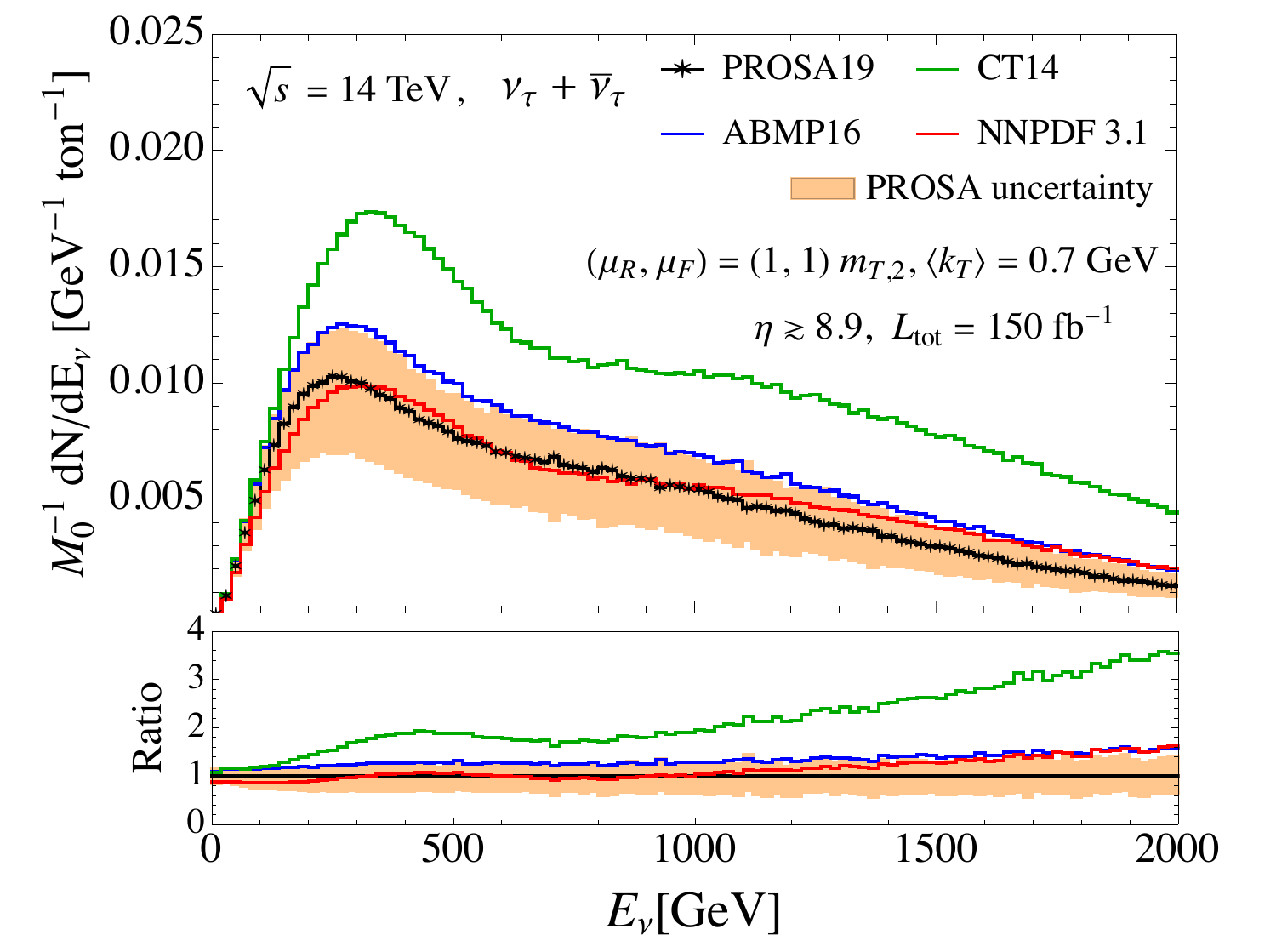}       
    \caption{The energy distribution of tau neutrino and antineutrino charged-current events per ton at a distance $D_d=480$ m from the collider interaction point, obtained with PDF sets by different collaborations, ABMP16 \cite{Alekhin:2018pai}, CT14 \cite{Dulat:2015mca}, NNPDF3.1 \cite{Ball:2017nwa} and PROSA FFNS. The uncertainty band accounts for PDF uncertainties computed with the PROSA PDFs. The predictions refer to the pseudorapidity ranges $7.2 <\eta_\nu < 8.6$ (upper) and $\eta_\nu > 8.9$ (lower). The corresponding ratios to the central predictions of energy distributions of charged-current event numbers with PROSA uncertainty bands and with alternative PDFs are shown in the ratio plots.}
    \label{fig:dnde-pdf}
\end{figure}

For the two neutrino rapidity ranges also considered in previous plots, in 
Fig.~\ref{fig:dnde-pdf} we compare the energy distributions of the $\nu_\tau + \bar{\nu}_\tau $ CC events  obtained using as input for charm production the PROSA PDFs with those computed by using as input the central NLO PDF sets provided by other groups, i.e. ABMP16 \cite{Alekhin:2018pai}, CT14 \cite{Dulat:2015mca} and NNPDF3.1 \cite{Ball:2017nwa}. 
The NNPDF3.1 PDFs yield results consistent with  PROSA , with differences between central predictions within 10\% for $E_\nu \lsim$ 1000 GeV. At higher energies, the discrepancy increases with energy, so that the NNPDF predictions reach the upper edge of the PROSA PDF uncertainty band at $E_\nu \simeq 1500 {~\rm GeV}$. 
The ABMP16 PDF predictions are almost on the edge of upper boundary of the PROSA PDF uncertainty band, while those with CT14 PDFs are out of the uncertainty ranges.
They are larger than the central prediction with the PROSA PDFs  by a factor of about 1.5 -- 2 for $E_\nu \simeq$ 200 -- 1000 GeV, and 
the discrepancies become even larger at higher energies, reaching a factor $\sim$ 3 -- 3.5 at $E_\nu$ = 2000 GeV. The difference in the charm mass value used in the simulations with PROSA and CT14 NLO PDFs (see Table \ref{tab:charm-masses}) plays a decreasing role with energy and is definitely not enough to explain such a large discrepancy.
Fig. \ref{fig:PDFun} shows that the gluon PDF for CT14 is more than a factor of $\sim 2$ larger than the gluon PDF for PROSA for $x=0.45$, with an increasing factor as $x$ gets larger, while at small $x$, the ratio of the CT14 gluon PDF to PROSA gluon PDF for $Q^2=10$ GeV$^2$ is closer to $\sim 1.1$. Considering Fig. \ref{fig:PDFun} together with the lower panel of Fig. \ref{fig:xregions}, we can infer that it is the large $x$ behavior of the PDFs that is responsible for the discrepancy between the CT14 and PROSA results for  neutrino energies above $\sim 1$ TeV.

\begin{table}[h]
	\vskip 0.35in
	\begin{center}		    
		\begin{tabular}{|c||c|c|c|c|c|c||c|c|c|}
			\hline
			${\cal L}=$ 150 fb$^{-1}$
                 &  $\nu_\tau$ &  $\bar{\nu}_\tau $  & 
            \multicolumn{4}{c||}{$\nu_\tau + \bar{\nu}_\tau $} &  $\nu_\tau$ &  $\bar{\nu}_\tau $  &   $\nu_\tau + \bar{\nu}_\tau $ \\
            			1 m
                 &   &    & 
            \multicolumn{4}{c||}{ } &  &    &  \\
                       \hline  
		      $ (\mu_R, \ \mu_F) $  & \multicolumn{3}{c|} {(1, 1)  $m_{T,2}$} 
		   & scale(u/l) & PDF(u/l) & $\sigma_{\rm int}$& 
		      \multicolumn{3}{c|} {(1, 2)  $m_{T}$} \\
		      \hline	                              
		      $\langle k_T \rangle $  & \multicolumn{6}{c||} {0.7 GeV} & \multicolumn{3}{c|} {1.2 GeV} \\
			\hline
		 	 $7.2 < \eta  < 8.6$  &  101 & 47 & $148^{+136}_{-118}$ & +133/-109 &  +30/-44 & $\pm$3.5 & 181 & 85 & 266 \\
			\hline 
		 	$\eta  > 8.9$  & 6.5  & 3.1 & $9.6^{+9.2}_{-7.8}$ & +8.9/-7.1 & +2.2/-3.1 & $\pm$0.2
		 	& 10.6  & 5.0 & 15.7 \\
			\hline
		\end{tabular}
	\end{center}
	\caption {The numbers of events induced at different pseudorapidities
	in 1 m length of tungsten  
	by the CC interactions of $\nu_\tau$ and $\bar{\nu}_\tau$
	arising from the 
	decay of $D_s^\pm$ mesons produced in $pp$ collisions at $\sqrt{s}$ = 14 TeV for an integrated luminosity ${\cal L}=150$ fb$^{-1}$.
	}
	\label{table:events1.5-m}
\end{table}

In Table \ref{table:events1.5-m}, the expected numbers of CC events induced by tau neutrinos, tau antineutrinos, as well as the sum of these two components, are shown, respectively for a 1 m length of tungsten detector. For the given rapidity ranges, the target mass is 29.56 ton for $7.2 < \eta  < 8.6$ and 0.95 ton for $\eta > 8.9$.
As mentioned above, we only consider the events originated by $D_s^\pm$ decays and assess their numbers for the two scale choices for charm production discussed in the previous section, $(\mu_R, \ \mu_F) = (1, 1)  m_{T,2}$ with $\langle k_T \rangle $ = 0.7 GeV and $(\mu_R, \ \mu_F) = (1, 2)  m_{T}$ with $\langle k_T \rangle $ = 1.2 GeV. 
The predictions with the second set of parameters are 80\% (64\%) larger than those with the first set (default) for $7.2 < \eta_\nu  < 8.6$ ($\eta_\nu > 8.9$).
For the results with the default set, we also present the total uncertainty in the number of ($\nu_\tau + \bar{\nu}_\tau$) induced CC events, as well as the uncertainty component due to QCD scale variation, the PDF uncertainties related to neutrino production, using the PROSA PDFs and those related to neutrino interactions, using the nCTEQ15 PDFs.  
The total CC event numbers range between values that are larger by a factor of $\sim$~2 and about 80\% smaller compared to the central prediction. 
As seen in Fig. \ref{fig:e-direct-chain-scale-tot}, the total uncertainty is dominated by the QCD scale variations, which yield 90\% higher and 70\% lower values than the central CC event numbers, for the upper and lower predictions.
The 40 different sets of tbe PROSA PDF fit in neutrino production impact the total events number by $+20$\% and $-30$\% while the 32 variants of nCTEQ15 in neutrino-nucleus interaction make a difference of only $\pm 2$\%.

\begin{table}[h]
	\vskip 0.35in
	\begin{center}		    
		\begin{tabular}{|c||c|c|c|c|c|c|}
		\hline 	${\cal L}=$ 150 fb$^{-1}$
            &  $\nu_\tau$ &  $\bar{\nu}_\tau $  &   $\nu_\tau + \bar{\nu}_\tau $ &
            \multicolumn{3}{c|}{$\nu_\tau + \bar{\nu}_\tau $}  \\	 
        \hline  
		      $(\mu_R, \ \mu_F)$, $\langle k_T \rangle $  & \multicolumn{6}{c|} {(1, 1)  $m_{T,2}$,  0.7 GeV}  \\
		 \hline	                              
		        & \multicolumn{3}{c|}{} &
		        scale(u/l) & PDF(u/l) &$\sigma_{\rm int}$ \\
		\hline
		 	 SND@LHC &  2.8 & 1.3 & $4.2^{+3.8}_{-3.3}$ & +3.7/-3.1 & +0.8/-1.2 &  $\pm$0.1 \\
		 	 $7.2<\eta_\nu<8.6$, 830 kg & & & & & & \\	
		\hline 
		 	FASER$\nu$  & 8.2  & 3.9 & $12.1^{+11.6}_{-9.8}$ & +11.3/-9.0 & +2.8/-3.9 & $\pm$0.3  \\
		 		 	 $\eta_\nu>8.9$, 1.2 ton & & & & & & \\
		\hline
%
         \hline  
              $(\mu_R, \ \mu_F)$, $\langle k_T \rangle $  &
              \multicolumn{3}{c|} {(1, 2)  $m_{T}$,  1.2 GeV} & \multicolumn{3}{c|} {(1, 1)  $m_{T,2}$,  0.7 GeV } \\
	     \hline
	        PDF &
            \multicolumn{3}{c|}{PROSA FFNS} &  NNPDF3.1 & CT14 & ABMP16\\
		\hline
		 	 SND@LHC  & 5.1 & 2.4 & 7.5 & 4.0 & 6.6 & 5.0 \\
			 	 $7.2<\eta_\nu<8.6$, 830 kg & & & & & & \\	 	 
		\hline 
		 	FASER$\nu$ & 13.5  & 6.4 & 19.9 &  12.8  & 23.5 & 15.6 \\
		 			 	 $\eta_\nu>8.9$, 1.2 ton & & & & & & \\
		\hline
		\end{tabular}		
	\end{center}
	\caption {
	The numbers of events induced in the FASER$\nu$ (1.2 tons of tungsten, $\eta_\nu>8.9$) and SND@LHC (830 kg of tungsten, 
	$7.2<\eta_\nu<8.6$)
	detectors, by the CC interactions of $\nu_\tau$ and $\bar{\nu}_\tau$
	arising from the decay of $D_s^\pm$ mesons produced in $pp$ collisions at $\sqrt{s}$ = 14 TeV for an integrated luminosity ${\cal L}=150$ fb$^{-1}$.
	    	The actual shape of the detectors is taken into account in an approximate way,
	by assuming that they are cylinders or a portion of a cylindrical shell as described in Sec. \ref{sec:evts}.}
	\label{table:events1p5}
\end{table}

In Table \ref{table:events1p5}, we estimate the numbers of $\nu_\tau$, $\bar{\nu}_\tau$ and $\nu_\tau + \bar{\nu}_\tau$ induced CC interaction events for the two experiments, FASER$\nu$ ($\eta_\nu>8.9)$ and SND@LHC $(7.2<\eta_\nu<8.6)$ which will be carried out during the Run 3 stage of the LHC by taking into account the different target masses of the two detectors. 
The expected total event numbers are in the range of $0.9-8.0$ for SND@LHC and in the range of $2.3-23.7$ for FASER$\nu$, according to the estimate with our default parameter set. 
The number of events is dominated by neutrinos with hundreds of GeV rather than TeV energies. The results from  different central PDF sets, although not always lying within the PROSA PDF uncertainty band, turn out to always lie within the scale uncertainty range in the evaluation.
With the $m_{T}$ dependent parameter set, the central prediction is 7.5 and 19.9 for SND@LHC and FASER$\nu$, respectively.

\subsubsection{During the High Luminosity-LHC stage}

The High-Luminosity (HL)-LHC stage is planned to start from 2027, 
and to lead to an overall integrated luminosity $\mathcal{L} = 3000 {\, \rm fb^{-1}}$. 
Possible forward experiments for the HL-LHC era are under study with upgrades of FASER$\nu$ and SND@LHC, and with more additional experiments. 
The detector specifics are not yet determined for the experiments at this stage, and it is also possible that the baseline will be changed. 
Thus, in order to estimate the event numbers, we use a hypothetical detector by assuming that both the radius and the length of detector is 1 m as in Ref. \cite{Bai:2020ukz}. The pseudorapidity range covered by this setup corresponds to $\eta \gsim 6.9$ for 480 m of the baseline.

\begin{figure}
    \centering
    \includegraphics[width=0.7\textwidth]{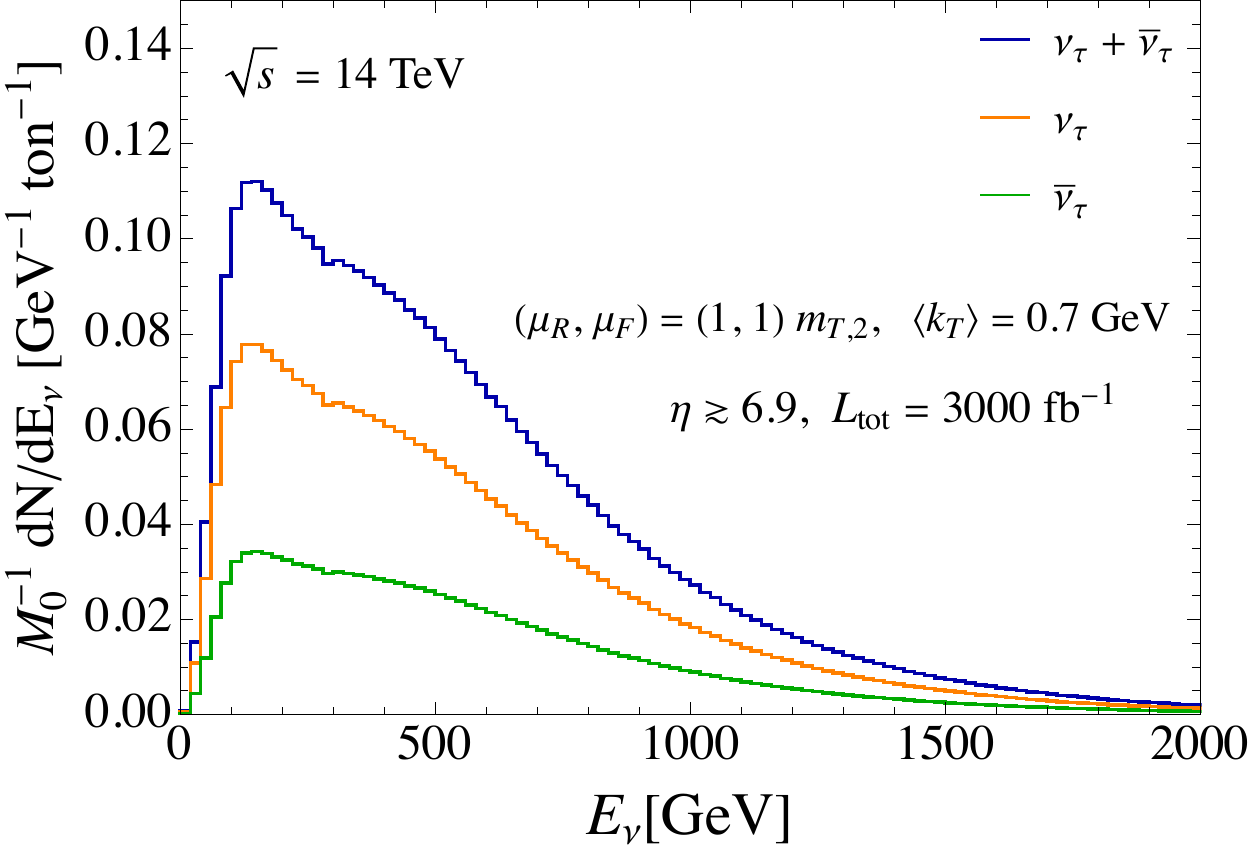}
    \caption{The same as Fig.~\ref{fig:dnde-run3}, except for the pseudorapidity range and the integrated luminosity, which in this plot are set to $\eta > 6.9$ and ${\cal L} = 3000 {\, \rm fb^{-1}}$, respectively.}
    \label{fig:dnde-6p9}
\end{figure}

\begin{figure}
    \centering
    \includegraphics[width=0.70\textwidth]{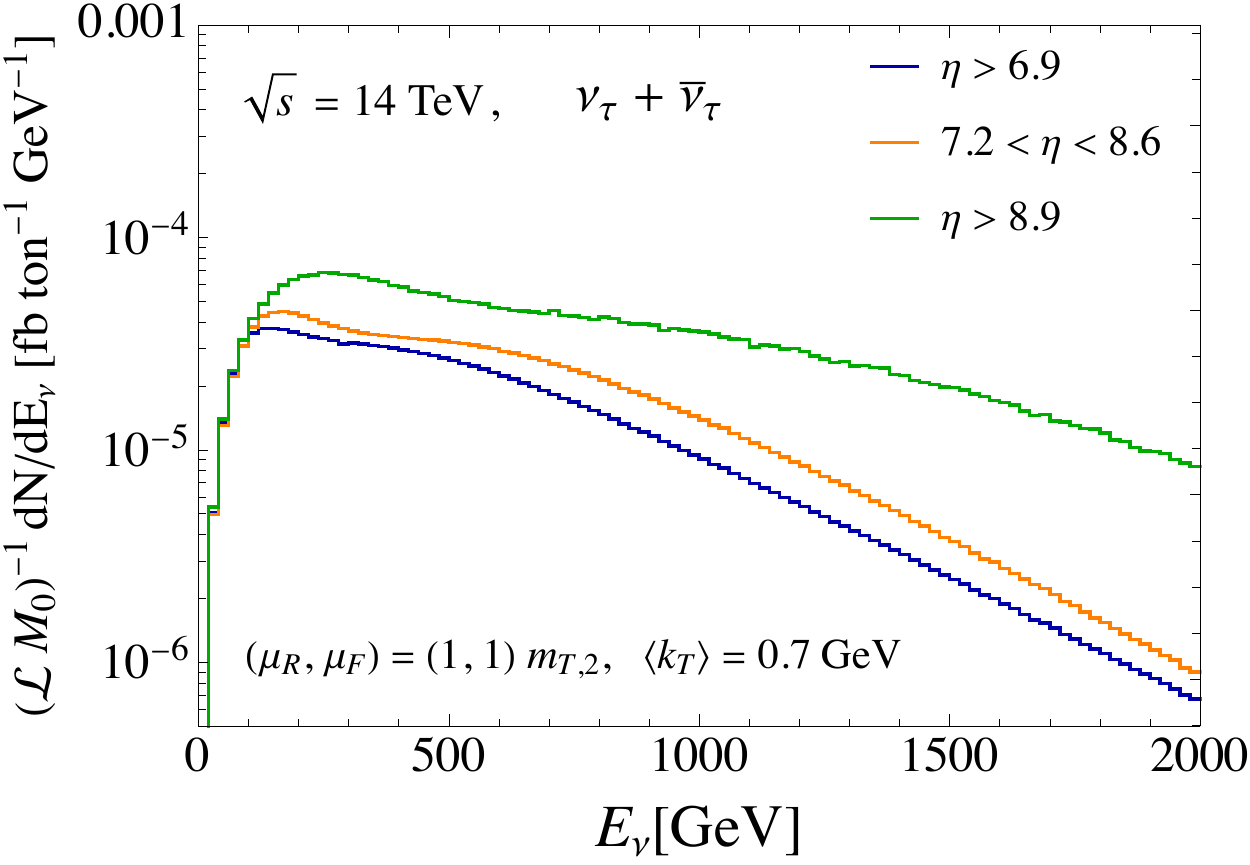}   
    \includegraphics[width=0.70\textwidth]{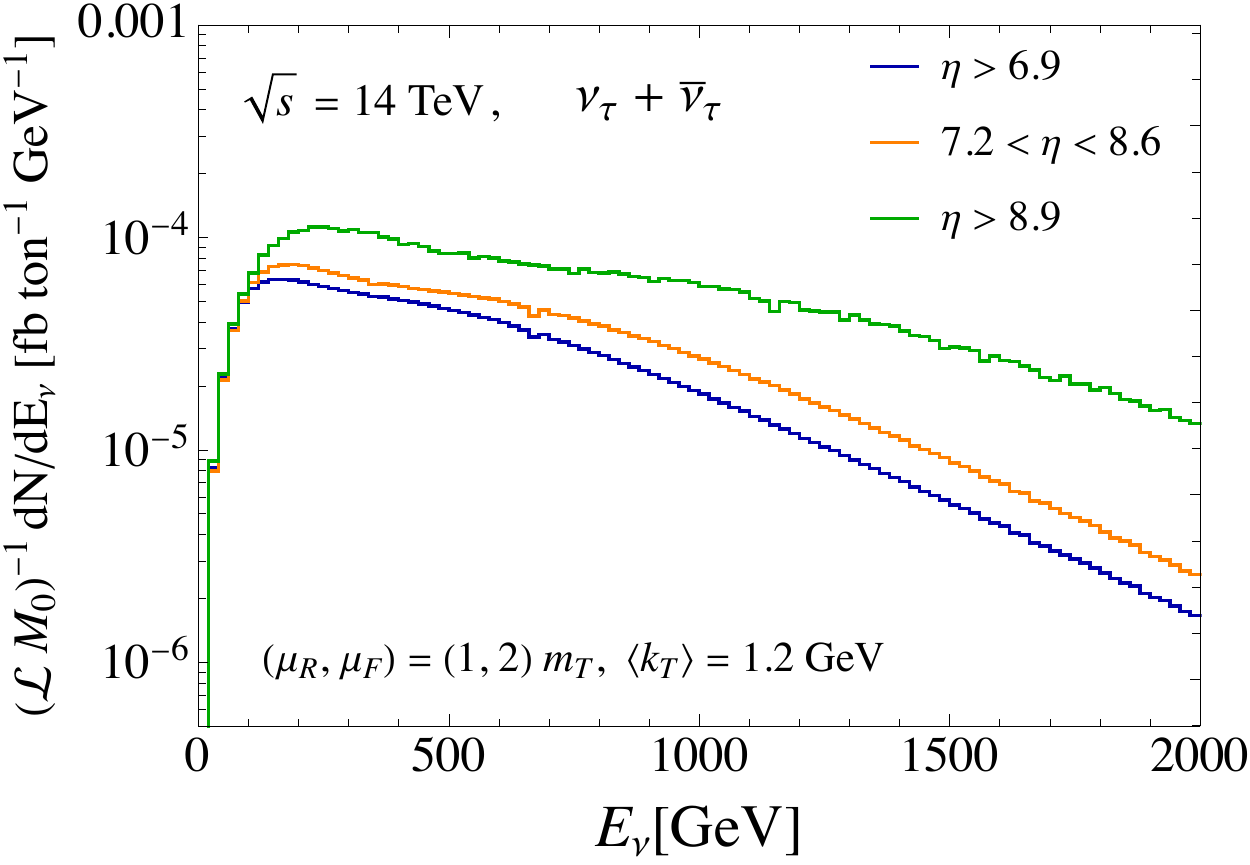}      
    \caption{The energy distributions of the total  charged-current events of tau neutrinos and antineutrinos per ton at a distance $D_d=480$ m from the collider interaction point for different rapidity ranges, using as input parameter sets  $(\mu_R, \mu_F) = (1, 1)\, m_{T,2}$ and $\langle k_T \rangle =$ 0.7 GeV (upper) and $(\mu_R, \mu_F) = (1, 2)\, m_T$ and $\langle k_T \rangle =$ 1.2 GeV (lower).
    } 
    \label{fig:dnde-rap}
\end{figure}

As in Fig.~\ref{fig:dnde-run3} related to  different rapidity ranges, the energy distribution of the number of  $\nu_\tau$, $\bar{\nu}_\tau$ and $\nu_\tau + \bar{\nu}_\tau$ induced CC events per energy per ton of tungsten detector for the pseudorapidity range $\eta_\nu > 6.9$ is shown in Fig.~\ref{fig:dnde-6p9} 
for the default parameter set, $(\mu_R, \mu_F) = (1, 1)\, m_{T,2}$ with $\langle k_T \rangle =$ 0.7 GeV.
Even in this case, as visible in the figure, the transition between direct and chain neutrino contributions to the number of events, here occurring at $E_\nu\sim 300$ GeV, is visible in the shape of the distribution.
Fig.~\ref{fig:dnde-rap} shows the energy spectrum of the total number of $\nu_\tau + \bar{\nu}_\tau$ events for the different rapidity range probed in this work for the default input parameter set, $(\mu_R, \mu_F) = (1, 1)\, m_{T,2}$ with $\langle k_T \rangle =$ 0.7 GeV (upper) and the alternative set $(\mu_R, \mu_F) = (1, 2)\, m_T$ and $\langle k_T \rangle =$ 1.2 GeV (lower). 
In order to compare the shape of the spectra, we present the results per unit luminosity. 
As can be inferred from Fig.~\ref{fig:xregions}, the neutrino events at high energies are mostly from charm production at large rapidity. This feature combined with the scaling behavior of the neutrino rapidity distributions makes the energy spectrum of the event number for $\eta>8.9$ harder at high energies compared to those for the lower rapidity ranges. As discussed above, one can also see that the predictions for the neutrino energy distributions for $\eta_\nu>6.9$ and $7.2<\eta_\nu<8.6$ are lower than for $\eta_\nu>8.9$ due to deviation from the scaling behaviour  with $\Omega_\nu(\eta_\nu)$ for the neutrino rapidity distribution when $\eta_\nu < 8.3$ (see Fig. \ref{fig:direct-chain-PROSAfits}). The figure also shows that the transition from the dominance of events from direct neutrinos to  chain neutrinos somewhat increases with energy as the minimum rapidity increases.

\begin{figure}
    \centering
    \includegraphics[width=0.7\textwidth]{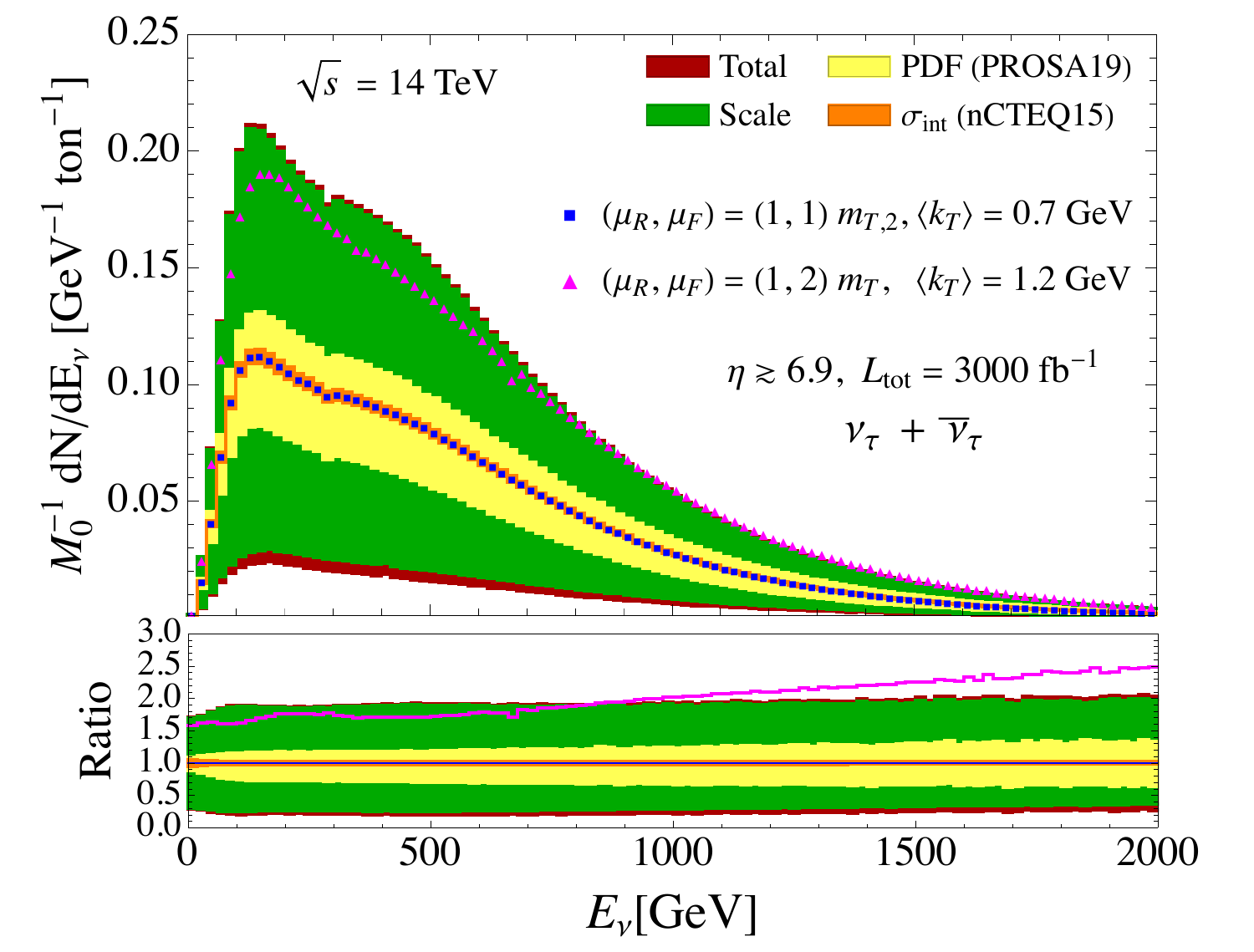}
    \caption{The same as Fig.~\ref{fig:dnde-uncert}, except for the pseudorapidity range and the integrated luminosity, which in this plot are set to $\eta_\nu > 6.9$ and ${\cal L} = 3000 {\, \rm fb^{-1}}$.}
    \label{fig:dnde-uncert-6p9}
\end{figure}

Fig.~\ref{fig:dnde-uncert-6p9} shows the uncertainties in the energy distribution of $\nu_\tau + \bar{\nu}_\tau$ events due to the QCD scale variations (green) and PROSA PDF eigenvectors (yellow) involved in charm production, the uncertainty of the interaction cross section from the nCTEQ15 PDF sets (orange), as well as the total combined uncertainties (red) using the quadrature formula.
The percentage size of uncertainties with respect to the central predictions are similar to Fig.~\ref{fig:dnde-uncert}. 
For the difference in the central predictions with the alternative parameter set $(\mu_R, \mu_F) = (1, 2)\, m_T$ with $\langle k_T \rangle =$ 1.2 GeV and the default parameter set $(\mu_R, \mu_F) = (1, 1)\, m_{T,2}$ with $\langle k_T \rangle =$ 0.7 GeV, the trend is similar to the case of $7.2 < \eta_\nu  < 8.6$. The result evaluated with the first one is 60 -- 70\% larger, for $E_\nu \lsim 500$ GeV, and the difference further increases at higher energies, approaching a factor of $\sim$ 2 (2.5) at $E_\nu \sim$ 1000 (2000) GeV.

\begin{figure}
    \centering
    \includegraphics[width=0.7\textwidth]{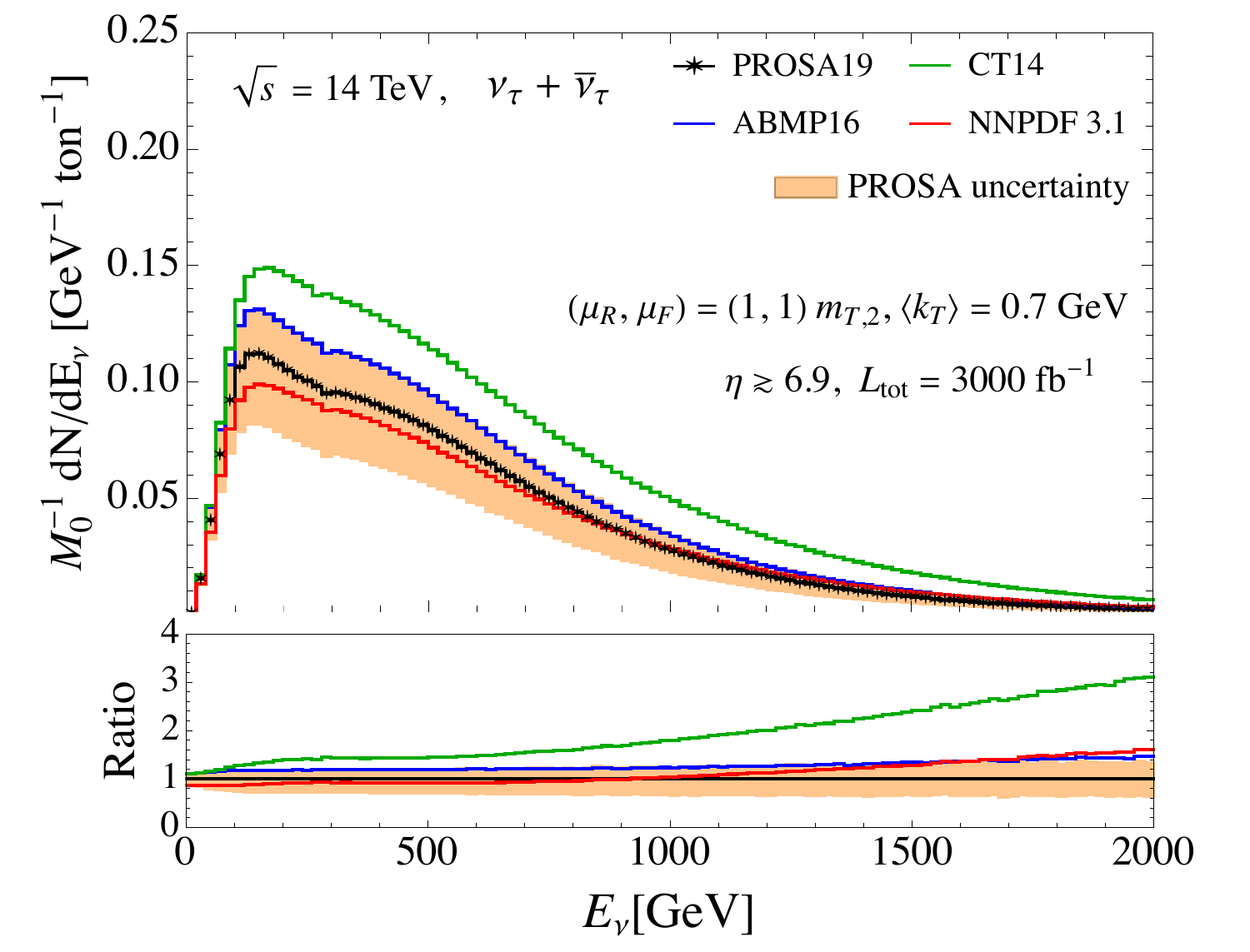}      
    \caption{The same as Fig.~\ref{fig:dnde-pdf} except for the pseudorapidity range and the integrated luminosity, which are set to $\eta > 6.9$ and ${\cal L} = 3000 {\, \rm fb^{-1}}$, respectively.}
    \label{fig:dnde-pdf-6p9}
\end{figure}

As in Fig.~\ref{fig:dnde-pdf}, Fig.~\ref{fig:dnde-pdf-6p9} presents the comparison of the energy distributions of the events for $\nu_\tau + \bar{\nu}_\tau $ for the PDFs by different groups.
While the results with the NNPDF3.1 and ABMP16 PDFs can be considered as being consistent with the predictions with the PROSA PDFs, being within the uncertainties of the latter, the central CT14 prediction is obviously out of range of the PROSA PDFs. The size of the difference is similar to the case of the $7.2 <\eta_\nu  < 8.6$ range. 
Unfortunately, the CT14 PDFs with three active flavours at all scales do not come with a group of variations out of which an uncertainty band can be computed.

\begin{table}[h]
	\begin{center}		    
		\begin{tabular}{|c||c|c|c|c|c|c|}
		\hline ${\cal L}=$ 3000 fb$^{-1}$, 1 m
            &  $\nu_\tau$ &  $\bar{\nu}_\tau $  &   $\nu_\tau + \bar{\nu}_\tau $ &
            \multicolumn{3}{c|}{$\nu_\tau + \bar{\nu}_\tau $}  \\	 
        \hline  
		      $(\mu_R, \ \mu_F)$, $\langle k_T \rangle $  & \multicolumn{6}{c|} {(1, 1)  $m_{T,2}$,  0.7 GeV}  \\
		 \hline	                              
		        & \multicolumn{3}{c|}{} &
		        scale (u/l) & PDF (u/l) &$\sigma_{\rm int}$ \\
		\hline 
		 	$\eta  \gsim 6.9$  & 3260  & 1515 & $4775^{+4307}_{-3763}$ & +4205/-3494 & +926/-1391 & $\pm$112  \\
		\hline
         \hline  
              $(\mu_R, \ \mu_F)$, $\langle k_T \rangle $  &
              \multicolumn{3}{c|} {(1, 2)  $m_{T}$,  1.2 GeV} & \multicolumn{3}{c|} {(1, 1)  $m_{T,2}$,  0.7 GeV } \\
	     \hline
	        PDF &
            \multicolumn{3}{c|}{PROSA FFNS} &  NNPDF3.1 & CT14 & ABMP16\\
		\hline 
		 	$\eta  \gsim 6.9$ & 5877  & 2739 & 8616 &  4545  & 7304 & 5735 \\
		\hline
		\end{tabular}	
	\end{center}
	\caption{The number of charged-current interaction events induced 
	in 1 meter length of tungsten 
	by  tau neutrinos and antineutrinos  
	from $D_s^\pm$ produced in $pp$ collisions at $\sqrt{s}$ = 14 TeV for an integrated luminosity ${\cal L}=3000$ fb$^{-1}$ and $\eta_\nu\gtrsim 6.9$. 
	}
	\label{table:events3ab}
\end{table}

Table \ref{table:events3ab} presents the prediction for the number of CC interaction events due to tau neutrinos, antineutrinos and their sum, measured in a 1 m long tungsten detector, in the pseudorapidity range $\eta \gsim 6.9$, which corresponds to a target mass of 60.63 ton.  
We present the results with our default set of input parameters, $(\mu_R, \ \mu_F) = (1, 1)  m_{T,2}$ with 
$\langle k_T \rangle $ = 0.7 GeV, together with their scale and PDF uncertainties affecting neutrino production and the PDF uncertainties affecting neutrino interactions,  
as well as central results with different PDFs and those from the
set of parameters $(\mu_R, \ \mu_F) = (1, 2)  m_{T}$ with $\langle k_T \rangle $ = 1.2 GeV. 
The total 
($\nu_\tau$ + $\bar{\nu}_\tau$)-induced CC interaction
event number with the default parameter set is predicted to be about 4800 with a variation in the range of 1000 -- 9000 due to the considered uncertainties. The central prediction evaluated with the $m_T$-related parameter set leads to about 8600 events, that is 80\% larger than the results from the default set, but still in its uncertainty bands. 

\subsubsection{Comparisons with previous computations}

In Ref. \cite{Bai:2020ukz}, we evaluated the number of events for $\eta_\nu>6.87$ in a lead detector of length 1 m. A volume of tungsten has $\sim 1.9\times$ more nucleons than the same volume of lead. Depending on the renormalization and factorization scales and $\langle k_T\rangle$, the number of $\nu_\tau+\bar\nu_\tau$ CC events for ${\cal L}=3000$ fb$^{-1}$, 
approximately converted to tungsten, range between $3600-7300$, with a wider uncertainty band associated with the scale uncertainties. Differences between our evaluations in the work presented here and in our prior work \cite{Bai:2020ukz} are three-fold: we used different scale, PDF and $\langle k_T\rangle$ inputs, charm quark fragmentation was implemented in the hadron center-of-mass frame in our earlier work, as compared to implementation in the parton center-of-mass frame in the present work, and the neutrino interaction cross section was evaluated in Ref. \cite{Bai:2020ukz} using the nCTEQ PDFs for lead. In our comparisons below, we convert our results from lead to tungsten by multiplying by the ratio of the number of targets in tungsten relative to lead and neglect nuclear effects that may arise in the different PDFs for tungsten and lead. 

The scale dependence considered in Ref. \cite{Bai:2020ukz} was based on $m_T$ rather than $m_{T,2}$ used here. For the central scales $(1,1)m_T$ and $\langle k_T\rangle$, Table 2 of Ref. \cite{Bai:2020ukz}, converted to a tungsten detector with the same volume, yields a central prediction of 4541 CC events from $\nu_\tau+\bar{\nu}_\tau$ from $D_s^\pm$ decay. This agrees well with the central value shown here in Table \ref{table:events3ab}, 4775 for $(1,1)m_{T,2}$ and $\langle k_T\rangle=0.7$ GeV. However, as noted in Sec. 2.2, the error bands from the scale dependence are more asymmetric for scales that depend on $m_T$ rather than $m_{T,2}$. Ref. \cite{Bai:2020ukz} shows that the upper edge of the envelope for the scale dependence around the number of events as a function of energy for $(1,1)m_{T}$ and $\langle k_T\rangle=0.7$ GeV is of order a factor of 3 larger than the central value, whereas here with $m_{T,2}$ scale dependence, the upper edge of the uncertainty envelope for the scale dependence is close to a factor of 2 larger than the central value. The lower edge of the scale uncertainty envelope in our previous work deviates less from the central value than what we find here for $m_{T,w}$ dependent scales. In Ref. \cite{Bai:2020ukz}, we found that the event number on the lower edge of the scale uncertainty envelope is $\sim 75\%$ of the central value for the event number. 

We also considered as a central scale $(1,1.5)m_T$ 
with $\langle k_T\rangle=0.7$ GeV, and, for better agreement with the LHCb data \cite{Aaij:2015bpa}, $\langle k_T\rangle=2.2$ GeV. A seven-point scale variation around $(1,1.5)m_T$ with $\langle k_T\rangle=0.7$ GeV yielded an event number \cite{Bai:2020ukz} which, when converted to tungsten, amount to $7284^{+14143}_{-3698}$ accounting for the scale variation. Again, this shows an asymmetric scale uncertainty band.  We found that for $(1,1.5)m_T$ 
with $\langle k_T\rangle=2.2$ GeV, the predicted number of events, converted to tungsten, is 5251 CC events, which can be compared with our predictions shown here in Table \ref{table:events3ab}, 8616 CC events for
$(1,2)m_T$ with $\langle k_T\rangle=1.2$ GeV.
Overall, however, the scale uncertainty band in Ref. \cite{Bai:2020ukz} leads to a higher minimum number of CC events than what we have presented here in Table \ref{table:events3ab}.


An estimate of the number of tau neutrino and antineutrino induced interaction events 
computed with various standalone event generators initially developed  
for simulations of cosmic ray-induced extended air showers appears in
Ref.~\cite{Kling:2021gos}. Event generators, used in Ref. \cite{Kling:2021gos} with their default options, include SIBYLL2.3c \cite{Riehn:2015oba,Riehn:2017mfm,Fedynitch:2018cbl} and DPMJET (version from 2017) \cite{Roesler:2000he,Fedynitch:2015kcn}.
As an alternative to SIBYLL and DPMJET, the PYTHIA8 \cite{Sjostrand:2014zea} code,
very popular for LHC phenomenology, was also adopted in Ref. \cite{Kling:2021gos}, using as the seed a tree-level description of $c\bar{c}$ quark pair production, on top of which parton shower, hadronization and other soft physics effects plus hadron decay are included. 

The theory frameworks implemented in the SIBYLL and DPMJET generators are 
oriented especially to the description of soft physics effects. Their use in the description of charm production, a process characterized by a hard scale even at small $p_T$, presents a number of challenges, the first one being related to the 
lack of radiative corrections in the description of the hard-scattering events. Radiative corrections affect not only the normalization but also the shape of differential distributions.  It is worth mentioning that, notwithstanding this lack, extensive and systematic comparisons of $D$-meson energy distributions obtained  using a QCD based approach including NLO QCD radiative corrections, parton shower and non-perturbative QCD effects, with those from SIBYLL, were carried out by one of us in close collaboration with the authors of this code, finding reasonable 
agreement on a very wide set of center-of-mass energies, at least as far as the $p-$Air collisions that matter most for cosmic ray interactions in the atmosphere are concerned.
Unfortunately these comparisons, although of practical interest for those experiments still relying on these Monte Carlo generators, less CPU-intensive than more complex calculations, do not allow for deep QCD insights, con\-si\-de\-ring that the uncertainties associated with leading-order evaluations of charm production are so large that leading-order calculations can provide, at most, an order of magnitude estimate for this process. By no means can they lead to an accurate evaluation and, therefore, should not be misunderstood as such.

The Monte Carlo evaluations in Ref. \cite{Kling:2021gos} that account for specific detector geometries (in contrast to our use of approximate rapidity ranges) yield estimates of the number of CC events in the range 10.1--22.4 for SND@LHC and 21.2--131 for FASER$\nu$, where the range of values arises from the use of different generators, and is far from being 
an estimate of the uncertainties within each event generator. 
The lower numbers are  
compatible with   
our central predictions with the $\mu_0=(1,2)m_T$ and $\langle k_T \rangle = 1.2$ GeV parameter set, which yield a good match to cross sections for $D_s^\pm$ production at LHCb. 
On the other hand, our central predictions with our default parameter set yield a smaller number of tau neutrinos, which enables a very cautious 
perspective concerning the possibility to detect a significant number of tau neutrinos in the Run-3 far-forward experiments. In any case, considering the large uncertainty due to scale variations affecting our predictions, and the possible enhancing effect of additional radiative corrections (in particular the NNLO ones are positive for inclusive cross sections), one can still hope for a more optimistic scenario. 
Furthermore, considering the even larger uncertainty affecting calculations as approximate as those in Ref.~\cite{Kling:2021gos}, which, as explained above, can provide at most just an order-of-magnitude estimate, one can still claim for consistency
between their predictions, the predictions of other authors using tools with similar accuracy (see, e.g., the evaluations conducted in Ref.~\cite{Beni:2020yfy} 
and in 
Ref. \cite{Ahdida:2020evc}) 
and ours.

\section{Conclusions}
\label{sec:conclu}

We have performed a QCD evaluation of the ($\nu_\tau+\bar\nu_\tau$) rapidity and energy distributions, as well as distributions of CC events induced by them, of interest to experiments in the forward region of the LHC. NLO QCD radiative corrections are included in our estimates of charm production and neutrino-induced DIS. Our focus is on PDF uncertainties. 
As for charm production, we use as a basis the 40 sets of the most recent PROSA NLO PDF fit, and, as an alternative, the central sets of other 3-flavour NLO PDFs. In particular, we consider those provided by the ABMP16, CT14, and NNPDF3.1 collaborations, together with their associated charm mass values, converted to the on-shell renormalization scheme for those PDFs accompanied by charm mass values in the $\overline{{\rm MS}}$ mass renormalization scheme, and their associated $\alpha_s(M_Z)$ values. 
All these sets have been widely used for LHC predictions at small and mid-rapidities and their effects have also already been studied in a more inclusive rapidity range in the context of prompt neutrino production in the atmosphere (see e.g. ~\cite{Zenaiev:2019ktw, Garzelli:2016xmx, Benzke:2017yjn, Gauld:2015kvh}). 
We investigate their effects in the computation of forward $\nu_\tau$ and $\bar{\nu}_\tau$ spectra at the LHC for the first time in this work.

Within the PROSA PDFs, the largest uncertainties turned out to come from the scale dependence in the theoretical predictions for charm production used for the fit, rather than from the fit procedure itself, the experimental data, and the parameterization ansatz specifying the form of the PDFs at the initial scale or the variations of this scale. These PDF features are reflected in our predictions, where the largest component of the PDF uncertainty turned out to be the so-called ``model" uncertainty from the PROSA fit. 
Our computation of charm hadroproduction has the same accuracy as in the PROSA fit, and, as default, we adopt the same renormalization and factorization scales ($m_{T,2}$).
While the scale uncertainty is reduced when setting the scale equal to $m_{T,2}$, rather than to the $m_T$ functional form adopted more often in the literature, the total number of events ranges between $\sim 0.2-2$ times the central prediction, depending on the scale. The percentage size of scale uncertainty with respect to central predictions is approximately uniform with rapidity. This large scale dependence is a signal that higher-order corrections to charm pair production are needed to improve our current estimates of $D$-meson production at the LHC. This applies to all rapidity ranges discussed in this paper.

For fixed input renormalization and factorization scales and $\langle\,  k_T \,\rangle$, the PROSA PDF uncertainty band does not include all the other PDF central predictions for the tau neutrino energy distribution and the corresponding number of events, although it is expected that the uncertainty bands for the other PDFs overlap those of the 40 PROSA sets. The central CT14 NLO prediction is the largest. It yields a number of events that is a factor of $\sim 1.6$ $(\sim 1.9)$ times higher than the central PROSA prediction for $7.2<\eta_\nu<8.6$ ($\eta_\nu>8.9$) as compared to the PROSA PDF uncertainty of $\sim \pm 20-30\%$. However, the scale uncertainty for the PROSA evaluation still covers the range of event number predictions from the other central PDF sets.

If we restrict our attention to $E_\nu\gsim 1$ TeV, the CT14 PDF predicted number of events are more than a factor of 2 larger than the PROSA ones. Predictions corresponding to high neutrino energies are sensitive to gluon PDFs in the $x\gsim 0.3$ region, where the constraints from data are not strong. Besides the invaluable role that will be played by future collider experiments, like the foreseen electron-ion collider (EIC)
\cite{AbdulKhalek:2021gbh}, even future measurements of tau neutrino and antineutrino interactions in this energy regime may help pin down the behavior of large-$x$ PDFs. At such high energies at forward rapidities, $\nu_e+\bar\nu_e$ fluxes from charm meson decays dominate with respect to the corresponding contributions from kaon decays 
\cite{Kling:2021gos}, thereby enhancing the statistics for CC events from high energy neutrinos from charm decays. Kaon decay contributions to $\nu_\mu+\bar\nu_\mu$ would have to be well modeled in order to exploit high energy fluxes of all three flavours of neutrinos to better understand large-$x$ PDFs at a Forward Physics Facility.

We have considered different renormalization and factorization scale dependencies and different $\langle k_T\rangle$ values.
The choice of the functional form of the scale dependence ($m_T $ or $m_{T,2}$) causes an offset to the energy distribution of $\nu_\tau+\bar\nu_\tau$ that changes slowly with energy above a few hundred GeV. The offset is largely independent of rapidity range.
On the other hand, the value of $\langle k_T\rangle$ impacts different rapidity ranges in a different way for neutrino energies above $\sim 500$ GeV. In particular,  the ratio of the energy distribution with $\langle k_T\rangle=1.2$ GeV to the energy distribution with $\langle k_T\rangle=0.7$ GeV is larger, and grows with energy,
for $\eta_\nu>6.9$ as compared to $\eta_\nu>8.9$.

The energy distributions of $\nu_\tau+\bar{\nu}_\tau$, shown in Figs. \ref{fig:energy-pdfuncertainties} and \ref{fig:e-direct-chain-scale-tot}, are reported in Tables \ref{tab:6.9}--\ref{tab:8.9a} in Appendix \ref{sec:tables} and in ancillary files on the \href{https://arxiv.org/src/2112.11605v1/anc}{arXiv}, distributed together with the preprint version of this manuscript.

Our $\nu_\tau+\bar{\nu}_\tau$ energy distributions differ somewhat from those estimated with Monte Carlo generators, as in, e.g., Ref. \cite{Kling:2021gos}. Detailed comparisons are difficult as the detector geometries are handled in different ways. SND@LHC reports that they expect 25 $\nu_\tau+\bar\nu_\tau$ CC events for their 850 kg tungsten detector for 150 fb$^{-1}$ \cite{Ahdida:2020evc}, much more than our prediction of $1-8$ events for 830 kg of tungsten for $7.2<\eta_\nu<8.6$. The estimate of 1-8 events accounts for our uncertainty band in Table 3, whereas the value of 25 events is not accompanied by an uncertainty. However, if it would, the corresponding band would be bigger than our one.  
Compared to the SIBYLL~2.3c result for FASER$\nu$ in Ref. \cite{Kling:2021gos}, our NLO neutrino energy distribution for $\eta_\nu>8.9$ is harder. For $(\mu_R,\mu_F)=(1,2)m_T$ and $\langle k_T\rangle=1.2$ GeV, our results for the neutrino energy distribution are larger by a factor as large as $\sim 2-6$ for
$E_\nu=10-300$ GeV, increasing to a factor of $\sim 10$ at 2 TeV neutrino energy, nearly matching the DPMJET3.2017 histogram in Ref. \cite{Kling:2021gos} at TeV neutrino energies. Our neutrino energy distribution evaluated using default parameter set with PROSA PDFs most resembles the PYTHIA8.2 predictions for $E_\nu\gsim $ a few hundred GeV. However, our distributions at low energies are higher than the PYTHIA8.2 ones. That the central predictions using perturbative QCD at NLO are larger than Monte Carlo evaluations is not surprising, since the Monte Carlos simulations use as a seed leading order evaluations of charm quark pair production. 
While performing all these comparisons, we should recall that 
leading order predictions for heavy-quark hadroproduction can, at most, provide an order-of-magnitude estimate, being accompanied by a huge scale uncertainty band, much larger than the corresponding NLO scale uncertainty band. 
A NNLO/NLO comparison will be 
interesting, and will become possible as soon as NNLO predictions for
differential distributions for charm pair and $D$-meson production become available.

For the total number of events, the uncertainties due to the scale dependence and the PROSA PDFs in the evaluation of charm production are similar in all the rapidity ranges considered here, namely rapidities larger than 6.9. 
With our defaults scales proportional to $m_{T,2}$ and $\langle k_T\rangle$ value of 0.7 GeV, the scale variations lead to approximately +90\% and -70\% uncertainty around the central value. The PROSA PDF uncertainties are between +20\% and -30\%. The NNPDF3.1, CT14 and ABMP16 predictions for the number of events with the default scale and  $\langle k_T\rangle$ value lie within these large error bands. The CT14 set yields the largest prediction for the number of events given the default inputs, depending on the rapidity interval, which is a factor of 1.5--1.8 larger than the prediction using the PROSA PDF central set with $(\mu_R,\mu_F)=(1,1)m_T$ and $\langle k_T\rangle=0.7$ GeV, as shown in Tables \ref{table:events1p5} and \ref{table:events3ab}.
For an integrated luminosity of 150 fb$^{-1}$ and nominal rapidity and mass values for SND@LHC ($7.2<\eta_\nu<8.6$, 830 kg) and FASER$\nu$ ($\eta_\nu>8.9$, 1.2 ton), the predicted total number of $\nu_\tau+\bar\nu_\tau$-induced charged-current events ranges in the intervals of 0.9--8.0 
and 2.3--23.7 events, respectively.

The high luminosity era and larger neutrino detectors will bring a significant increase in the number of $\nu_\tau+\bar\nu_\tau$ charged current events. Our predictions with central PDF set range in the interval $\sim 4,500-8,600$ events for a tungsten detector with a 1~m radius and a 1~m length. 
 The high statistics of neutrino events from charm in experiments in a Forward Physics Facility will enable better constraints on the small-$x$ and large-$x$ PDFs, as LHCb measurements of charm hadrons for $y=2-4.5$ has already done for gluon distributions with $x$ as small as~$\gtrsim 10^{-6}$. \cite{Zenaiev:2019ktw,PROSA:2015yid,Gauld:2016kpd}. Meanwhile, it will be important to develop differential calculations including NNLO and higher-order corrections (see e.g. the recent work of Ref.~\cite{Catani:2020kkl}), to refine the theoretical predictions by reducing the large scale uncertainties that are inherent to NLO, regardless of the PDF set.\\

\noindent{\bf Note added:} Additional data tables with the double-differential neutrino energy and pseudo-rapidity distributions of $\nu_\tau+\bar\nu_\tau$ from $D_s^\pm$ production and decay, and of $\nu_e+\bar\nu_e$ from $D^\pm$ production and decay, appear on the arXiv as ancillary files for Ref. \cite{Bai:2022jcs}.

\acknowledgments

This work is closely related to a Snowmass Letter of Interest, that
we submitted in summer 2020 to the Snowmass process. 
We are indebted with the conveners of the Snowmass EF06 group for continuous support and stimulating discussions.
We are grateful to members of the PROSA collaboration for useful discussions on various aspects of the PROSA and of the other PDF fits of which they are authors.
This work is supported in part by US Department of Energy grants DE-SC-0010113, DE-SC-0012704, by  German BMBF contract 05H21GUCCA, and the Korean Research Foundation (KRF) through the CERN-Korea Fellowship program and the National Research Foundation of Korea (NRF) grant funded by the 
Korea government(MSIT\footnote{MSIT : Ministry of Science and ICT}) (No. 2021R1A2C1009296 and NRF-2017R1A2B4004308). 

\appendix
\section{PDF Uncertainty Bands}
\label{sec:errorbands}

The uncertainty bands inherent the PROSA PDF fit (sets $i=1-15$), the underlying theoretical models used for theory predictions (sets $i=16-29$) and the PROSA parameterization ($i=30-40$) are each evaluated according to a prescription that depends on whether the differential distribution for set $i$, generically denoted d$\sigma(i)$ is larger or smaller that the differential distribution for the same bin(s) with the best-fit set d$\sigma(0)$. In terms of 
$\Delta(i)={\rm d}\sigma(i)-{\rm d}\sigma(0)$, for each category of uncertainty $a=$~fit or model, the error $\Sigma_{a\pm}$ above or below the best fit value is determined according to
\begin{eqnarray}
\Sigma_{a+}^2 &=& \sum_{i, \Delta(i)>0} [\Delta(i)]^2\\
\Sigma_{a-}^2 &=& \sum_{i, \Delta(i)<0} [\Delta(i)]^2\,.
    \label{eq:pdf-uncertain}
\end{eqnarray}
For the parameterization uncertainty ($i=30-40$),
\begin{eqnarray}
\Sigma_{a={\rm param}+}^2 &=& \Bigl[{\rm max}\bigl(\Delta(i)\bigr)\Bigr]^2, \quad\Delta(i)>0\\
\Sigma_{a={\rm param}-}^2 &=& 
\Bigl[{\rm max}\bigl(-\Delta(i)\bigr)\Bigr]^2,\quad \Delta(i)<0\,.
    \label{eq:uncertain-param}
\end{eqnarray}
The total PROSA PDF uncertainty comes from 
\begin{equation}
    \Sigma_\pm = \Bigl[\Sigma^2_{{\rm model}\pm}+\Sigma^2_{{\rm fit}\pm}+
    \Sigma^2_{{\rm param}\pm}\Bigr]^{1/2}\,.
\end{equation}

For the nCTEQ15 PDFs, the prescription for the PDF uncertainty comes from $n=32$ sets. The PDF uncertainty bars, symmetric around the central PDF set, come from evaluating the differential cross sections \cite{Kovarik:2015cma}
\begin{equation}
\label{eq:ncteqerror}
    \Sigma_{\pm}=\pm \frac{1}{2}\sqrt{\sum_{k=1,16} (d\sigma(2k)-d\sigma(2k-1))^2}\,.
\end{equation}

\section{Tables for the energy distributions and charged-current cross sections of tau neutrinos plus antineutrinos}
\label{sec:tables}

Tables 5--10 show the energy distributions and errors that correspond to Fig.~\ref{fig:energy-pdfuncertainties} and \ref{fig:e-direct-chain-scale-tot}. Table 11 shows the tau neutrino and antineutrino charged-current cross section per nucleon with PDF uncertainties for the nCTEQ15 PDF sets for a tungsten target and Table 12 shows the tau neutrino charged-current cross section per nucleon for the VFNS PROSA PDF set,
as in Fig. \ref{fig:sigma-nuc}. The data tables are also available at
\href{https://arxiv.org/src/2112.11605v1/anc}{https://arxiv.org/src/2112.11605v1/anc}.

\begin{table}[]
    \centering
    \begin{tabular}{|c|c|c|c|c|c|c|}
    \hline
   $E_\nu$ & Bin & PROSA $m_{T,2}$ &
   $\Delta_{\rm scale}^{\rm upper}$ &
     $\Delta_{\rm scale}^{\rm lower}$ &$\Delta_{\rm PDF}^{\rm upper}$
     &$\Delta_{\rm PDF}^{\rm lower}$\\
     \hline\hline
     [GeV] & [GeV] & [$\mu$b/GeV] & 
     [$\mu$b/GeV] & [$\mu$b/GeV] & [$\mu$b/GeV] & [$\mu$b/GeV]\\
    \hline
    \hline
   10 &  20 & $   1.703\times 10^{-3}$ & $   1.196\times 10^{-3}$ & $  -1.183\times 10^{-3}$ & $   1.585\times 10^{-4}$ & $  -2.008\times 10^{-4}$ \\ \hline
   30 &  20 & $   3.574\times 10^{-3}$ & $   2.595\times 10^{-3}$ & $  -2.545\times 10^{-3}$ & $   4.099\times 10^{-4}$ & $  -5.697\times 10^{-4}$ \\ \hline
   50 &  20 & $   4.536\times 10^{-3}$ & $   3.555\times 10^{-3}$ & $  -3.313\times 10^{-3}$ & $   5.640\times 10^{-4}$ & $  -9.084\times 10^{-4}$ \\ \hline
   70 &  20 & $   4.963\times 10^{-3}$ & $   4.155\times 10^{-3}$ & $  -3.703\times 10^{-3}$ & $   6.911\times 10^{-4}$ & $  -1.145\times 10^{-3}$ \\ \hline
   90 &  20 & $   4.863\times 10^{-3}$ & $   4.231\times 10^{-3}$ & $  -3.660\times 10^{-3}$ & $   7.532\times 10^{-4}$ & $  -1.211\times 10^{-3}$ \\ \hline
  110 &  20 & $   4.399\times 10^{-3}$ & $   3.834\times 10^{-3}$ & $  -3.310\times 10^{-3}$ & $   6.877\times 10^{-4}$ & $  -1.151\times 10^{-3}$ \\ \hline
  130 &  20 & $   3.803\times 10^{-3}$ & $   3.323\times 10^{-3}$ & $  -2.845\times 10^{-3}$ & $   6.168\times 10^{-4}$ & $  -1.019\times 10^{-3}$ \\ \hline
  150 &  20 & $   3.238\times 10^{-3}$ & $   2.800\times 10^{-3}$ & $  -2.407\times 10^{-3}$ & $   5.541\times 10^{-4}$ & $  -8.692\times 10^{-4}$ \\ \hline
  170 &  20 & $   2.763\times 10^{-3}$ & $   2.372\times 10^{-3}$ & $  -2.029\times 10^{-3}$ & $   4.714\times 10^{-4}$ & $  -7.436\times 10^{-4}$ \\ \hline
  190 &  20 & $   2.374\times 10^{-3}$ & $   2.037\times 10^{-3}$ & $  -1.745\times 10^{-3}$ & $   4.236\times 10^{-4}$ & $  -6.383\times 10^{-4}$ \\ \hline
  220 &  40 & $   1.948\times 10^{-3}$ & $   1.656\times 10^{-3}$ & $  -1.429\times 10^{-3}$ & $   3.415\times 10^{-4}$ & $  -5.268\times 10^{-4}$ \\ \hline
  260 &  40 & $   1.558\times 10^{-3}$ & $   1.334\times 10^{-3}$ & $  -1.148\times 10^{-3}$ & $   2.698\times 10^{-4}$ & $  -4.298\times 10^{-4}$ \\ \hline
  300 &  40 & $   1.279\times 10^{-3}$ & $   1.105\times 10^{-3}$ & $  -9.422\times 10^{-4}$ & $   2.274\times 10^{-4}$ & $  -3.574\times 10^{-4}$ \\ \hline
  340 &  40 & $   1.102\times 10^{-3}$ & $   9.639\times 10^{-4}$ & $  -8.190\times 10^{-4}$ & $   1.984\times 10^{-4}$ & $  -3.089\times 10^{-4}$ \\ \hline
  380 &  40 & $   9.532\times 10^{-4}$ & $   8.413\times 10^{-4}$ & $  -7.106\times 10^{-4}$ & $   1.714\times 10^{-4}$ & $  -2.716\times 10^{-4}$ \\ \hline
  420 &  40 & $   8.280\times 10^{-4}$ & $   7.360\times 10^{-4}$ & $  -6.139\times 10^{-4}$ & $   1.524\times 10^{-4}$ & $  -2.382\times 10^{-4}$ \\ \hline
  460 &  40 & $   7.212\times 10^{-4}$ & $   6.501\times 10^{-4}$ & $  -5.399\times 10^{-4}$ & $   1.345\times 10^{-4}$ & $  -2.108\times 10^{-4}$ \\ \hline
  500 &  40 & $   6.296\times 10^{-4}$ & $   5.610\times 10^{-4}$ & $  -4.710\times 10^{-4}$ & $   1.163\times 10^{-4}$ & $  -1.878\times 10^{-4}$ \\ \hline
  540 &  40 & $   5.464\times 10^{-4}$ & $   4.878\times 10^{-4}$ & $  -4.075\times 10^{-4}$ & $   1.056\times 10^{-4}$ & $  -1.620\times 10^{-4}$ \\ \hline
  580 &  40 & $   4.753\times 10^{-4}$ & $   4.248\times 10^{-4}$ & $  -3.528\times 10^{-4}$ & $   9.474\times 10^{-5}$ & $  -1.433\times 10^{-4}$ \\ \hline
  650 & 100 & $   3.733\times 10^{-4}$ & $   3.295\times 10^{-4}$ & $  -2.750\times 10^{-4}$ & $   7.604\times 10^{-5}$ & $  -1.141\times 10^{-4}$ \\ \hline
  750 & 100 & $   2.614\times 10^{-4}$ & $   2.293\times 10^{-4}$ & $  -1.909\times 10^{-4}$ & $   5.679\times 10^{-5}$ & $  -8.145\times 10^{-5}$ \\ \hline
  850 & 100 & $   1.834\times 10^{-4}$ & $   1.622\times 10^{-4}$ & $  -1.323\times 10^{-4}$ & $   4.161\times 10^{-5}$ & $  -5.851\times 10^{-5}$ \\ \hline
  950 & 100 & $   1.289\times 10^{-4}$ & $   1.145\times 10^{-4}$ & $  -9.169\times 10^{-5}$ & $   3.039\times 10^{-5}$ & $  -4.161\times 10^{-5}$ \\ \hline
 1050 & 100 & $   9.105\times 10^{-5}$ & $   8.183\times 10^{-5}$ & $  -6.423\times 10^{-5}$ & $   2.199\times 10^{-5}$ & $  -3.032\times 10^{-5}$ \\ \hline
 1150 & 100 & $   6.421\times 10^{-5}$ & $   5.857\times 10^{-5}$ & $  -4.474\times 10^{-5}$ & $   1.684\times 10^{-5}$ & $  -2.111\times 10^{-5}$ \\ \hline
 1250 & 100 & $   4.574\times 10^{-5}$ & $   4.216\times 10^{-5}$ & $  -3.164\times 10^{-5}$ & $   1.249\times 10^{-5}$ & $  -1.528\times 10^{-5}$ \\ \hline
 1350 & 100 & $   3.273\times 10^{-5}$ & $   3.019\times 10^{-5}$ & $  -2.240\times 10^{-5}$ & $   9.190\times 10^{-6}$ & $  -1.112\times 10^{-5}$ \\ \hline
 1450 & 100 & $   2.340\times 10^{-5}$ & $   2.176\times 10^{-5}$ & $  -1.584\times 10^{-5}$ & $   6.673\times 10^{-6}$ & $  -8.022\times 10^{-6}$ \\ \hline
 1550 & 100 & $   1.681\times 10^{-5}$ & $   1.610\times 10^{-5}$ & $  -1.121\times 10^{-5}$ & $   5.177\times 10^{-6}$ & $  -5.602\times 10^{-6}$ \\ \hline
 1650 & 100 & $   1.212\times 10^{-5}$ & $   1.166\times 10^{-5}$ & $  -8.008\times 10^{-6}$ & $   3.845\times 10^{-6}$ & $  -4.186\times 10^{-6}$ \\ \hline
 1750 & 100 & $   8.752\times 10^{-6}$ & $   8.339\times 10^{-6}$ & $  -5.736\times 10^{-6}$ & $   2.863\times 10^{-6}$ & $  -2.998\times 10^{-6}$ \\ \hline
 1850 & 100 & $   6.329\times 10^{-6}$ & $   6.047\times 10^{-6}$ & $  -4.095\times 10^{-6}$ & $   2.080\times 10^{-6}$ & $  -2.173\times 10^{-6}$ \\ \hline
 1950 & 100 & $   4.595\times 10^{-6}$ & $   4.530\times 10^{-6}$ & $  -2.945\times 10^{-6}$ & $   1.513\times 10^{-6}$ & $  -1.636\times 10^{-6}$ \\ \hline
    \end{tabular}
    \caption{Sum of $\nu_\tau$ and $\bar\nu_\tau$ energy distributions $d\sigma/dE_\nu$ for $\sqrt{s}=14$ TeV and $\eta_\nu>6.9$.
    The predictions are shown for the PROSA NLO PDF set with scales $(\mu_R,\mu_F)=(1,1)m_{T,2}$ with $\langle k_T\rangle = 0.7$ GeV.
    This table with 20 GeV energy bins for the whole energy range is available as an ancillary file on the \href{https://arxiv.org/src/2112.11605v1/anc}{arXiv link} to this paper.}
    \label{tab:6.9}
\end{table}

\begin{table}[]
    \centering
    \begin{tabular}{|c|c|c|c|c|c|c|}
    \hline
   $E_\nu$  & Bin &PROSA $m_{T,2}$ &
   $\Delta_{\rm scale}^{\rm upper}$ &
     $\Delta_{\rm scale}^{\rm lower}$ &$\Delta_{\rm PDF}^{\rm upper}$
     &$\Delta_{\rm PDF}^{\rm lower}$\\
     \hline\hline
     [GeV] & [GeV]& [$\mu$b/GeV] & [$\mu$b/GeV]
     &[$\mu$b/GeV] & [$\mu$b/GeV] &[$\mu$b/GeV] \\
    \hline
    \hline
   10 &  20 & $   8.215\times 10^{-4}$ & $   5.807\times 10^{-4}$ & $  -5.717\times 10^{-4}$ & $   7.903\times 10^{-5}$ & $  -9.584\times 10^{-5}$ \\ \hline
   30 &  20 & $   1.708\times 10^{-3}$ & $   1.243\times 10^{-3}$ & $  -1.216\times 10^{-3}$ & $   1.995\times 10^{-4}$ & $  -2.666\times 10^{-4}$ \\ \hline
   50 &  20 & $   2.135\times 10^{-3}$ & $   1.647\times 10^{-3}$ & $  -1.552\times 10^{-3}$ & $   2.664\times 10^{-4}$ & $  -4.149\times 10^{-4}$ \\ \hline
   70 &  20 & $   2.342\times 10^{-3}$ & $   1.933\times 10^{-3}$ & $  -1.737\times 10^{-3}$ & $   3.171\times 10^{-4}$ & $  -5.250\times 10^{-4}$ \\ \hline
   90 &  20 & $   2.383\times 10^{-3}$ & $   2.065\times 10^{-3}$ & $  -1.787\times 10^{-3}$ & $   3.676\times 10^{-4}$ & $  -5.804\times 10^{-4}$ \\ \hline
  110 &  20 & $   2.302\times 10^{-3}$ & $   2.035\times 10^{-3}$ & $  -1.740\times 10^{-3}$ & $   3.649\times 10^{-4}$ & $  -6.053\times 10^{-4}$ \\ \hline
  130 &  20 & $   2.124\times 10^{-3}$ & $   1.895\times 10^{-3}$ & $  -1.607\times 10^{-3}$ & $   3.522\times 10^{-4}$ & $  -5.879\times 10^{-4}$ \\ \hline
  150 &  20 & $   1.884\times 10^{-3}$ & $   1.684\times 10^{-3}$ & $  -1.420\times 10^{-3}$ & $   3.409\times 10^{-4}$ & $  -5.224\times 10^{-4}$ \\ \hline
  170 &  20 & $   1.648\times 10^{-3}$ & $   1.462\times 10^{-3}$ & $  -1.235\times 10^{-3}$ & $   2.921\times 10^{-4}$ & $  -4.659\times 10^{-4}$ \\ \hline
  190 &  20 & $   1.426\times 10^{-3}$ & $   1.260\times 10^{-3}$ & $  -1.060\times 10^{-3}$ & $   2.686\times 10^{-4}$ & $  -4.014\times 10^{-4}$ \\ \hline
  220 &  40 & $   1.159\times 10^{-3}$ & $   1.002\times 10^{-3}$ & $  -8.559\times 10^{-4}$ & $   2.117\times 10^{-4}$ & $  -3.293\times 10^{-4}$ \\ \hline
  260 &  40 & $   8.970\times 10^{-4}$ & $   7.726\times 10^{-4}$ & $  -6.612\times 10^{-4}$ & $   1.646\times 10^{-4}$ & $  -2.541\times 10^{-4}$ \\ \hline
  300 &  40 & $   7.245\times 10^{-4}$ & $   6.249\times 10^{-4}$ & $  -5.340\times 10^{-4}$ & $   1.324\times 10^{-4}$ & $  -2.057\times 10^{-4}$ \\ \hline
  340 &  40 & $   6.073\times 10^{-4}$ & $   5.316\times 10^{-4}$ & $  -4.498\times 10^{-4}$ & $   1.145\times 10^{-4}$ & $  -1.703\times 10^{-4}$ \\ \hline
  380 &  40 & $   5.279\times 10^{-4}$ & $   4.647\times 10^{-4}$ & $  -3.937\times 10^{-4}$ & $   9.438\times 10^{-5}$ & $  -1.520\times 10^{-4}$ \\ \hline
  420 &  40 & $   4.651\times 10^{-4}$ & $   4.151\times 10^{-4}$ & $  -3.433\times 10^{-4}$ & $   8.669\times 10^{-5}$ & $  -1.353\times 10^{-4}$ \\ \hline
  460 &  40 & $   4.144\times 10^{-4}$ & $   3.783\times 10^{-4}$ & $  -3.115\times 10^{-4}$ & $   7.945\times 10^{-5}$ & $  -1.225\times 10^{-4}$ \\ \hline
  500 &  40 & $   3.725\times 10^{-4}$ & $   3.389\times 10^{-4}$ & $  -2.814\times 10^{-4}$ & $   7.058\times 10^{-5}$ & $  -1.134\times 10^{-4}$ \\ \hline
  540 &  40 & $   3.330\times 10^{-4}$ & $   3.052\times 10^{-4}$ & $  -2.515\times 10^{-4}$ & $   6.512\times 10^{-5}$ & $  -1.007\times 10^{-4}$ \\ \hline
  580 &  40 & $   2.981\times 10^{-4}$ & $   2.738\times 10^{-4}$ & $  -2.245\times 10^{-4}$ & $   6.137\times 10^{-5}$ & $  -9.185\times 10^{-5}$ \\ \hline
  650 & 100 & $   2.449\times 10^{-4}$ & $   2.233\times 10^{-4}$ & $  -1.838\times 10^{-4}$ & $   5.074\times 10^{-5}$ & $  -7.678\times 10^{-5}$ \\ \hline
  750 & 100 & $   1.814\times 10^{-4}$ & $   1.637\times 10^{-4}$ & $  -1.353\times 10^{-4}$ & $   3.969\times 10^{-5}$ & $  -5.869\times 10^{-5}$ \\ \hline
  850 & 100 & $   1.319\times 10^{-4}$ & $   1.191\times 10^{-4}$ & $  -9.714\times 10^{-5}$ & $   3.112\times 10^{-5}$ & $  -4.284\times 10^{-5}$ \\ \hline
  950 & 100 & $   9.535\times 10^{-5}$ & $   8.569\times 10^{-5}$ & $  -6.937\times 10^{-5}$ & $   2.304\times 10^{-5}$ & $  -3.155\times 10^{-5}$ \\ \hline
 1050 & 100 & $   6.849\times 10^{-5}$ & $   6.159\times 10^{-5}$ & $  -4.924\times 10^{-5}$ & $   1.671\times 10^{-5}$ & $  -2.328\times 10^{-5}$ \\ \hline
 1150 & 100 & $   4.867\times 10^{-5}$ & $   4.457\times 10^{-5}$ & $  -3.449\times 10^{-5}$ & $   1.297\times 10^{-5}$ & $  -1.627\times 10^{-5}$ \\ \hline
 1250 & 100 & $   3.471\times 10^{-5}$ & $   3.193\times 10^{-5}$ & $  -2.438\times 10^{-5}$ & $   9.370\times 10^{-6}$ & $  -1.181\times 10^{-5}$ \\ \hline
 1350 & 100 & $   2.461\times 10^{-5}$ & $   2.282\times 10^{-5}$ & $  -1.703\times 10^{-5}$ & $   6.995\times 10^{-6}$ & $  -8.389\times 10^{-6}$ \\ \hline
 1450 & 100 & $   1.739\times 10^{-5}$ & $   1.618\times 10^{-5}$ & $  -1.184\times 10^{-5}$ & $   5.023\times 10^{-6}$ & $  -6.004\times 10^{-6}$ \\ \hline
 1550 & 100 & $   1.225\times 10^{-5}$ & $   1.177\times 10^{-5}$ & $  -8.204\times 10^{-6}$ & $   3.861\times 10^{-6}$ & $  -4.022\times 10^{-6}$ \\ \hline
 1650 & 100 & $   8.687\times 10^{-6}$ & $   8.315\times 10^{-6}$ & $  -5.734\times 10^{-6}$ & $   2.737\times 10^{-6}$ & $  -3.012\times 10^{-6}$ \\ \hline
 1750 & 100 & $   6.112\times 10^{-6}$ & $   5.850\times 10^{-6}$ & $  -3.994\times 10^{-6}$ & $   2.006\times 10^{-6}$ & $  -2.097\times 10^{-6}$ \\ \hline
 1850 & 100 & $   4.306\times 10^{-6}$ & $   4.050\times 10^{-6}$ & $  -2.774\times 10^{-6}$ & $   1.386\times 10^{-6}$ & $  -1.481\times 10^{-6}$ \\ \hline
 1950 & 100 & $   3.040\times 10^{-6}$ & $   2.949\times 10^{-6}$ & $  -1.932\times 10^{-6}$ & $   1.017\times 10^{-6}$ & $  -1.093\times 10^{-6}$ \\ \hline
    \end{tabular}
    \caption{Sum of $\nu_\tau$ and $\bar\nu_\tau$ energy distributions $d\sigma/dE_\nu$ for $\sqrt{s}=14$ TeV, $7.2<\eta_\nu<8.6$. 
The predictions are shown for the PROSA PDF set with scales $(\mu_R,\mu_F)=(1,1)m_{T,2}$ with $\langle k_T\rangle~=~0.7$ GeV.
    This table with 20 GeV energy bins for the whole energy range is available as an ancillary file on the \href{https://arxiv.org/src/2112.11605v1/anc}{arXiv link} to this paper.    }
    \label{tab:7.2}
\end{table}

\begin{table}[]
    \centering
    \begin{tabular}{|c|c|c|c|c|c|c|}
    \hline
   $E_\nu$ & Bin  & PROSA $m_{T,2}$ &
   $\Delta_{\rm scale}^{\rm upper}$ &
     $\Delta_{\rm scale}^{\rm lower}$ &$\Delta_{\rm PDF}^{\rm upper}$
     &$\Delta_{\rm PDF}^{\rm lower}$\\
     \hline\hline
     [GeV] & [GeV] & [$\mu$b/GeV] & 
     [$\mu$b/GeV] & [$\mu$b/GeV] &[$\mu$b/GeV]  & [$\mu$b/GeV] \\
    \hline
    \hline
   10 &  20 & $   2.926\times 10^{-5}$ & $   2.009\times 10^{-5}$ & $  -2.026\times 10^{-5}$ & $   4.037\times 10^{-6}$ & $  -4.653\times 10^{-6}$ \\ \hline
   30 &  20 & $   5.933\times 10^{-5}$ & $   4.387\times 10^{-5}$ & $  -4.205\times 10^{-5}$ & $   7.516\times 10^{-6}$ & $  -8.640\times 10^{-6}$ \\ \hline
   50 &  20 & $   7.324\times 10^{-5}$ & $   5.659\times 10^{-5}$ & $  -5.291\times 10^{-5}$ & $   1.043\times 10^{-5}$ & $  -1.286\times 10^{-5}$ \\ \hline
   70 &  20 & $   7.989\times 10^{-5}$ & $   6.484\times 10^{-5}$ & $  -5.874\times 10^{-5}$ & $   1.259\times 10^{-5}$ & $  -1.665\times 10^{-5}$ \\ \hline
   90 &  20 & $   8.135\times 10^{-5}$ & $   6.697\times 10^{-5}$ & $  -6.035\times 10^{-5}$ & $   1.062\times 10^{-5}$ & $  -1.952\times 10^{-5}$ \\ \hline
  110 &  20 & $   8.047\times 10^{-5}$ & $   6.889\times 10^{-5}$ & $  -6.065\times 10^{-5}$ & $   1.326\times 10^{-5}$ & $  -2.025\times 10^{-5}$ \\ \hline
  130 &  20 & $   7.749\times 10^{-5}$ & $   6.953\times 10^{-5}$ & $  -5.829\times 10^{-5}$ & $   1.347\times 10^{-5}$ & $  -1.978\times 10^{-5}$ \\ \hline
  150 &  20 & $   7.431\times 10^{-5}$ & $   6.645\times 10^{-5}$ & $  -5.618\times 10^{-5}$ & $   1.250\times 10^{-5}$ & $  -2.123\times 10^{-5}$ \\ \hline
  170 &  20 & $   7.008\times 10^{-5}$ & $   6.415\times 10^{-5}$ & $  -5.350\times 10^{-5}$ & $   1.166\times 10^{-5}$ & $  -2.007\times 10^{-5}$ \\ \hline
  190 &  20 & $   6.558\times 10^{-5}$ & $   6.124\times 10^{-5}$ & $  -4.999\times 10^{-5}$ & $   1.230\times 10^{-5}$ & $  -1.968\times 10^{-5}$ \\ \hline
  220 &  40 & $   5.838\times 10^{-5}$ & $   5.364\times 10^{-5}$ & $  -4.437\times 10^{-5}$ & $   1.136\times 10^{-5}$ & $  -1.760\times 10^{-5}$ \\ \hline
  260 &  40 & $   5.021\times 10^{-5}$ & $   4.522\times 10^{-5}$ & $  -3.819\times 10^{-5}$ & $   9.776\times 10^{-6}$ & $  -1.608\times 10^{-5}$ \\ \hline
  300 &  40 & $   4.211\times 10^{-5}$ & $   3.851\times 10^{-5}$ & $  -3.159\times 10^{-5}$ & $   8.884\times 10^{-6}$ & $  -1.299\times 10^{-5}$ \\ \hline
  340 &  40 & $   3.524\times 10^{-5}$ & $   3.205\times 10^{-5}$ & $  -2.632\times 10^{-5}$ & $   8.008\times 10^{-6}$ & $  -1.102\times 10^{-5}$ \\ \hline
  380 &  40 & $   2.977\times 10^{-5}$ & $   2.645\times 10^{-5}$ & $  -2.211\times 10^{-5}$ & $   6.903\times 10^{-6}$ & $  -9.382\times 10^{-6}$ \\ \hline
  420 &  40 & $   2.530\times 10^{-5}$ & $   2.295\times 10^{-5}$ & $  -1.877\times 10^{-5}$ & $   5.461\times 10^{-6}$ & $  -7.881\times 10^{-6}$ \\ \hline
  460 &  40 & $   2.190\times 10^{-5}$ & $   2.000\times 10^{-5}$ & $  -1.617\times 10^{-5}$ & $   4.418\times 10^{-6}$ & $  -6.709\times 10^{-6}$ \\ \hline
  500 &  40 & $   1.901\times 10^{-5}$ & $   1.684\times 10^{-5}$ & $  -1.409\times 10^{-5}$ & $   4.236\times 10^{-6}$ & $  -5.602\times 10^{-6}$ \\ \hline
  540 &  40 & $   1.685\times 10^{-5}$ & $   1.542\times 10^{-5}$ & $  -1.238\times 10^{-5}$ & $   3.565\times 10^{-6}$ & $  -5.195\times 10^{-6}$ \\ \hline
  580 &  40 & $   1.501\times 10^{-5}$ & $   1.352\times 10^{-5}$ & $  -1.122\times 10^{-5}$ & $   3.454\times 10^{-6}$ & $  -4.686\times 10^{-6}$ \\ \hline
  650 & 100 & $   1.269\times 10^{-5}$ & $   1.205\times 10^{-5}$ & $  -9.589\times 10^{-6}$ & $   3.110\times 10^{-6}$ & $  -3.966\times 10^{-6}$ \\ \hline
  750 & 100 & $   1.042\times 10^{-5}$ & $   9.695\times 10^{-6}$ & $  -7.949\times 10^{-6}$ & $   2.417\times 10^{-6}$ & $  -3.541\times 10^{-6}$ \\ \hline
  850 & 100 & $   8.669\times 10^{-6}$ & $   8.137\times 10^{-6}$ & $  -6.612\times 10^{-6}$ & $   2.124\times 10^{-6}$ & $  -3.104\times 10^{-6}$ \\ \hline
  950 & 100 & $   7.124\times 10^{-6}$ & $   6.847\times 10^{-6}$ & $  -5.396\times 10^{-6}$ & $   1.967\times 10^{-6}$ & $  -2.447\times 10^{-6}$ \\ \hline
 1050 & 100 & $   5.957\times 10^{-6}$ & $   5.622\times 10^{-6}$ & $  -4.505\times 10^{-6}$ & $   1.555\times 10^{-6}$ & $  -2.233\times 10^{-6}$ \\ \hline
 1150 & 100 & $   4.841\times 10^{-6}$ & $   4.618\times 10^{-6}$ & $  -3.599\times 10^{-6}$ & $   1.456\times 10^{-6}$ & $  -1.819\times 10^{-6}$ \\ \hline
 1250 & 100 & $   3.975\times 10^{-6}$ & $   3.844\times 10^{-6}$ & $  -2.926\times 10^{-6}$ & $   1.265\times 10^{-6}$ & $  -1.409\times 10^{-6}$ \\ \hline
 1350 & 100 & $   3.297\times 10^{-6}$ & $   3.127\times 10^{-6}$ & $  -2.403\times 10^{-6}$ & $   1.012\times 10^{-6}$ & $  -1.230\times 10^{-6}$ \\ \hline
 1450 & 100 & $   2.650\times 10^{-6}$ & $   2.598\times 10^{-6}$ & $  -1.929\times 10^{-6}$ & $   7.990\times 10^{-7}$ & $  -9.300\times 10^{-7}$ \\ \hline
 1550 & 100 & $   2.184\times 10^{-6}$ & $   2.127\times 10^{-6}$ & $  -1.560\times 10^{-6}$ & $   6.545\times 10^{-7}$ & $  -8.266\times 10^{-7}$ \\ \hline
 1650 & 100 & $   1.731\times 10^{-6}$ & $   1.748\times 10^{-6}$ & $  -1.221\times 10^{-6}$ & $   6.160\times 10^{-7}$ & $  -6.468\times 10^{-7}$ \\ \hline
 1750 & 100 & $   1.384\times 10^{-6}$ & $   1.372\times 10^{-6}$ & $  -9.640\times 10^{-7}$ & $   4.698\times 10^{-7}$ & $  -5.099\times 10^{-7}$ \\ \hline
 1850 & 100 & $   1.091\times 10^{-6}$ & $   1.114\times 10^{-6}$ & $  -7.461\times 10^{-7}$ & $   4.140\times 10^{-7}$ & $  -3.764\times 10^{-7}$ \\ \hline
 1950 & 100 & $   8.694\times 10^{-7}$ & $   9.003\times 10^{-7}$ & $  -5.861\times 10^{-7}$ & $   2.954\times 10^{-7}$ & $  -3.301\times 10^{-7}$ \\ \hline
    \end{tabular}
    \caption{Sum of $\nu_\tau$ and $\bar\nu_\tau$ energy distributions $d\sigma/dE_\nu$ for $\sqrt{s}=14$ TeV, $\eta_\nu>8.9$.
    The predictions are shown for the PROSA PDF set with scales $(\mu_R,\mu_F)=(1,1)m_{T,2}$ with $\langle k_T\rangle = 0.7$~GeV.
    This table with 20 GeV energy bins for the whole energy range is available as an ancillary file on the \href{https://arxiv.org/src/2112.11605v1/anc}{arXiv link} to this paper.}
    \label{tab:8.9}
\end{table}

\begin{table}[]
    \centering
    \begin{tabular}{|c|c|c|c|c|c|}
    \hline
    $E_\nu$ & Bin      & PROSA $m_{T,1}$ & CT14 $m_{T,2}$ & 
         ABMP16 $m_{T,2}$ & NNPDF3.1 $m_{T,2}$\\
     \hline\hline
 [GeV]& [GeV]  &  [$\mu$b/GeV] & [$\mu$b/GeV] & [$\mu$b/GeV]  & [$\mu$b/GeV ] \\
    \hline
    \hline
   10 &  20 & $   2.685\times 10^{-3}$ & $   1.868\times 10^{-3}$ & $   1.891\times 10^{-3}$ & $   1.456\times 10^{-3}$ \\ \hline
   30 &  20 & $   5.778\times 10^{-3}$ & $   4.014\times 10^{-3}$ & $   4.028\times 10^{-3}$ & $   3.106\times 10^{-3}$ \\ \hline
   50 &  20 & $   7.409\times 10^{-3}$ & $   5.236\times 10^{-3}$ & $   5.161\times 10^{-3}$ & $   3.945\times 10^{-3}$ \\ \hline
   70 &  20 & $   7.998\times 10^{-3}$ & $   5.936\times 10^{-3}$ & $   5.719\times 10^{-3}$ & $   4.298\times 10^{-3}$ \\ \hline
   90 &  20 & $   7.806\times 10^{-3}$ & $   6.029\times 10^{-3}$ & $   5.662\times 10^{-3}$ & $   4.208\times 10^{-3}$ \\ \hline
  110 &  20 & $   7.125\times 10^{-3}$ & $   5.585\times 10^{-3}$ & $   5.136\times 10^{-3}$ & $   3.809\times 10^{-3}$ \\ \hline
  130 &  20 & $   6.296\times 10^{-3}$ & $   4.941\times 10^{-3}$ & $   4.442\times 10^{-3}$ & $   3.320\times 10^{-3}$ \\ \hline
  150 &  20 & $   5.498\times 10^{-3}$ & $   4.291\times 10^{-3}$ & $   3.790\times 10^{-3}$ & $   2.858\times 10^{-3}$ \\ \hline
  170 &  20 & $   4.770\times 10^{-3}$ & $   3.733\times 10^{-3}$ & $   3.235\times 10^{-3}$ & $   2.464\times 10^{-3}$ \\ \hline
  190 &  20 & $   4.175\times 10^{-3}$ & $   3.259\times 10^{-3}$ & $   2.791\times 10^{-3}$ & $   2.145\times 10^{-3}$ \\ \hline
  220 &  40 & $   3.440\times 10^{-3}$ & $   2.718\times 10^{-3}$ & $   2.298\times 10^{-3}$ & $   1.778\times 10^{-3}$ \\ \hline
  260 &  40 & $   2.736\times 10^{-3}$ & $   2.184\times 10^{-3}$ & $   1.842\times 10^{-3}$ & $   1.434\times 10^{-3}$ \\ \hline
  300 &  40 & $   2.247\times 10^{-3}$ & $   1.841\times 10^{-3}$ & $   1.517\times 10^{-3}$ & $   1.180\times 10^{-3}$ \\ \hline
  340 &  40 & $   1.883\times 10^{-3}$ & $   1.570\times 10^{-3}$ & $   1.310\times 10^{-3}$ & $   1.015\times 10^{-3}$ \\ \hline
  380 &  40 & $   1.629\times 10^{-3}$ & $   1.357\times 10^{-3}$ & $   1.135\times 10^{-3}$ & $   8.745\times 10^{-4}$ \\ \hline
  420 &  40 & $   1.415\times 10^{-3}$ & $   1.181\times 10^{-3}$ & $   9.863\times 10^{-4}$ & $   7.563\times 10^{-4}$ \\ \hline
  460 &  40 & $   1.233\times 10^{-3}$ & $   1.032\times 10^{-3}$ & $   8.590\times 10^{-4}$ & $   6.576\times 10^{-4}$ \\ \hline
  500 &  40 & $   1.081\times 10^{-3}$ & $   9.042\times 10^{-4}$ & $   7.497\times 10^{-4}$ & $   5.725\times 10^{-4}$ \\ \hline
  540 &  40 & $   9.498\times 10^{-4}$ & $   7.952\times 10^{-4}$ & $   6.505\times 10^{-4}$ & $   4.973\times 10^{-4}$ \\ \hline
  580 &  40 & $   8.378\times 10^{-4}$ & $   6.966\times 10^{-4}$ & $   5.687\times 10^{-4}$ & $   4.355\times 10^{-4}$ \\ \hline
  650 & 100 & $   6.628\times 10^{-4}$ & $   5.605\times 10^{-4}$ & $   4.471\times 10^{-4}$ & $   3.445\times 10^{-4}$ \\ \hline
  750 & 100 & $   4.830\times 10^{-4}$ & $   4.100\times 10^{-4}$ & $   3.149\times 10^{-4}$ & $   2.469\times 10^{-4}$ \\ \hline
  850 & 100 & $   3.523\times 10^{-4}$ & $   3.024\times 10^{-4}$ & $   2.220\times 10^{-4}$ & $   1.789\times 10^{-4}$ \\ \hline
  950 & 100 & $   2.559\times 10^{-4}$ & $   2.238\times 10^{-4}$ & $   1.576\times 10^{-4}$ & $   1.303\times 10^{-4}$ \\ \hline
 1050 & 100 & $   1.855\times 10^{-4}$ & $   1.674\times 10^{-4}$ & $   1.128\times 10^{-4}$ & $   9.576\times 10^{-5}$ \\ \hline
 1150 & 100 & $   1.341\times 10^{-4}$ & $   1.254\times 10^{-4}$ & $   8.058\times 10^{-5}$ & $   7.085\times 10^{-5}$ \\ \hline
 1250 & 100 & $   9.727\times 10^{-5}$ & $   9.426\times 10^{-5}$ & $   5.815\times 10^{-5}$ & $   5.241\times 10^{-5}$ \\ \hline
 1350 & 100 & $   7.121\times 10^{-5}$ & $   7.120\times 10^{-5}$ & $   4.235\times 10^{-5}$ & $   3.924\times 10^{-5}$ \\ \hline
 1450 & 100 & $   5.207\times 10^{-5}$ & $   5.417\times 10^{-5}$ & $   3.092\times 10^{-5}$ & $   2.937\times 10^{-5}$ \\ \hline
 1550 & 100 & $   3.813\times 10^{-5}$ & $   4.138\times 10^{-5}$ & $   2.259\times 10^{-5}$ & $   2.219\times 10^{-5}$ \\ \hline
 1650 & 100 & $   2.808\times 10^{-5}$ & $   3.142\times 10^{-5}$ & $   1.661\times 10^{-5}$ & $   1.672\times 10^{-5}$ \\ \hline
 1750 & 100 & $   2.077\times 10^{-5}$ & $   2.393\times 10^{-5}$ & $   1.224\times 10^{-5}$ & $   1.268\times 10^{-5}$ \\ \hline
 1850 & 100 & $   1.526\times 10^{-5}$ & $   1.826\times 10^{-5}$ & $   9.025\times 10^{-6}$ & $   9.586\times 10^{-6}$ \\ \hline
 1950 & 100 & $   1.128\times 10^{-5}$ & $   1.395\times 10^{-5}$ & $   6.625\times 10^{-6}$ & $   7.198\times 10^{-6}$ \\ \hline
    \end{tabular}
    \caption{Sum of $\nu_\tau$ and $\bar\nu_\tau$ energy distributions $d\sigma/dE_\nu$ from $D_s^\pm$ for $\sqrt{s}=14$ TeV, $\eta_\nu>6.9$. The predictions are shown for the PROSA PDF set with scales $(\mu_R,\mu_F)=(1,2)m_T$ with $\langle k_T\rangle = 1.2$ GeV, and the
    CT14, ABPM16 and NNPDF3.1 with  $(\mu_R,\mu_F)=(1,1)m_{T,2}$ and $\langle k_T\rangle = 0.7$ GeV.
    This table with 20 GeV energy bins for the whole energy range is available as an ancillary file on the \href{https://arxiv.org/src/2112.11605v1/anc}{arXiv link} to this paper.}
    \label{tab:6.9a}
\end{table}

\begin{table}[]
    \centering
    \begin{tabular}{|c|c|c|c|c|c|}
    \hline
     $E_\nu$ & Bin          & PROSA $m_{T,1}$ & CT14 $m_{T,2}$ & 
         ABMP16 $m_{T,2}$ & NNPDF3.1 $m_{T,2}$\\
     \hline\hline
  [GeV]& [GeV]       &  [$\mu$b/GeV] & [$\mu$b/GeV] & [$\mu$b/GeV]  & [$\mu$b/GeV]  \\
    \hline
    \hline
   10 &  20 & $   1.290\times 10^{-3}$ & $   8.994\times 10^{-4}$ & $   9.151\times 10^{-4}$ & $   7.042\times 10^{-4}$ \\ \hline
   30 &  20 & $   2.743\times 10^{-3}$ & $   1.915\times 10^{-3}$ & $   1.925\times 10^{-3}$ & $   1.487\times 10^{-3}$ \\ \hline
   50 &  20 & $   3.509\times 10^{-3}$ & $   2.432\times 10^{-3}$ & $   2.423\times 10^{-3}$ & $   1.852\times 10^{-3}$ \\ \hline
   70 &  20 & $   3.849\times 10^{-3}$ & $   2.738\times 10^{-3}$ & $   2.681\times 10^{-3}$ & $   2.023\times 10^{-3}$ \\ \hline
   90 &  20 & $   3.897\times 10^{-3}$ & $   2.900\times 10^{-3}$ & $   2.759\times 10^{-3}$ & $   2.058\times 10^{-3}$ \\ \hline
  110 &  20 & $   3.724\times 10^{-3}$ & $   2.921\times 10^{-3}$ & $   2.693\times 10^{-3}$ & $   1.991\times 10^{-3}$ \\ \hline
  130 &  20 & $   3.429\times 10^{-3}$ & $   2.802\times 10^{-3}$ & $   2.497\times 10^{-3}$ & $   1.856\times 10^{-3}$ \\ \hline
  150 &  20 & $   3.088\times 10^{-3}$ & $   2.571\times 10^{-3}$ & $   2.230\times 10^{-3}$ & $   1.671\times 10^{-3}$ \\ \hline
  170 &  20 & $   2.728\times 10^{-3}$ & $   2.308\times 10^{-3}$ & $   1.955\times 10^{-3}$ & $   1.475\times 10^{-3}$ \\ \hline
  190 &  20 & $   2.418\times 10^{-3}$ & $   2.041\times 10^{-3}$ & $   1.698\times 10^{-3}$ & $   1.299\times 10^{-3}$ \\ \hline
  220 &  40 & $   2.009\times 10^{-3}$ & $   1.691\times 10^{-3}$ & $   1.381\times 10^{-3}$ & $   1.071\times 10^{-3}$ \\ \hline
  260 &  40 & $   1.589\times 10^{-3}$ & $   1.329\times 10^{-3}$ & $   1.072\times 10^{-3}$ & $   8.424\times 10^{-4}$ \\ \hline
  300 &  40 & $   1.288\times 10^{-3}$ & $   1.074\times 10^{-3}$ & $   8.633\times 10^{-4}$ & $   6.823\times 10^{-4}$ \\ \hline
  340 &  40 & $   1.061\times 10^{-3}$ & $   8.952\times 10^{-4}$ & $   7.272\times 10^{-4}$ & $   5.717\times 10^{-4}$ \\ \hline
  380 &  40 & $   9.234\times 10^{-4}$ & $   7.681\times 10^{-4}$ & $   6.303\times 10^{-4}$ & $   4.922\times 10^{-4}$ \\ \hline
  420 &  40 & $   8.041\times 10^{-4}$ & $   6.745\times 10^{-4}$ & $   5.577\times 10^{-4}$ & $   4.305\times 10^{-4}$ \\ \hline
  460 &  40 & $   7.097\times 10^{-4}$ & $   6.027\times 10^{-4}$ & $   4.963\times 10^{-4}$ & $   3.821\times 10^{-4}$ \\ \hline
  500 &  40 & $   6.331\times 10^{-4}$ & $   5.441\times 10^{-4}$ & $   4.471\times 10^{-4}$ & $   3.414\times 10^{-4}$ \\ \hline
  540 &  40 & $   5.647\times 10^{-4}$ & $   4.937\times 10^{-4}$ & $   4.000\times 10^{-4}$ & $   3.050\times 10^{-4}$ \\ \hline
  580 &  40 & $   5.088\times 10^{-4}$ & $   4.467\times 10^{-4}$ & $   3.603\times 10^{-4}$ & $   2.744\times 10^{-4}$ \\ \hline
  650 & 100 & $   4.131\times 10^{-4}$ & $   3.773\times 10^{-4}$ & $   2.970\times 10^{-4}$ & $   2.277\times 10^{-4}$ \\ \hline
  750 & 100 & $   3.160\times 10^{-4}$ & $   2.929\times 10^{-4}$ & $   2.213\times 10^{-4}$ & $   1.721\times 10^{-4}$ \\ \hline
  850 & 100 & $   2.402\times 10^{-4}$ & $   2.246\times 10^{-4}$ & $   1.625\times 10^{-4}$ & $   1.300\times 10^{-4}$ \\ \hline
  950 & 100 & $   1.799\times 10^{-4}$ & $   1.703\times 10^{-4}$ & $   1.181\times 10^{-4}$ & $   9.711\times 10^{-5}$ \\ \hline
 1050 & 100 & $   1.344\times 10^{-4}$ & $   1.285\times 10^{-4}$ & $   8.575\times 10^{-5}$ & $   7.224\times 10^{-5}$ \\ \hline
 1150 & 100 & $   9.965\times 10^{-5}$ & $   9.618\times 10^{-5}$ & $   6.149\times 10^{-5}$ & $   5.379\times 10^{-5}$ \\ \hline
 1250 & 100 & $   7.334\times 10^{-5}$ & $   7.188\times 10^{-5}$ & $   4.425\times 10^{-5}$ & $   3.977\times 10^{-5}$ \\ \hline
 1350 & 100 & $   5.420\times 10^{-5}$ & $   5.361\times 10^{-5}$ & $   3.186\times 10^{-5}$ & $   2.950\times 10^{-5}$ \\ \hline
 1450 & 100 & $   4.008\times 10^{-5}$ & $   3.988\times 10^{-5}$ & $   2.286\times 10^{-5}$ & $   2.186\times 10^{-5}$ \\ \hline
 1550 & 100 & $   2.942\times 10^{-5}$ & $   2.967\times 10^{-5}$ & $   1.643\times 10^{-5}$ & $   1.616\times 10^{-5}$ \\ \hline
 1650 & 100 & $   2.168\times 10^{-5}$ & $   2.189\times 10^{-5}$ & $   1.177\times 10^{-5}$ & $   1.201\times 10^{-5}$ \\ \hline
 1750 & 100 & $   1.602\times 10^{-5}$ & $   1.615\times 10^{-5}$ & $   8.407\times 10^{-6}$ & $   8.870\times 10^{-6}$ \\ \hline
 1850 & 100 & $   1.165\times 10^{-5}$ & $   1.187\times 10^{-5}$ & $   6.021\times 10^{-6}$ & $   6.521\times 10^{-6}$ \\ \hline
 1950 & 100 & $   8.572\times 10^{-6}$ & $   8.761\times 10^{-6}$ & $   4.287\times 10^{-6}$ & $   4.779\times 10^{-6}$ \\ \hline
    \end{tabular}
    \caption{Sum of $\nu_\tau$ and $\bar\nu_\tau$ energy distributions $d\sigma/dE_\nu$ from $D_s^\pm$ for $\sqrt{s}=14$ TeV, $7.2<\eta_\nu<8.6$. The predictions are shown for the PROSA NLO PDF set with scales $(\mu_R,\mu_F)=(1,2)m_T$ with $\langle k_T\rangle = 1.2$ GeV, and the
    CT14, ABPM16 and NNPDF3.1 NLO PDFs with  $(\mu_R,\mu_F)=(1,1)m_{T,2}$ and $\langle k_T\rangle = 0.7$ GeV.
    This table with 20 GeV energy bins for the whole energy range is available as an ancillary file on the \href{https://arxiv.org/src/2112.11605v1/anc}{arXiv link} to this paper.}
    \label{tab:7.2a}
\end{table}

\begin{table}[]
    \centering
    \begin{tabular}{|c|c|c|c|c|c|}
    \hline
     $E_\nu$ & Bin          & PROSA $m_{T,1}$ & CT14 $m_{T,2}$ & 
         ABMP16 $m_{T,2}$ & NNPDF3.1 $m_{T,2}$\\
     \hline\hline
  [GeV]& [GeV]       &  [$\mu$b/GeV] & [$\mu$b/GeV] & [$\mu$b/GeV]  & [$\mu$b/GeV]  \\    
    \hline
    \hline
   10 &  20 & $   4.607\times 10^{-5}$ & $   3.137\times 10^{-5}$ & $   3.170\times 10^{-5}$ & $   2.561\times 10^{-5}$ \\ \hline
   30 &  20 & $   9.772\times 10^{-5}$ & $   6.840\times 10^{-5}$ & $   6.789\times 10^{-5}$ & $   5.242\times 10^{-5}$ \\ \hline
   50 &  20 & $   1.196\times 10^{-4}$ & $   8.448\times 10^{-5}$ & $   8.402\times 10^{-5}$ & $   6.401\times 10^{-5}$ \\ \hline
   70 &  20 & $   1.319\times 10^{-4}$ & $   9.141\times 10^{-5}$ & $   9.106\times 10^{-5}$ & $   6.869\times 10^{-5}$ \\ \hline
   90 &  20 & $   1.341\times 10^{-4}$ & $   9.598\times 10^{-5}$ & $   9.312\times 10^{-5}$ & $   6.956\times 10^{-5}$ \\ \hline
  110 &  20 & $   1.320\times 10^{-4}$ & $   9.661\times 10^{-5}$ & $   9.313\times 10^{-5}$ & $   6.872\times 10^{-5}$ \\ \hline
  130 &  20 & $   1.315\times 10^{-4}$ & $   9.472\times 10^{-5}$ & $   8.983\times 10^{-5}$ & $   6.726\times 10^{-5}$ \\ \hline
  150 &  20 & $   1.244\times 10^{-4}$ & $   9.548\times 10^{-5}$ & $   8.751\times 10^{-5}$ & $   6.390\times 10^{-5}$ \\ \hline
  170 &  20 & $   1.159\times 10^{-4}$ & $   9.342\times 10^{-5}$ & $   8.253\times 10^{-5}$ & $   6.105\times 10^{-5}$ \\ \hline
  190 &  20 & $   1.096\times 10^{-4}$ & $   9.147\times 10^{-5}$ & $   7.801\times 10^{-5}$ & $   5.804\times 10^{-5}$ \\ \hline
  220 &  40 & $   9.687\times 10^{-5}$ & $   8.613\times 10^{-5}$ & $   6.987\times 10^{-5}$ & $   5.333\times 10^{-5}$ \\ \hline
  260 &  40 & $   8.190\times 10^{-5}$ & $   7.876\times 10^{-5}$ & $   6.109\times 10^{-5}$ & $   4.754\times 10^{-5}$ \\ \hline
  300 &  40 & $   6.808\times 10^{-5}$ & $   7.188\times 10^{-5}$ & $   5.223\times 10^{-5}$ & $   4.135\times 10^{-5}$ \\ \hline
  340 &  40 & $   5.809\times 10^{-5}$ & $   6.359\times 10^{-5}$ & $   4.454\times 10^{-5}$ & $   3.599\times 10^{-5}$ \\ \hline
  380 &  40 & $   4.879\times 10^{-5}$ & $   5.542\times 10^{-5}$ & $   3.774\times 10^{-5}$ & $   3.122\times 10^{-5}$ \\ \hline
  420 &  40 & $   4.117\times 10^{-5}$ & $   4.823\times 10^{-5}$ & $   3.225\times 10^{-5}$ & $   2.698\times 10^{-5}$ \\ \hline
  460 &  40 & $   3.551\times 10^{-5}$ & $   4.162\times 10^{-5}$ & $   2.794\times 10^{-5}$ & $   2.333\times 10^{-5}$ \\ \hline
  500 &  40 & $   3.097\times 10^{-5}$ & $   3.588\times 10^{-5}$ & $   2.456\times 10^{-5}$ & $   2.025\times 10^{-5}$ \\ \hline
  540 &  40 & $   2.795\times 10^{-5}$ & $   3.117\times 10^{-5}$ & $   2.109\times 10^{-5}$ & $   1.733\times 10^{-5}$ \\ \hline
  580 &  40 & $   2.545\times 10^{-5}$ & $   2.699\times 10^{-5}$ & $   1.919\times 10^{-5}$ & $   1.518\times 10^{-5}$ \\ \hline
  650 & 100 & $   2.120\times 10^{-5}$ & $   2.211\times 10^{-5}$ & $   1.600\times 10^{-5}$ & $   1.237\times 10^{-5}$ \\ \hline
  750 & 100 & $   1.698\times 10^{-5}$ & $   1.772\times 10^{-5}$ & $   1.301\times 10^{-5}$ & $   9.863\times 10^{-6}$ \\ \hline
  850 & 100 & $   1.435\times 10^{-5}$ & $   1.537\times 10^{-5}$ & $   1.078\times 10^{-5}$ & $   8.296\times 10^{-6}$ \\ \hline
  950 & 100 & $   1.202\times 10^{-5}$ & $   1.332\times 10^{-5}$ & $   9.104\times 10^{-6}$ & $   7.215\times 10^{-6}$ \\ \hline
 1050 & 100 & $   9.962\times 10^{-6}$ & $   1.195\times 10^{-5}$ & $   7.762\times 10^{-6}$ & $   6.308\times 10^{-6}$ \\ \hline
 1150 & 100 & $   7.825\times 10^{-6}$ & $   1.051\times 10^{-5}$ & $   6.419\times 10^{-6}$ & $   5.425\times 10^{-6}$ \\ \hline
 1250 & 100 & $   6.468\times 10^{-6}$ & $   9.124\times 10^{-6}$ & $   5.359\times 10^{-6}$ & $   4.595\times 10^{-6}$ \\ \hline
 1350 & 100 & $   5.454\times 10^{-6}$ & $   7.896\times 10^{-6}$ & $   4.488\times 10^{-6}$ & $   3.957\times 10^{-6}$ \\ \hline
 1450 & 100 & $   4.248\times 10^{-6}$ & $   6.864\times 10^{-6}$ & $   3.746\times 10^{-6}$ & $   3.357\times 10^{-6}$ \\ \hline
 1550 & 100 & $   3.418\times 10^{-6}$ & $   5.938\times 10^{-6}$ & $   3.085\times 10^{-6}$ & $   2.863\times 10^{-6}$ \\ \hline
 1650 & 100 & $   2.752\times 10^{-6}$ & $   5.087\times 10^{-6}$ & $   2.513\times 10^{-6}$ & $   2.360\times 10^{-6}$ \\ \hline
 1750 & 100 & $   2.190\times 10^{-6}$ & $   4.273\times 10^{-6}$ & $   2.068\times 10^{-6}$ & $   1.994\times 10^{-6}$ \\ \hline
 1850 & 100 & $   1.776\times 10^{-6}$ & $   3.617\times 10^{-6}$ & $   1.682\times 10^{-6}$ & $   1.651\times 10^{-6}$ \\ \hline
 1950 & 100 & $   1.378\times 10^{-6}$ & $   3.025\times 10^{-6}$ & $   1.334\times 10^{-6}$ & $   1.356\times 10^{-6}$ \\ \hline
    \end{tabular}
    \caption{Sum of $\nu_\tau$ and $\bar\nu_\tau$ energy distribution $d\sigma/dE_\nu$ from $D_s^\pm$ for $\sqrt{s}=14$ TeV, $\eta_\nu>8.9$. The results are shown for the PROSA NLO PDF set with scales $(\mu_R,\mu_F)=(1,2)m_T$ with $\langle k_T\rangle = 1.2$ GeV, and the
    CT14, ABPM16 and NNPDF3.1 NLO PDF sets with  $(\mu_R,\mu_F)=(1,1)m_{T,2}$ and $\langle k_T\rangle = 0.7$ GeV.
    This table with 20 GeV energy bins for the whole energy range is available as an ancillary file on the \href{https://arxiv.org/src/2112.11605v1/anc}{arXiv link} to this paper.}
    \label{tab:8.9a}
\end{table}

\vfil\break

\begin{table}[]
    \centering
    \begin{tabular}{|c|c|c|c|c|}
    \hline
         $E$     & $\sigma_{\rm CC}^{\nu_\tau A}/A$ & $\Delta_{{\nu}_\tau}/A$ &
      $\sigma_{\rm CC}^{\bar\nu_\tau A}/A$  & $\Delta_{\bar{\nu}_\tau}/A$ \\ \hline\hline
      [GeV] & [$10^{-38}$ cm$^2$]  & [$10^{-38}$ cm$^2$]  & [$10^{-38}$ cm$^2$] & [$10^{-38}$ cm$^2$] \\
     \hline\hline
      5 &$   8.076 \times 10^{-2} $ & $   3.987 \times 10^{-3} $ & $   2.581 \times 10^{-2} $ & $   2.642 \times 10^{-3} $ \\ \hline
     6 &$   2.182 \times 10^{-1} $ & $   1.005 \times 10^{-2} $ & $   7.440 \times 10^{-2} $ & $   6.681 \times 10^{-3} $ \\ \hline
     7 &$   4.044 \times 10^{-1} $ & $   1.759 \times 10^{-2} $ & $   1.435 \times 10^{-1} $ & $   1.187 \times 10^{-2} $ \\ \hline
     8 &$   6.339 \times 10^{-1} $ & $   2.658 \times 10^{-2} $ & $   2.322 \times 10^{-1} $ & $   1.804 \times 10^{-2} $ \\ \hline
     9 &$   9.028 \times 10^{-1} $ & $   3.671 \times 10^{-2} $ & $   3.362 \times 10^{-1} $ & $   2.486 \times 10^{-2} $ \\ \hline
    10 &$   1.204  $ & $   4.764 \times 10^{-2} $ & $   4.569 \times 10^{-1} $ & $   3.231 \times 10^{-2} $ \\ \hline
    11 &$   1.536  $ & $   5.978 \times 10^{-2} $ & $   5.904 \times 10^{-1} $ & $   3.993 \times 10^{-2} $ \\ \hline
    12 &$   1.893  $ & $   7.212 \times 10^{-2} $ & $   7.337 \times 10^{-1} $ & $   4.774 \times 10^{-2} $ \\ \hline
    13 &$   2.270  $ & $   8.488 \times 10^{-2} $ & $   8.873 \times 10^{-1} $ & $   5.589 \times 10^{-2} $ \\ \hline
    14 &$   2.672  $ & $   9.830 \times 10^{-2} $ & $   1.049  $ & $   6.411 \times 10^{-2} $ \\ \hline
    15 &$   3.091  $ & $   1.120 \times 10^{-1} $ & $   1.220  $ & $   7.256 \times 10^{-2} $ \\ \hline
    16 &$   3.514  $ & $   1.259 \times 10^{-1} $ & $   1.393  $ & $   8.077 \times 10^{-2} $ \\ \hline
    17 &$   3.963  $ & $   1.402 \times 10^{-1} $ & $   1.577  $ & $   8.931 \times 10^{-2} $ \\ \hline
    18 &$   4.427  $ & $   1.549 \times 10^{-1} $ & $   1.768  $ & $   9.803 \times 10^{-2} $ \\ \hline
    19 &$   4.891  $ & $   1.692 \times 10^{-1} $ & $   1.958  $ & $   1.059 \times 10^{-1} $ \\ \hline
    20 &$   5.374  $ & $   1.842 \times 10^{-1} $ & $   2.155  $ & $   1.140 \times 10^{-1} $ \\ \hline
    30 &$   1.053 \times 10^{1} $ & $   3.339 \times 10^{-1} $ & $   4.306  $ & $   1.971 \times 10^{-1} $ \\ \hline
    40 &$   1.616 \times 10^{1} $ & $   4.881 \times 10^{-1} $ & $   6.691  $ & $   2.797 \times 10^{-1} $ \\ \hline
    50 &$   2.198 \times 10^{1} $ & $   6.391 \times 10^{-1} $ & $   9.178  $ & $   3.600 \times 10^{-1} $ \\ \hline
    60 &$   2.797 \times 10^{1} $ & $   7.883 \times 10^{-1} $ & $   1.176 \times 10^{1} $ & $   4.404 \times 10^{-1} $ \\ \hline
    70 &$   3.400 \times 10^{1} $ & $   9.359 \times 10^{-1} $ & $   1.439 \times 10^{1} $ & $   5.159 \times 10^{-1} $ \\ \hline
    80 &$   4.017 \times 10^{1} $ & $   1.077  $ & $   1.709 \times 10^{1} $ & $   5.916 \times 10^{-1} $ \\ \hline
    90 &$   4.628 \times 10^{1} $ & $   1.221  $ & $   1.984 \times 10^{1} $ & $   6.682 \times 10^{-1} $ \\ \hline
   100 &$   5.253 \times 10^{1} $ & $   1.358  $ & $   2.265 \times 10^{1} $ & $   7.392 \times 10^{-1} $ \\ \hline
   150 &$   8.386 \times 10^{1} $ & $   2.033  $ & $   3.690 \times 10^{1} $ & $   1.110  $ \\ \hline
   200 &$   1.157 \times 10^{2} $ & $   2.685  $ & $   5.169 \times 10^{1} $ & $   1.507  $ \\ \hline
   250 &$   1.471 \times 10^{2} $ & $   3.332  $ & $   6.644 \times 10^{1} $ & $   1.882  $ \\ \hline
   300 &$   1.783 \times 10^{2} $ & $   3.962  $ & $   8.141 \times 10^{1} $ & $   2.266  $ \\ \hline
   400 &$   2.406 \times 10^{2} $ & $   5.178  $ & $   1.113 \times 10^{2} $ & $   3.011  $ \\ \hline
   500 &$   3.026 \times 10^{2} $ & $   6.386  $ & $   1.419 \times 10^{2} $ & $   3.763  $ \\ \hline
   750 &$   4.552 \times 10^{2} $ & $   9.215  $ & $   2.188 \times 10^{2} $ & $   5.472  $ \\ \hline
  1000 &$   6.035 \times 10^{2} $ & $   1.193 \times 10^{1} $ & $   2.940 \times 10^{2} $ & $   7.194  $ \\ \hline
  2000 &$   1.162 \times 10^{3} $ & $   2.159 \times 10^{1} $ & $   5.923 \times 10^{2} $ & $   1.353 \times 10^{1} $ \\ \hline
\end{tabular}
    \caption{The $\nu_\tau$ and $\bar\nu_\tau$ charged current cross section per nucleon  for interactions with tungsten, evaluated using the nCTEQ PDFs. The PDF error is also shown, where $(\sigma^{\rm CC}\pm \Delta)/A$ represents the PDF error band for a tungsten target. Tables for incident tau neutrino and antineutrino energies between 5 GeV and 5 TeV are available as ancillary files on the \href{https://arxiv.org/src/2112.11605v1/anc}{arXiv link} to this paper.}
    \label{tab:sigcc}
\end{table}

\vfil\break

\begin{table}[]
    \centering
    \begin{tabular}{|c|c|c|}
    \hline
         $E$     & $\sigma_{\rm CC}^{\nu_\tau A}/A$ &
      $\sigma_{\rm CC}^{\bar\nu_\tau A}/A$   \\ \hline\hline
      [GeV] & [$10^{-38}$ cm$^2$]  & [$10^{-38}$ cm$^2$] \\
     \hline\hline
     5 &$   9.452 \times   10^{-2}  $ & $   2.709 \times   10^{-2}  $ \\ \hline
     6 &$   2.491 \times   10^{-1}  $ & $   7.651 \times   10^{-2}  $ \\ \hline
     7 &$   4.542 \times   10^{-1}  $ & $   1.457 \times   10^{-1}  $ \\ \hline
     8 &$   7.027 \times   10^{-1}  $ & $   2.337 \times   10^{-1}  $ \\ \hline
     9 &$   9.908 \times   10^{-1}  $ & $   3.363 \times   10^{-1}  $ \\ \hline
    10 &$   1.309    $ & $   4.540 \times   10^{-1}  $ \\ \hline
    11 &$   1.659    $ & $   5.840 \times   10^{-1}  $ \\ \hline
    12 &$   2.032    $ & $   7.218 \times   10^{-1}  $ \\ \hline
    13 &$   2.421    $ & $   8.688 \times   10^{-1}  $ \\ \hline
    14 &$   2.836    $ & $   1.023    $ \\ \hline
    15 &$   3.263    $ & $   1.184    $ \\ \hline
    16 &$   3.694    $ & $   1.347    $ \\ \hline
    17 &$   4.146    $ & $   1.518    $ \\ \hline
    18 &$   4.615    $ & $   1.696    $ \\ \hline
    19 &$   5.082    $ & $   1.872    $ \\ \hline
    20 &$   5.566    $ & $   2.053    $ \\ \hline
    30 &$   1.066 \times   10^{1}  $ & $   3.998    $ \\ \hline
    40 &$   1.614 \times   10^{1}  $ & $   6.120    $ \\ \hline
    50 &$   2.177 \times   10^{1}  $ & $   8.316    $ \\ \hline
    60 &$   2.757 \times   10^{1}  $ & $   1.059 \times   10^{1}  $ \\ \hline
    70 &$   3.337 \times   10^{1}  $ & $   1.290 \times   10^{1}  $ \\ \hline
    80 &$   3.933 \times   10^{1}  $ & $   1.529 \times   10^{1}  $ \\ \hline
    90 &$   4.523 \times   10^{1}  $ & $   1.772 \times   10^{1}  $ \\ \hline
   100 &$   5.125 \times   10^{1}  $ & $   2.020 \times   10^{1}  $ \\ \hline
   150 &$   8.145 \times   10^{1}  $ & $   3.287 \times   10^{1}  $ \\ \hline
   200 &$   1.122 \times   10^{2}  $ & $   4.609 \times   10^{1}  $ \\ \hline
   250 &$   1.425 \times   10^{2}  $ & $   5.935 \times   10^{1}  $ \\ \hline
   300 &$   1.727 \times   10^{2}  $ & $   7.288 \times   10^{1}  $ \\ \hline
   400 &$   2.331 \times   10^{2}  $ & $   9.998 \times   10^{1}  $ \\ \hline
   500 &$   2.933 \times   10^{2}  $ & $   1.280 \times   10^{2}  $ \\ \hline
   750 &$   4.419 \times   10^{2}  $ & $   1.992 \times   10^{2}  $ \\ \hline
  1000 &$   5.860 \times   10^{2}  $ & $   2.687 \times   10^{2}  $ \\ \hline
  2000 &$   1.131 \times   10^{3}  $ & $   5.484 \times   10^{2}  $ \\ \hline
\end{tabular}
    \caption{The $\nu_\tau$ and $\bar\nu_\tau$ charged current cross section per nucleon  for interactions with tungsten, evaluated using the PROSA VFNS PDFs. Isospin symmetry is used for the neutron PDFs. Tables for incident tau neutrino and antineutrino energies between 5 GeV and 5 TeV are available as ancillary files on the \href{https://arxiv.org/src/2112.11605v1/anc}{arXiv link} to this paper.}
    \label{tab:sigcc-prosa}
\end{table}

\bibliography{LHC-PDF}

\providecommand{\href}[2]{#2}\begingroup\raggedright\begin{thebibliography}{10}

\bibitem{DeRujula:1984pg}
A.~De~Rujula and R.~Ruckl, \emph{{Neutrino and muon physics in the collider
  mode of future accelerators}},  in \emph{{ECFA-CERN Workshop on large hadron
  collider in the LEP tunnel, Lausanne and CERN, Geneva, Switzerland, 21-27 Mar
  1984: Proceedings. 1.}}, pp.~571--596, 1984.
\newblock \href{http://dx.doi.org/10.5170/CERN-1984-010-V-2.571}{DOI}.

\bibitem{Winter:1990ry}
K.~Winter, \emph{{Detection of the tau-neutrino at the LHC}},  in \emph{{ECFA
  Large Hadron Collider Workshop, Aachen, Germany, 4-9 Oct 1990:
  Proceedings.2.}}, pp.~37--49, 1990.

\bibitem{DeRujula:1992sn}
A.~De~Rujula, E.~Fernandez and J.~J. Gomez-Cadenas, \emph{{Neutrino fluxes at
  future hadron colliders}},
  \href{http://dx.doi.org/10.1016/0550-3213(93)90427-Q}{\emph{Nucl. Phys.} {\bf
  B405} (1993) 80--108}.

\bibitem{Vannucci:1993ud}
F.~Vannucci, \emph{{Neutrino physics at LHC / SSC}},  in \emph{{5th
  International Symposium on Neutrino Telescopes Venice, Italy, March 2-4,
  1993}}, pp.~57--68, 1993.

\bibitem{Abreu:2019yak}
{\scshape FASER} collaboration, H.~Abreu et~al., \emph{{Detecting and Studying
  High-Energy Collider Neutrinos with FASER at the LHC}},
  \href{http://arxiv.org/abs/1908.02310}{{\tt 1908.02310}}.

\bibitem{Abreu:2020ddv}
{\scshape FASER} collaboration, H.~Abreu et~al., \emph{{Technical Proposal:
  FASERnu}},  \href{http://arxiv.org/abs/2001.03073}{{\tt 2001.03073}}.

\bibitem{Buontempo:2018gta}
S.~Buontempo, G.~M. Dallavalle, G.~De~Lellis, D.~Lazic and F.~L. Navarria,
  \emph{{CMS-XSEN: LHC Neutrinos at CMS. Experiment Feasibility Study}},
  \href{http://arxiv.org/abs/1804.04413}{{\tt 1804.04413}}.

\bibitem{Beni:2019pyp}
{\scshape XSEN} collaboration, N.~Beni et~al., \emph{{XSEN: a $\nu$N Cross
  Section Measurement using High Energy Neutrinos from pp collisions at the
  LHC}},  \href{http://arxiv.org/abs/1910.11340}{{\tt 1910.11340}}.

\bibitem{Ahdida:2020evc}
{\scshape SHiP} collaboration, C.~Ahdida et~al., \emph{{SND@LHC}},
  \href{http://arxiv.org/abs/2002.08722}{{\tt 2002.08722}}.

\bibitem{FASER:2021mtu}
{\scshape FASER} collaboration, H.~Abreu et~al., \emph{{First neutrino
  interaction candidates at the LHC}},
  \href{http://dx.doi.org/10.1103/PhysRevD.104.L091101}{\emph{Phys. Rev. D}
  {\bf 104} (2021) L091101}, [\href{http://arxiv.org/abs/2105.06197}{{\tt
  2105.06197}}].

\bibitem{Anchordoqui:2021ghd}
L.~A. Anchordoqui et~al., \emph{{The Forward Physics Facility: Sites,
  Experiments, and Physics Potential}},
  \href{http://arxiv.org/abs/2109.10905}{{\tt 2109.10905}}.

\bibitem{Feng:2022inv}
J.~L. Feng et~al., \emph{{The Forward Physics Facility at the High-Luminosity
  LHC}},  in \emph{{2022 Snowmass Summer Study}}, 3, 2022.
\newblock \href{http://arxiv.org/abs/2203.05090}{{\tt 2203.05090}}.

\bibitem{E949:2005efl}
{\scshape E949} collaboration, A.~V. Artamonov et~al., \emph{{Upper Limit on
  the Branching Ratio for the Decay $\pi^0 \to \nu \overline {\nu}$}},
  \href{http://dx.doi.org/10.1103/PhysRevD.72.091102}{\emph{Phys. Rev. D} {\bf
  72} (2005) 091102}, [\href{http://arxiv.org/abs/hep-ex/0506028}{{\tt
  hep-ex/0506028}}].

\bibitem{NA62:2021zjw}
{\scshape NA62} collaboration, E.~Cortina~Gil et~al., \emph{{Measurement of the
  very rare K$^{+}$\textrightarrow{}$ {\pi}^{+}\nu \overline{\nu} $ decay}},
  \href{http://dx.doi.org/10.1007/JHEP06(2021)093}{\emph{JHEP} {\bf 06} (2021)
  093}, [\href{http://arxiv.org/abs/2103.15389}{{\tt 2103.15389}}].

\bibitem{Bai:2020ukz}
W.~Bai, M.~Diwan, M.~V. Garzelli, Y.~S. Jeong and M.~H. Reno,
  \emph{{Far-forward neutrinos at the Large Hadron Collider}},
  \href{http://dx.doi.org/10.1007/JHEP06(2020)032}{\emph{JHEP} {\bf 06} (2020)
  032}, [\href{http://arxiv.org/abs/2002.03012}{{\tt 2002.03012}}].

\bibitem{Park:2011gh}
H.~Park, \emph{{The estimation of neutrino fluxes produced by proton-proton
  collisions at $\sqrt{s}=14$ TeV of the LHC}},
  \href{http://dx.doi.org/10.1007/JHEP10(2011)092}{\emph{JHEP} {\bf 10} (2011)
  092}, [\href{http://arxiv.org/abs/1110.1971}{{\tt 1110.1971}}].

\bibitem{Beni:2020yfy}
N.~Beni et~al., \emph{{Further studies on the physics potential of an
  experiment using LHC neutrinos}},
  \href{http://dx.doi.org/10.1088/1361-6471/aba7ad}{\emph{J. Phys. G} {\bf 47}
  (2020) 125004}, [\href{http://arxiv.org/abs/2004.07828}{{\tt 2004.07828}}].

\bibitem{Aaij:2016jht}
{\scshape LHCb} collaboration, R.~Aaij et~al., \emph{{Measurements of prompt
  charm production cross-sections in pp collisions at $ \sqrt{s}=5 $ TeV}},
  \href{http://dx.doi.org/10.1007/JHEP06(2017)147}{\emph{JHEP} {\bf 06} (2017)
  147}, [\href{http://arxiv.org/abs/1610.02230}{{\tt 1610.02230}}].

\bibitem{LHCb:2013xam}
{\scshape LHCb} collaboration, R.~Aaij et~al., \emph{{Prompt charm production
  in pp collisions at sqrt(s)=7 TeV}},
  \href{http://dx.doi.org/10.1016/j.nuclphysb.2013.02.010}{\emph{Nucl. Phys. B}
  {\bf 871} (2013) 1--20}, [\href{http://arxiv.org/abs/1302.2864}{{\tt
  1302.2864}}].

\bibitem{Aaij:2015bpa}
{\scshape LHCb} collaboration, R.~Aaij et~al., \emph{{Measurements of prompt
  charm production cross-sections in $pp$ collisions at $ \sqrt{s}=13 $ TeV}},
  \href{http://dx.doi.org/10.1007/JHEP03(2016)159}{\emph{JHEP} {\bf 03} (2016)
  159}, [\href{http://arxiv.org/abs/1510.01707}{{\tt 1510.01707}}].

\bibitem{Nason:1987xz}
P.~Nason, S.~Dawson and R.~K. Ellis, \emph{{The Total Cross-Section for the
  Production of Heavy Quarks in Hadronic Collisions}},
  \href{http://dx.doi.org/10.1016/0550-3213(88)90422-1}{\emph{Nucl. Phys.} {\bf
  B303} (1988) 607--633}.

\bibitem{Nason:1989zy}
P.~Nason, S.~Dawson and R.~K. Ellis, \emph{{The One Particle Inclusive
  Differential Cross-Section for Heavy Quark Production in Hadronic
  Collisions}}, \href{http://dx.doi.org/10.1016/0550-3213(90)90180-L,
  10.1016/0550-3213(89)90286-1}{\emph{Nucl. Phys.} {\bf B327} (1989) 49--92}.

\bibitem{Mangano:1991jk}
M.~L. Mangano, P.~Nason and G.~Ridolfi, \emph{{Heavy quark correlations in
  hadron collisions at next-to-leading order}},
  \href{http://dx.doi.org/10.1016/0550-3213(92)90435-E}{\emph{Nucl. Phys.} {\bf
  B373} (1992) 295--345}.

\bibitem{H1:2015ubc}
{\scshape H1, ZEUS} collaboration, H.~Abramowicz et~al., \emph{{Combination of
  measurements of inclusive deep inelastic ${e^{\pm }p}$ scattering cross
  sections and QCD analysis of HERA data}},
  \href{http://dx.doi.org/10.1140/epjc/s10052-015-3710-4}{\emph{Eur. Phys. J.
  C} {\bf 75} (2015) 580}, [\href{http://arxiv.org/abs/1506.06042}{{\tt
  1506.06042}}].

\bibitem{H1:2018flt}
{\scshape H1, ZEUS} collaboration, H.~Abramowicz et~al., \emph{{Combination and
  QCD analysis of charm and beauty production cross-section measurements in
  deep inelastic $ep$ scattering at HERA}},
  \href{http://dx.doi.org/10.1140/epjc/s10052-018-5848-3}{\emph{Eur. Phys. J.
  C} {\bf 78} (2018) 473}, [\href{http://arxiv.org/abs/1804.01019}{{\tt
  1804.01019}}].

\bibitem{Zenaiev:2019ktw}
{\scshape PROSA} collaboration, O.~Zenaiev, M.~Garzelli, K.~Lipka, S.~Moch,
  A.~Cooper-Sarkar, F.~Olness et~al., \emph{{Improved constraints on parton
  distributions using LHCb, ALICE and HERA heavy-flavour measurements and
  implications for the predictions for prompt atmospheric-neutrino fluxes}},
  \href{http://dx.doi.org/10.1007/JHEP04(2020)118}{\emph{JHEP} {\bf 04} (2020)
  118}, [\href{http://arxiv.org/abs/1911.13164}{{\tt 1911.13164}}].

\bibitem{PROSA:2015yid}
{\scshape PROSA} collaboration, O.~Zenaiev et~al., \emph{{Impact of
  heavy-flavour production cross sections measured by the LHCb experiment on
  parton distribution functions at low x}},
  \href{http://dx.doi.org/10.1140/epjc/s10052-015-3618-z}{\emph{Eur. Phys. J.
  C} {\bf 75} (2015) 396}, [\href{http://arxiv.org/abs/1503.04581}{{\tt
  1503.04581}}].

\bibitem{Gauld:2015kvh}
R.~Gauld, J.~Rojo, L.~Rottoli, S.~Sarkar and J.~Talbert, \emph{{The prompt
  atmospheric neutrino flux in the light of LHCb}},
  \href{http://dx.doi.org/10.1007/JHEP02(2016)130}{\emph{JHEP} {\bf 02} (2016)
  130}, [\href{http://arxiv.org/abs/1511.06346}{{\tt 1511.06346}}].

\bibitem{Gauld:2016kpd}
R.~Gauld and J.~Rojo, \emph{{Precision determination of the small-$x$ gluon
  from charm production at LHCb}},
  \href{http://dx.doi.org/10.1103/PhysRevLett.118.072001}{\emph{Phys. Rev.
  Lett.} {\bf 118} (2017) 072001}, [\href{http://arxiv.org/abs/1610.09373}{{\tt
  1610.09373}}].

\bibitem{Bertone:2018dse}
V.~Bertone, R.~Gauld and J.~Rojo, \emph{{Neutrino Telescopes as QCD
  Microscopes}}, \href{http://dx.doi.org/10.1007/JHEP01(2019)217}{\emph{JHEP}
  {\bf 01} (2019) 217}, [\href{http://arxiv.org/abs/1808.02034}{{\tt
  1808.02034}}].

\bibitem{NNPDF:2014otw}
{\scshape NNPDF} collaboration, R.~D. Ball et~al., \emph{{Parton distributions
  for the LHC Run II}},
  \href{http://dx.doi.org/10.1007/JHEP04(2015)040}{\emph{JHEP} {\bf 04} (2015)
  040}, [\href{http://arxiv.org/abs/1410.8849}{{\tt 1410.8849}}].

\bibitem{Ball:2017nwa}
{\scshape NNPDF} collaboration, R.~D. Ball et~al., \emph{{Parton distributions
  from high-precision collider data}},
  \href{http://dx.doi.org/10.1140/epjc/s10052-017-5199-5}{\emph{Eur. Phys. J.
  C} {\bf 77} (2017) 663}, [\href{http://arxiv.org/abs/1706.00428}{{\tt
  1706.00428}}].

\bibitem{Dulat:2015mca}
S.~Dulat, T.-J. Hou, J.~Gao, M.~Guzzi, J.~Huston, P.~Nadolsky et~al.,
  \emph{{New parton distribution functions from a global analysis of quantum
  chromodynamics}},
  \href{http://dx.doi.org/10.1103/PhysRevD.93.033006}{\emph{Phys. Rev. D} {\bf
  93} (2016) 033006}, [\href{http://arxiv.org/abs/1506.07443}{{\tt
  1506.07443}}].

\bibitem{Alekhin:2018pai}
S.~Alekhin, J.~Blümlein and S.~Moch, \emph{{NLO PDFs from the ABMP16 fit}},
  \href{http://dx.doi.org/10.1140/epjc/s10052-018-5947-1}{\emph{Eur. Phys. J.
  C} {\bf 78} (2018) 477}, [\href{http://arxiv.org/abs/1803.07537}{{\tt
  1803.07537}}].

\bibitem{Harland-Lang:2015qea}
L.~A. Harland-Lang, A.~D. Martin, P.~Motylinski and R.~S. Thorne, \emph{{Charm
  and beauty quark masses in the MMHT2014 global PDF analysis}},
  \href{http://dx.doi.org/10.1140/epjc/s10052-015-3843-5}{\emph{Eur. Phys. J.
  C} {\bf 76} (2016) 10}, [\href{http://arxiv.org/abs/1510.02332}{{\tt
  1510.02332}}].

\bibitem{Frixione:2007vw}
S.~Frixione, P.~Nason and C.~Oleari, \emph{{Matching NLO QCD computations with
  Parton Shower simulations: the POWHEG method}},
  \href{http://dx.doi.org/10.1088/1126-6708/2007/11/070}{\emph{JHEP} {\bf 11}
  (2007) 070}, [\href{http://arxiv.org/abs/0709.2092}{{\tt 0709.2092}}].

\bibitem{Sjostrand:2019zhc}
T.~Sj{\"o}strand, \emph{{The PYTHIA Event Generator: Past, Present and
  Future}}, \href{http://dx.doi.org/10.1016/j.cpc.2019.106910}{\emph{Comput.
  Phys. Commun.} {\bf 246} (2020) 106910},
  [\href{http://arxiv.org/abs/1907.09874}{{\tt 1907.09874}}].

\bibitem{Peterson:1982ak}
C.~Peterson, D.~Schlatter, I.~Schmitt and P.~M. Zerwas, \emph{{Scaling
  Violations in Inclusive $e^+ e^-$ Annihilation Spectra}},
  \href{http://dx.doi.org/10.1103/PhysRevD.27.105}{\emph{Phys. Rev.} {\bf D27}
  (1983) 105}.

\bibitem{Buckley:2014ana}
A.~Buckley, J.~Ferrando, S.~Lloyd, K.~Nordström, B.~Page, M.~Rüfenacht
  et~al., \emph{{LHAPDF6: parton density access in the LHC precision era}},
  \href{http://dx.doi.org/10.1140/epjc/s10052-015-3318-8}{\emph{Eur. Phys. J.
  C} {\bf 75} (2015) 132}, [\href{http://arxiv.org/abs/1412.7420}{{\tt
  1412.7420}}].

\bibitem{Benzke:2017yjn}
M.~Benzke, M.~V. Garzelli, B.~Kniehl, G.~Kramer, S.~Moch and G.~Sigl,
  \emph{{Prompt neutrinos from atmospheric charm in the general-mass
  variable-flavor-number scheme}},
  \href{http://dx.doi.org/10.1007/JHEP12(2017)021}{\emph{JHEP} {\bf 12} (2017)
  021}, [\href{http://arxiv.org/abs/1705.10386}{{\tt 1705.10386}}].

\bibitem{Accardi:2016ndt}
A.~Accardi et~al., \emph{{A Critical Appraisal and Evaluation of Modern PDFs}},
  \href{http://dx.doi.org/10.1140/epjc/s10052-016-4285-4}{\emph{Eur. Phys. J.
  C} {\bf 76} (2016) 471}, [\href{http://arxiv.org/abs/1603.08906}{{\tt
  1603.08906}}].

\bibitem{Chetyrkin:1999qi}
K.~Chetyrkin and M.~Steinhauser, \emph{{The Relation between the MS-bar and the
  on-shell quark mass at order alpha(s)**3}},
  \href{http://dx.doi.org/10.1016/S0550-3213(99)00784-1}{\emph{Nucl. Phys. B}
  {\bf 573} (2000) 617--651}, [\href{http://arxiv.org/abs/hep-ph/9911434}{{\tt
  hep-ph/9911434}}].

\bibitem{Marquard:2015qpa}
P.~Marquard, A.~V. Smirnov, V.~A. Smirnov and M.~Steinhauser, \emph{{Quark Mass
  Relations to Four-Loop Order in Perturbative QCD}},
  \href{http://dx.doi.org/10.1103/PhysRevLett.114.142002}{\emph{Phys. Rev.
  Lett.} {\bf 114} (2015) 142002}, [\href{http://arxiv.org/abs/1502.01030}{{\tt
  1502.01030}}].

\bibitem{Ball:2016qeg}
R.~D. Ball, \emph{{Charm Production: Pole Mass or Running Mass?}},
  \href{http://dx.doi.org/10.1063/1.4977120}{\emph{AIP Conf. Proc.} {\bf 1819}
  (2017) 030002}, [\href{http://arxiv.org/abs/1612.03790}{{\tt 1612.03790}}].

\bibitem{Cacciari:2012ny}
M.~Cacciari, S.~Frixione, N.~Houdeau, M.~L. Mangano, P.~Nason and G.~Ridolfi,
  \emph{{Theoretical predictions for charm and bottom production at the LHC}},
  \href{http://dx.doi.org/10.1007/JHEP10(2012)137}{\emph{JHEP} {\bf 10} (2012)
  137}, [\href{http://arxiv.org/abs/1205.6344}{{\tt 1205.6344}}].

\bibitem{Tanabashi:2018oca}
{\scshape Particle Data Group} collaboration, M.~Tanabashi et~al.,
  \emph{{Review of Particle Physics}},
  \href{http://dx.doi.org/10.1103/PhysRevD.98.030001}{\emph{Phys. Rev.} {\bf
  D98} (2018) 030001}.

\bibitem{Apanasevich:1998ki}
L.~Apanasevich et~al., \emph{{$k_{T}$ effects in direct photon production}},
  \href{http://dx.doi.org/10.1103/PhysRevD.59.074007}{\emph{Phys. Rev.} {\bf
  D59} (1999) 074007}, [\href{http://arxiv.org/abs/hep-ph/9808467}{{\tt
  hep-ph/9808467}}].

\bibitem{Miu:1998ju}
G.~Miu and T.~Sjostrand, \emph{{$W$ production in an improved parton shower
  approach}},
  \href{http://dx.doi.org/10.1016/S0370-2693(99)00068-4}{\emph{Phys. Lett.}
  {\bf B449} (1999) 313--320}, [\href{http://arxiv.org/abs/hep-ph/9812455}{{\tt
  hep-ph/9812455}}].

\bibitem{Balazs:2000sz}
C.~Balazs, J.~Huston and I.~Puljak, \emph{{Higgs production: A Comparison of
  parton showers and resummation}},
  \href{http://dx.doi.org/10.1103/PhysRevD.63.014021}{\emph{Phys.\ Rev.\ D}
  {\bf 63} (2001) 014021}, [\href{http://arxiv.org/abs/hep-ph/0002032}{{\tt
  hep-ph/0002032}}].

\bibitem{Amoroso:2022eow}
S.~Amoroso et~al., \emph{{Snowmass 2021 whitepaper: Proton structure at the
  precision frontier}},  \href{http://arxiv.org/abs/2203.13923}{{\tt
  2203.13923}}.

\bibitem{H1:2021xxi}
{\scshape H1, ZEUS} collaboration, I.~Abt et~al., \emph{{Impact of
  jet-production data on the next-to-next-to-leading-order determination of
  HERAPDF2.0 parton distributions}},
  \href{http://dx.doi.org/10.1140/epjc/s10052-022-10083-9}{\emph{Eur. Phys. J.
  C} {\bf 82} (2022) 243}, [\href{http://arxiv.org/abs/2112.01120}{{\tt
  2112.01120}}].

\bibitem{Czakon:2016olj}
M.~Czakon, N.~P. Hartland, A.~Mitov, E.~R. Nocera and J.~Rojo, \emph{{Pinning
  down the large-x gluon with NNLO top-quark pair differential distributions}},
  \href{http://dx.doi.org/10.1007/JHEP04(2017)044}{\emph{JHEP} {\bf 04} (2017)
  044}, [\href{http://arxiv.org/abs/1611.08609}{{\tt 1611.08609}}].

\bibitem{Bailey:2019yze}
S.~Bailey and L.~Harland-Lang, \emph{{Differential Top Quark Pair Production at
  the LHC: Challenges for PDF Fits}},
  \href{http://dx.doi.org/10.1140/epjc/s10052-020-7633-3}{\emph{Eur. Phys. J.
  C} {\bf 80} (2020) 60}, [\href{http://arxiv.org/abs/1909.10541}{{\tt
  1909.10541}}].

\bibitem{Czakon:2019yrx}
M.~Czakon, S.~Dulat, T.-J. Hou, J.~Huston, A.~Mitov, A.~S. Papanastasiou
  et~al., \emph{{An exploratory study of the impact of CMS double-differential
  top distributions on the gluon parton distribution function}},
  \href{http://dx.doi.org/10.1088/1361-6471/abb1b6}{\emph{J. Phys. G} {\bf 48}
  (2020) 015003}, [\href{http://arxiv.org/abs/1912.08801}{{\tt 1912.08801}}].

\bibitem{Kadir:2020yml}
M.~Kadir, A.~Ablat, S.~Dulat, T.-J. Hou and I.~Sitiwaldi, \emph{{The impact of
  ATLAS and CMS single differential top-quark pair measurements at $\sqrt {s}$
  = 8 TeV on CTEQ-TEA PDFs}},
  \href{http://dx.doi.org/10.1088/1674-1137/abce10}{\emph{Chin. Phys. C} {\bf
  45} (2021) 023111}, [\href{http://arxiv.org/abs/2003.13740}{{\tt
  2003.13740}}].

\bibitem{Guzzi:2022dis}
M.~Guzzi, A.~Ablat, C.-P. Yuan, I.~Sitiwaldi, K.~Xie, S.~Dulat et~al.,
  \emph{{Impact of Top-quark pair production on CTEQ PDFs}},  in
  \emph{DIS2022:XXIX International Workshop on Deep-Inelastic Scattering and
  Related Subjects}, 2022.

\bibitem{CMS:2017iqf}
{\scshape CMS} collaboration, A.~M. Sirunyan et~al., \emph{{Measurement of
  double-differential cross sections for top quark pair production in pp
  collisions at $\sqrt{s} = 8$ $\,\text {TeV}$ and impact on parton
  distribution functions}},
  \href{http://dx.doi.org/10.1140/epjc/s10052-017-4984-5}{\emph{Eur. Phys. J.
  C} {\bf 77} (2017) 459}, [\href{http://arxiv.org/abs/1703.01630}{{\tt
  1703.01630}}].

\bibitem{Reno:2022dis}
M.~H. Reno, W.~Bai, M.~V. Diwan, M.~V. Garzelli, Y.~S. Jeong and K.~Kumar,
  \emph{N{eutrinos from charm in the forward region}},  in \emph{DIS2022:XXIX
  International Workshop on Deep-Inelastic Scattering and Related Subjects},
  2022.

\bibitem{Kling:2021gos}
F.~Kling, \emph{{Forward Neutrino Fluxes at the LHC}},
  \href{http://arxiv.org/abs/2105.08270}{{\tt 2105.08270}}.

\bibitem{Kovarik:2015cma}
K.~Kovarik et~al., \emph{{nCTEQ15 - Global analysis of nuclear parton
  distributions with uncertainties in the CTEQ framework}},
  \href{http://dx.doi.org/10.1103/PhysRevD.93.085037}{\emph{Phys. Rev.} {\bf
  D93} (2016) 085037}, [\href{http://arxiv.org/abs/1509.00792}{{\tt
  1509.00792}}].

\bibitem{Kretzer:2002fr}
S.~Kretzer and M.~H. Reno, \emph{{Tau neutrino deep inelastic charged current
  interactions}},
  \href{http://dx.doi.org/10.1103/PhysRevD.66.113007}{\emph{Phys. Rev.} {\bf
  D66} (2002) 113007}, [\href{http://arxiv.org/abs/hep-ph/0208187}{{\tt
  hep-ph/0208187}}].

\bibitem{Kretzer:2004wk}
S.~Kretzer and M.~H. Reno, \emph{{sigma DIS (nu N), NLO perturbative QCD and
  O(1 GeV) mass corrections}},
  \href{http://dx.doi.org/10.1016/j.nuclphysbps.2004.11.237}{\emph{Nucl. Phys.
  Proc. Suppl.} {\bf 139} (2005) 134--139},
  [\href{http://arxiv.org/abs/hep-ph/0410184}{{\tt hep-ph/0410184}}].

\bibitem{Jeong:2010nt}
Y.~S. Jeong and M.~H. Reno, \emph{{Tau neutrino and antineutrino cross
  sections}}, \href{http://dx.doi.org/10.1103/PhysRevD.82.033010}{\emph{Phys.
  Rev.} {\bf D82} (2010) 033010}, [\href{http://arxiv.org/abs/1007.1966}{{\tt
  1007.1966}}].

\bibitem{Reno:2006hj}
M.~H. Reno, \emph{{Electromagnetic structure functions and neutrino nucleon
  scattering}}, \href{http://dx.doi.org/10.1103/PhysRevD.74.033001}{\emph{Phys.
  Rev.} {\bf D74} (2006) 033001},
  [\href{http://arxiv.org/abs/hep-ph/0605295}{{\tt hep-ph/0605295}}].

\bibitem{Capella:1994cr}
A.~Capella, A.~Kaidalov, C.~Merino and J.~Tran Thanh~Van, \emph{{Structure
  functions and low x physics}},
  \href{http://dx.doi.org/10.1016/0370-2693(94)90988-1}{\emph{Phys. Lett. B}
  {\bf 337} (1994) 358--366}, [\href{http://arxiv.org/abs/hep-ph/9405338}{{\tt
  hep-ph/9405338}}].

\bibitem{Bodek:2002ps}
A.~Bodek and U.~K. Yang, \emph{{Higher twist, xi(omega) scaling, and effective
  LO PDFs for lepton scattering in the few GeV region}},
  \href{http://dx.doi.org/10.1088/0954-3899/29/8/369}{\emph{J. Phys. G} {\bf
  29} (2003) 1899--1906}, [\href{http://arxiv.org/abs/hep-ex/0210024}{{\tt
  hep-ex/0210024}}].

\bibitem{Bodek:2002vp}
A.~Bodek and U.~Yang, \emph{{Modeling deep inelastic cross-sections in the few
  GeV region}},
  \href{http://dx.doi.org/10.1016/S0920-5632(02)01755-3}{\emph{Nucl.Phys.Proc.Suppl.}
  {\bf 112} (2002) 70--76}, [\href{http://arxiv.org/abs/hep-ex/0203009}{{\tt
  hep-ex/0203009}}].

\bibitem{Bodek:2004pc}
A.~Bodek, I.~Park and U.-k. Yang, \emph{{Improved low Q**2 model for neutrino
  and electron nucleon cross sections in few GeV region}},
  \href{http://dx.doi.org/10.1016/j.nuclphysbps.2004.11.208}{\emph{Nucl. Phys.
  B Proc. Suppl.} {\bf 139} (2005) 113--118},
  [\href{http://arxiv.org/abs/hep-ph/0411202}{{\tt hep-ph/0411202}}].

\bibitem{Bodek:2005de}
A.~Bodek and U.~Yang, \emph{A unified model for inelastic e - n and v - n cross
  sections at all q2}, \href{http://dx.doi.org/10.1063/1.2122031}{\emph{AIP
  Conference Proceedings} {\bf 792} (2005) 257--260}.

\bibitem{Bodek:2021bde}
A.~Bodek, U.~K. Yang and Y.~Xu, \emph{{Inelastic Axial and Vector Structure
  Functions for Lepton-Nucleon Scattering 2021 Update}},
  \href{http://arxiv.org/abs/2108.09240}{{\tt 2108.09240}}.

\bibitem{Jeong:2022tba}
Y.~S. Jeong and M.~H. Reno, ``Interface of shallow- and deep-inelastic
  neutrino-nucleon scattering at the forward physics facility.'' in
  preparation, 2022.

\bibitem{Reno:2021xx}
M.~H. Reno, \emph{{Evolution of the electroweak structure functions of
  nucleons}},
  \href{http://dx.doi.org/10.1140/epjs/s11734-021-00288-6}{\emph{Eur. Phys. J.
  Spec. Top.} (2021) }.

\bibitem{Ariga:2020pyj}
{\scshape Faser} collaboration, A.~Ariga, \emph{{Detecting and studying
  high-energy neutrinos with FASER$\nu$ at the LHC}},
  \href{http://dx.doi.org/10.22323/1.390.0112}{\emph{PoS} {\bf ICHEP2020}
  (2021) 112}.

\bibitem{Riehn:2015oba}
F.~Riehn, R.~Engel, A.~Fedynitch, T.~K. Gaisser and T.~Stanev, \emph{{A new
  version of the event generator Sibyll}},
  \href{http://dx.doi.org/10.22323/1.236.0558}{\emph{PoS} {\bf ICRC2015} (2016)
  558}, [\href{http://arxiv.org/abs/1510.00568}{{\tt 1510.00568}}].

\bibitem{Riehn:2017mfm}
F.~Riehn, H.~P. Dembinski, R.~Engel, A.~Fedynitch, T.~K. Gaisser and T.~Stanev,
  \emph{{The hadronic interaction model SIBYLL 2.3c and Feynman scaling}},
  \href{http://dx.doi.org/10.22323/1.301.0301}{\emph{PoS} {\bf ICRC2017} (2018)
  301}, [\href{http://arxiv.org/abs/1709.07227}{{\tt 1709.07227}}].

\bibitem{Fedynitch:2018cbl}
A.~Fedynitch, F.~Riehn, R.~Engel, T.~K. Gaisser and T.~Stanev, \emph{{Hadronic
  interaction model sibyll 2.3c and inclusive lepton fluxes}},
  \href{http://dx.doi.org/10.1103/PhysRevD.100.103018}{\emph{Phys. Rev.} {\bf
  D100} (2019) 103018}, [\href{http://arxiv.org/abs/1806.04140}{{\tt
  1806.04140}}].

\bibitem{Roesler:2000he}
S.~Roesler, R.~Engel and J.~Ranft, \emph{{The Monte Carlo event generator
  DPMJET-III}},  in \emph{{International Conference on Advanced Monte Carlo for
  Radiation Physics, Particle Transport Simulation and Applications (MC
  2000)}}, pp.~1033--1038, 12, 2000.
\newblock \href{http://arxiv.org/abs/hep-ph/0012252}{{\tt hep-ph/0012252}}.
\newblock \href{http://dx.doi.org/10.1007/978-3-642-18211-2_166}{DOI}.

\bibitem{Fedynitch:2015kcn}
A.~Fedynitch, \emph{{Cascade equations and hadronic interactions at very high
  energies}}.
\newblock PhD thesis, KIT, Karlsruhe, Dept. Phys., 11, 2015.
\newblock 10.5445/IR/1000055433.

\bibitem{Sjostrand:2014zea}
T.~Sj{\"o}strand, S.~Ask, J.~R. Christiansen, R.~Corke, N.~Desai, P.~Ilten
  et~al., \emph{{An Introduction to PYTHIA 8.2}},
  \href{http://dx.doi.org/10.1016/j.cpc.2015.01.024}{\emph{Comput. Phys.
  Commun.} {\bf 191} (2015) 159--177},
  [\href{http://arxiv.org/abs/1410.3012}{{\tt 1410.3012}}].

\bibitem{Garzelli:2016xmx}
{\scshape PROSA} collaboration, M.~V. Garzelli, S.~Moch, O.~Zenaiev,
  A.~Cooper-Sarkar, A.~Geiser, K.~Lipka et~al., \emph{{Prompt neutrino fluxes
  in the atmosphere with PROSA parton distribution functions}},
  \href{http://dx.doi.org/10.1007/JHEP05(2017)004}{\emph{JHEP} {\bf 05} (2017)
  004}, [\href{http://arxiv.org/abs/1611.03815}{{\tt 1611.03815}}].

\bibitem{AbdulKhalek:2021gbh}
R.~Abdul~Khalek et~al., \emph{{Science Requirements and Detector Concepts for
  the Electron-Ion Collider: EIC Yellow Report}},
  \href{http://arxiv.org/abs/2103.05419}{{\tt 2103.05419}}.

\bibitem{Catani:2020kkl}
S.~Catani, S.~Devoto, M.~Grazzini, S.~Kallweit and J.~Mazzitelli,
  \emph{{Bottom-quark production at hadron colliders: fully differential
  predictions in NNLO QCD}},
  \href{http://dx.doi.org/10.1007/JHEP03(2021)029}{\emph{JHEP} {\bf 03} (2021)
  029}, [\href{http://arxiv.org/abs/2010.11906}{{\tt 2010.11906}}].

\bibitem{Bai:2022jcs}
W.~Bai, M.~V. Diwan, M.~V. Garzelli, Y.~S. Jeong, K.~Kumar and M.~H. Reno,
  \emph{{Prompt electron and tau neutrinos and antineutrinos in the forward
  region at the LHC}},  in \emph{{2022 Snowmass Summer Study}}, 3, 2022.
\newblock \href{http://arxiv.org/abs/2203.07212}{{\tt 2203.07212}}.

\end{thebibliography}\endgroup

\end{document}